\documentclass[%
 aip,
 amsmath,amssymb,
 reprint,%
]{revtex4-1}

\usepackage{graphicx}
\usepackage{dcolumn}
\usepackage{bm}

\usepackage[utf8]{inputenc}
\usepackage[T1]{fontenc}
\usepackage{mathptmx}
\usepackage{etoolbox}
\usepackage{graphicx}
\usepackage{xcolor}
\usepackage{epstopdf, epsfig}
\usepackage{float}

\usepackage{caption}
\usepackage{subcaption}

\makeatletter
\def\@email#1#2{%
 \endgroup
 \patchcmd{\titleblock@produce}
  {\frontmatter@RRAPformat}
  {\frontmatter@RRAPformat{\produce@RRAP{*#1\href{mailto:#2}{#2}}}\frontmatter@RRAPformat}
  {}{}
}%
\makeatother
\begin{document}

\preprint{AIP/123-QED}

\title[Richtmyer-Meshkov instability with destabilising-stabilsing non-Newtonian effects]{Richtmyer-Meshkov Instability at high Mach Number: \\ Non-Newtonian Effects}
\author{U. Rana}
\affiliation{Department of Chemical Engineering, Imperial College London, SW7 2AZ, UK}
\author{T. Abadie}
\affiliation{School of Chemical Engineering, University of Birmingham, Birmingham B15 2TT, UK}
\author{D. Chapman}
\author{N. Joiner}
\affiliation{First Light Fusion Ltd., Oxford Pioneer Park,  Yarnton, Kidlington OX5 1QU, UK}
\author{O. K. Matar}
 \email{o.matar@imperial.ac.uk.}
\affiliation{Department of Chemical Engineering, Imperial College London, SW7 2AZ, UK}


\date{\today}

\begin{abstract}
The Richtmyer-Meshkov instability (RMI) occurs when a shock wave passes through an interface between fluids of different densities, a phenomenon prevalent in a variety of scenarios including supersonic combustion, supernovae, and inertial confinement fusion.
In the most advanced current numerical modelling of RMI, a multitude of secondary physical phenomena are typically neglected that may crucially change in silico predictions. In this study, we investigate the effects of shear-thinning behaviour of a fluid on the RMI at negative Atwood numbers via numerical simulations. A parametric study is carried out over a wide range of Atwood and Mach numbers that probes the flow dynamics following the impact on the interface of the initial shock wave and subsequent, reflected shocks. We demonstrate agreement between our numerical results and analytical predictions, which are valid during the early stages of the flow, and examine the effect of the system parameters on the vorticity distribution near the interface. We also carry out an analysis of the rate of vorticity production and dissipation budget which pinpoints the physical mechanisms leading to instability due to the initial and reflected shocks. Our findings indicate that the shear-thinning effects have a significant impact on instability growth and the development of secondary instabilities, which manifest themselves through the formation of Kelvin-Helmholtz waves. Specifically, we demonstrate that these effects influence vorticity generation and damping, which, in turn, affect the RMI growth. These insights have important implications for a range of applications, including inertial confinement fusion and bubble collapse within non-Newtonian materials. 
\end{abstract}

\maketitle


\section{Introduction}

The Richtmyer-Meshkov instability (RMI) results from the misalignment of pressure and density gradients at the interface of two materials with different densities when a shock wave passes through. This instability generates vorticity, i.e., a rotational eddy in the local fluid flow, which leads to the growth of interface perturbations. Immediately following the passage of the shock, in the linear growth phase, the interface perturbations grow quickly, driven by their initial amplitude, wave number and Atwood number, whilst roughly maintaining their shape. Later on, however, the perturbations grow nonlinearly and couple to additional hydrodynamic instabilities, most notably the Kelvin-Helmholtz instability (KHI), resulting in the onset and development of turbulence. The Atwood number $At$,  a dimensionless parameter that represents the ratio of the density difference to the sum of the densities of two fluids, significantly influences the progression of the RMI. This parameter quantifies the density contrast between the interacting fluids, and can be experimentally adjusted to access different growth regimes.

The RMI was first proposed by Richtmyer \cite{richtmyer_taylor_1960}, and the theory was later confirmed by the shock-tube experiments of Meshkov \cite{meshkov_instability_1969} and the numerical simulation of Meyer and Blewett \cite{Meyer_Blewett}. Vandenboomgaerde et al. \cite{Vandenboomgaerde} derived a general formula for the growth rate of the RMI within the framework of the impulsive model. This formula enables the prediction of the growth rate in both `heavy-to-light' and `light-to-heavy' configurations. The RMI has important implications for a wide range of natural and engineering problems, including supersonic combustion, supernovae, and inertial confinement fusion \cite{ICF_Supernova_Combustion}. For example, the occurrence of RMIs in supersonic combustion triggers better mixing and thus enhanced combustion efficiency \cite{ICF_Supernova_Combustion}. In supernova explosions \cite{ICF_Supernova_Combustion}, RMIs break the spherical symmetry of the progenitor star, and create conditions for the synthesis of heavy elements. In inertial confinement fusion, RMIs initiate mixing between the hotter and colder layers, resulting in a reduction of temperature that inhibits the thermonuclear burn \cite{ICF_Supernova_Combustion}.

The impulsive model proposed by Richtmyer \cite{richtmyer_taylor_1960} only analyzed the light-to-heavy flow configuration, where the shock travels in the direction of the denser fluid. However, Meshkov \cite{meshkov_instability_1969} ran shock-tube experiments with both flow configurations, suggesting that the impulsive model could be applied for both light-to-heavy and heavy-to-light flow configurations. The Richtmyer impulsive model, grounded in a linear framework, posits linear growth for the perturbation amplitude, reflecting its reliance on linear theory assumptions. This model is only valid for early times when the perturbation amplitude is considerably smaller than its wavelength. As the amplitude exceeds the wavelength, nonlinear effects become significant rendering the impulsive model invalid. In response to these limitations, Yang et al. \cite{Yang_Zhang_Sharpe} introduced a refined version of Richtmyer's model to incorporate compressibility effects, which provides a more comprehensive description of perturbation evolution under shock compression. 

\begin{figure*}
  \includegraphics[keepaspectratio=true,scale=0.45]{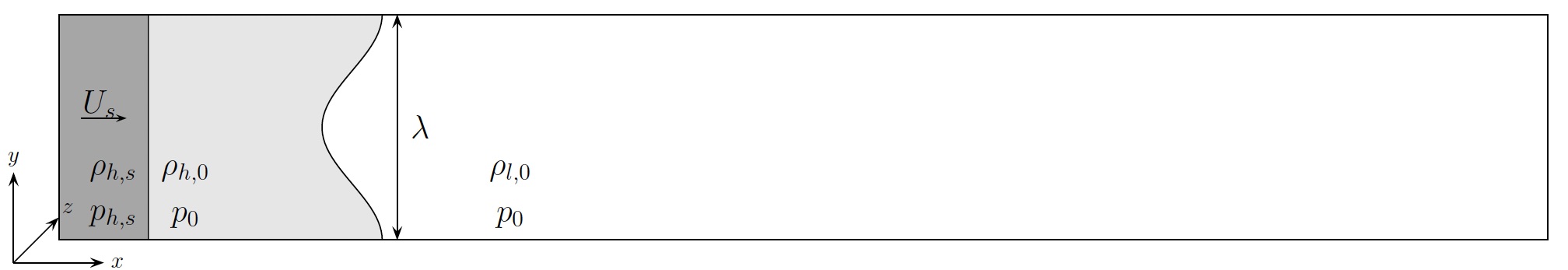}
\caption{Schematic of the flow configuration in which a perturbation of wavelength $\lambda$ is applied to an interface separating two phases at pressure $p_0$ of initial densities $\rho_{h,0}$ and $\rho_{l,0}$ shown by the light grey and white regions, respectively, where $\rho_{h,0} > \rho_{l,0}$. A shock is initiated in the heavier phase (dark grey region) with density $\rho_{h,s}$ and pressure $p_{h,s}$, which travels in the direction of the lighter phase at constant speed $U_s$.}
\label{fig:RMsketch1}
\end{figure*}
%
%

Additionally, the Zhang and Sohn \cite{Zhang_Sohn_Nonlinear} model offers an alternative approach to simulate the late-time behaviour of the perturbation, distinguished by its utilization of Padé approximants. However, this model has its limitation as well, notably its tendency to underestimate the late-time average growth rate of the perturbation. To address this issue, Sadot et al. \cite{Sadot_Study} introduced a model developed using data from shock-tube experiments with air and sulfur hexafluoride. Another significant model for the late-time stage is the Dimonte and Schneider \cite{dimonte_density_2000} model, which, unlike that of Richtmyer \cite{richtmyer_taylor_1960}, uses a drag coefficient to account for the reduced growth rate at the late-time stage. The predictions obtained from both these  
models will be compared to those provided by our numerical computations. The Nova laser experiments by Dimonte and Remington \cite{dimonte_richtmyer-meshkov_1993}, Dimonte et al. \cite{dimonte_richtmyermeshkov_1996} and Farley et al. \cite{,farley_high_1999} showed reduced growth rates as compared to the impulsive model, particularly at high Mach numbers. Building upon the foundational work in this field, Thornber et al. \cite{Thornber_1} studied the critical influence of three-dimensional multi-mode initial conditions on the growth of turbulent mixing layers in the RMI. Their use of high-resolution, large-eddy simulations underscored the significant impact of initial conditions on the evolution of these layers. 

Further advancing our understanding of the RMI, Thornber et al. \cite{Thornber_2} also explored the reshock phase, i.e., the period where the initial shock wave is reflected back towards the evolving interface, leading to further, more complex shock-interface interactions. Through high-order accurate Implicit Large-Eddy-Simulations, they extended theoretical models to accurately predict the behaviour of re-shocked mixing layers, offering invaluable insights for scenarios involving broadband initial perturbations.
Numerical simulations have also been used to study the effects of viscosity on the RMI. Mikaelian \cite{mikaelian_effect_1993} proposed a theoretical model in which viscosity dampens the growth of the instability, while  \citeauthor{carles_viscous_2001}\cite{carles_viscous_2001} used direct numerical simulations as a benchmark to derive a theoretical model for the RMI of viscous flows. \citeauthor{movahed_solution-adaptive_2013}\cite{movahed_solution-adaptive_2013} performed numerical simulations using a solution-adaptive method, which revealed that viscosity dampens the growth of the instability in agreement with the work of \citeauthor{mikaelian_effect_1993}\cite{mikaelian_effect_1993}. The study of vorticity evolution within RMI has also undergone considerable progress with \citeauthor{Kotelnikov_ReacceleratedInterfaces}\cite{Kotelnikov_ReacceleratedInterfaces} shedding light on the dynamics of secondary baroclinic vorticity deposition.

Building upon this work, \citeauthor{peng_vortex-accelerated_2003} \cite{peng_vortex-accelerated_2003} investigated the generation of secondary baroclinic vorticity accelerated by vortices. Their study focused on the effects of this phenomenon on the flow evolution and features of the RMI at various stages, which include amplitude, neck, span, and stream. These aspects are integral to the validation studies presented later in this paper.
Baroclinic vorticity has traditionally been viewed as the exclusive source of vorticity evolution, often overlooking the impact of viscosity. To investigate the effects of viscosity \citeauthor{Wang_VortexBreakdown}\cite{Wang_VortexBreakdown} studied a shock-bubble-interaction problem. They introduced a dimensionless number for the baroclinic-to-viscous source ratio in the vorticity evolution. \citeauthor{liu_contribution_2020}\cite{liu_contribution_2020} demonstrated that viscosity gradients, which emerge due to temperature variations within strong shock waves, make a positive contribution to the generation of vorticity, particularly at high Mach numbers.

Despite the extensive research that has been conducted on RMI and associated  phenomena, there are still many unresolved questions, particularly regarding the non-Newtonian effects on the instability. While previous studies have mainly focused on the effects of viscosity on the instability, non-Newtonian fluids exhibit additional complexities, such as shear-thinning and elasticity, which could have a significant impact on the instability growth. This gap in the literature will be addressed in the present work. We investigate the influence of non-Newtonian effects on the generation and damping of vorticity that influence the dynamics of the RMI. By incorporating these effects into our numerical simulations, we aim to provide insights into the dynamics associated with shock impact wherein at least one of the fluids is a non-Newtonian material; this can be relevant to nuclear fusion-type applications such as in the shock-driven fusion application of \citeauthor{Derentowicz_ThermonuclearFusionNeutrons}\cite{Derentowicz_ThermonuclearFusionNeutrons}. 

In this study, we use numerical simulations to investigate how shear-thinning properties in the more dense phase (with the lighter one maintained as Newtonian) impact the development of vorticity and subsequently influence the progression of both the RMI and KHI. The non-Newtonian fluid is modelled as a Bird-Carreau-Yasuda fluid; elasto/plastic effects are neglected in this study. We  carry out a parametric study to examine the effect of the Mach and Atwood numbers on the emergent dynamics and provide a breakdown of all the terms that contribute to vorticity production and damping to pinpoint the destabilising mechanisms unambiguously over the various stages of the flow; these include the dynamics following the passage of the initial shock as well as subsequent reflected shocks. The results are compared with those obtained when both fluids are either inviscid or Newtonian.  

The remainder of this paper is organised as follows. In Sec. 2, we provide details of the problem formulation and the numerical methods used to carry out the computations; a breakdown of the terms contributing to the rate of change of vorticity is also included in this section. In Sec. 3, we present the numerical results of our parameteric study, which include comparisons with  theoretical predictions, and a discussion of the destabilising mechanisms highlighting the role of shear-thinning. Finally, we provide concluding remarks in Sec. 4. 

\vspace{-0.15in}


\section{Problem formulation}

We consider an interface perturbation of wavelength $\lambda$ separating two fluids of different densities 
corresponding to a more dense and a less dense fluid, to which we  herein refer as `heavy' and `light', respectively. 
The initial densities and viscosities of the light and heavy fluids are $\rho_{l,0}$ and $\rho_{h,0}$, and $\mu_{l,0}$ and $\mu_{h,0}$, respectively, and both fluids are at a pressure $p_0$. 
As shown in Figure \ref{fig:RMsketch1}, a shock wave resulting from a jump in pressure and density within the heavy phase, propagates towards the lighter phase and impacts the interface, leading to the development of an RMI on the interface. 

The pressure and density ratios associated with the shock are denoted by $p_{h,s}/p_{0}$ and $\rho_{h,s}/\rho_{h,0}$, where the subscripts $s$ and $0$ designate the shocked left state (away from the interface) and the right state (close to the interface), respectively; interfacial tension effects are neglected. This assumption is motivated by practical considerations, as seen in the interactions at plastic-gas interfaces in ICF experiments such as those conducted by Derentowicz et al. \cite{Derentowicz_ThermonuclearFusionNeutrons}. The gas phase rheology is considered to have minimal impact on the dynamics as compared to that of the plastic phase, where non-Newtonian effects are more pronounced and relevant. We consider here that the heavy phase can be either inviscid or viscous but with a non-Newtonian rheology featuring a shear rate-dependent apparent viscosity. 
In contrast, the lighter phase is either considered to be inviscid or Newtonian, with a constant viscosity.

\subsection{Mathematical Model}
The shock and interface dynamics are governed by the mass conservation of each phase and the one-fluid formulation of the compressible momentum, and energy equations, respectively expressed by
 \begin{equation}
\centering
\frac{\partial  {\rho_h}}{\partial {t}} + \boldsymbol{\nabla} \cdot ( {\rho_h} {\mathbf{u}}) = 0,
\label{glg:continuity}
\end{equation}
 \begin{equation}
\centering
\frac{\partial  {\rho_l}}{\partial {t}} + \boldsymbol{\nabla} \cdot ( {\rho_l} {\mathbf{u}}) = 0,
\label{glg:continuity2}
\end{equation}
 \begin{eqnarray}
\centering
\frac{\partial {{\rho} \mathbf{u}}}{\partial {t}} 
 +   \boldsymbol{\nabla} \cdot \left({\rho}{\mathbf{u}} {\mathbf{u}}\right) 
 = 
 &-& \boldsymbol{\nabla} {p} + \rho \mathbf{g}\nonumber\\
 &+& \boldsymbol{\nabla} \cdot \left( \mu \left[ {\boldsymbol{\nabla}\mathbf{u}} + \left(\boldsymbol{\nabla}{\mathbf{u}}\right)^T\right]\right)\nonumber\\ 
 &-& \dfrac{2}{3} \boldsymbol{\nabla}\cdot\left(\mu\left[\left(\boldsymbol{\nabla} \cdot {\mathbf{u}}\right) \mathbf{I} \right] \right),
\label{glg:momentum}
\end{eqnarray}
\begin{equation}
\frac{\partial (\rho e)}{\partial t} + \boldsymbol{\nabla} \cdot [( \rho e + p) \mathbf{u}] = \boldsymbol{\nabla} \cdot (\mu \boldsymbol{\nabla} \cdot \mathbf{u} \mathbf{u}) + \boldsymbol{\nabla} \cdot ({\kappa} \boldsymbol{\nabla} {T}) - \rho \mathbf{g} \cdot \mathbf{u},
\label{glg:energy}
\end{equation}
where $t$, $\rho_h$, $\rho_l$, $\rho$, $\mu$, $\mathbf{u}$, $p$, $\mathbf{g}$, $e$, $T$, and $\kappa$ correspond to the time, heavy fluid density, light fluid density, local mixture density and viscosity, mixture velocity, mixture pressure, acceleration due to gravity, mixture total specific energy, temperature, and thermal conductivity, respectively. Using the wavelength of the interface perturbation $\lambda$ as the length scale and the shock speed $U_S$ as the velocity scale, the corresponding time scale is $t_S = \lambda/U_S$; in addition, we use the dynamic pressure scale $\rho_{h,0} U_S^2$, the energy per unit mass scale $U_S^2$, and the density and viscosity scales as $\rho_{h,0}$ and $\mu_{h,0}$, respectively, the above governing equations are written in dimensionless form as follows:
 \begin{equation}
\centering
\boldsymbol{\tilde{\nabla}} \cdot \tilde{\mathbf{u}} = - \dfrac{1}{\tilde{\rho_h}} \dfrac{D \tilde{\rho_h}}{D \tilde{t}} = - \dfrac{Ma^2}{\tilde{\rho_h}} \dfrac{D \tilde{p_h}}{D \tilde{t}},
\label{glg:continuity_nonDim}
\end{equation}
 \begin{equation}
\centering
\boldsymbol{\tilde{\nabla}} \cdot \tilde{\mathbf{u}} = - \dfrac{1}{\tilde{\rho_l}} \dfrac{D \tilde{\rho_l}}{D \tilde{t}} = - \dfrac{C_h^2}{{C_l^2}} \dfrac{Ma^2}{\tilde{\rho_l}} \dfrac{D \tilde{p_l}}{D \tilde{t}},
\label{glg:continuity_nonDim2}
\end{equation}
 \begin{eqnarray}
\frac{\partial {\tilde{\rho} \mathbf{\tilde{u}}}}{\partial {\tilde{t}}} 
 +  \boldsymbol{\tilde{\nabla}} \cdot \left(\tilde{\rho}\mathbf{\tilde{u}} \mathbf{\tilde{u}} \right)
 = 
 &-& \boldsymbol{\tilde{\nabla}} \tilde{p} 
 + \dfrac{1}{Fr^2}\tilde{\rho} \mathbf{\tilde{g}}\nonumber\\
 &+& \dfrac{1}{Re}\boldsymbol{\tilde{\nabla}} \cdot \left( \tilde{\mu} \left[ {\boldsymbol{\tilde{\nabla}}\mathbf{\tilde{u}}} + \left(\boldsymbol{\tilde{\nabla}}{\mathbf{\tilde{u}}}\right)^T\right]\right)\nonumber\\ 
 &-& \dfrac{2}{3~Re} \boldsymbol{\tilde{\nabla}}\cdot \left(\tilde{\mu}\left[\left(\boldsymbol{\tilde{\nabla}} \cdot {\mathbf{\tilde{u}}}\right) \mathbf{I} \right] \right) 
  \ ,
\label{glg:momentum_nonDim}
\end{eqnarray}

\begin{equation}
\begin{split}
\centering
\frac{\partial (\tilde{\rho} \tilde{e})}{\partial \tilde{t}} + \boldsymbol{\tilde{\nabla}} \cdot [(\tilde{\rho} \tilde{e} + \tilde{p}) \tilde{\mathbf{u}}]  = & 
 \frac{1}{Re} \boldsymbol{\tilde{\nabla}} \cdot (\tilde{\mu} \boldsymbol{\tilde{\nabla}} \cdot \tilde{\mathbf{u}} \tilde{\mathbf{u}}) + \frac{1}{ReBr} \boldsymbol{\tilde{\nabla}} \cdot (\tilde{\kappa} \boldsymbol{\tilde{\nabla}} \tilde{T}) \\
&- \frac{1}{Fr^2} \tilde{\rho} \tilde{\mathbf{g}} \cdot \tilde{\mathbf{u}},
\end{split}
\label{glg:energy_nonDim}
\end{equation}
where we have scaled $T$ on $T_{h,o}-T_{l,o}$, the initial temperature difference between the heavy and light phases. 
Using  mass conservation, Eq. (\ref{glg:momentum_nonDim}) can also be re-written as
 \begin{equation}
   \begin{split}
\centering
\frac{\partial {\tilde{\rho} \mathbf{\tilde{u}}}}{\partial {\tilde{t}}} 
 & +  \boldsymbol{\tilde{\nabla}} \cdot \left(\tilde{\rho}\mathbf{\tilde{u}} \mathbf{\tilde{u}} \right) 
  = 
   - \boldsymbol{\tilde{\nabla}} \tilde{p} \\
& 
 + \dfrac{1}{Re}\boldsymbol{\tilde{\nabla}} \cdot \left( \tilde{\mu} \left[ {\boldsymbol{\tilde{\nabla}}\mathbf{\tilde{u}}} + \left(\boldsymbol{\tilde{\nabla}}{\mathbf{\tilde{u}}}\right)^T \right] \right) 
 + \dfrac{Ma^2}{Re}\boldsymbol{\tilde{\nabla}} \cdot \left( \dfrac{2 \, \tilde{\mu}}{3 \, \tilde{\rho}} \ \dfrac{D \tilde{p}}{D \tilde{t}} \ \mathbf{I} \right) \\
 &
 + \dfrac{1}{Fr^2}\tilde{\rho} \mathbf{\tilde{g}}. \ 
 \end{split}
\label{glg:momentum_nonDim2}
\end{equation}

In the above equations,  $C_h$ and $C_l$ are the speeds of sound in the heavy and light phases, and $Ma = U_S/C_h$, $Re = \rho_{h,0} \, U_S \, \lambda / \mu_{h,0}$, and $Fr = U_S/\sqrt{g \, \lambda}$ stand for the Mach, Reynolds, and Froude numbers, respectively. Furthermore, $Br=\mu_{h,0}U^2_s/\kappa (T_{h,0}-T_{l,0})$ is the Brinkman number, and $ReBr$, which characterises the conduction term in Eq. (\ref{glg:energy_nonDim}) can be re-expressed 
as $RePrEc=PeEc$ where $Pe$, $Pr=C_{p \, h,0} \mu_{h,0}/\kappa$, and $Ec=U^2_s/C_{p \, h,0} (T_{h,0}-T_{l,0})$ correspond to the P\'eclet, Prandtl, and Eckert numbers, respectively, wherein $C_{p \, h,0}$ is the specific heat capacity of the heavy phase. 
In the context of melting plastic in ICF applications, such as those studied by \citeauthor{Derentowicz_ThermonuclearFusionNeutrons}\cite{Derentowicz_ThermonuclearFusionNeutrons}, the high Prandtl number characterises a regime where viscous diffusion is dominant over thermal diffusion, leading us to concentrate on the viscous effects while neglecting conductive heat flow.
Thus, in our study, we investigate cases where the influence of conduction is considered negligible corresponding to scenarios where the Péclet number is large, i.e., the \(Pe \to \infty\) limit. We also go one step further and consider the isothermal case to focus attention on the non-Newtonian effects on the RMI.  
%
%

Additionally, 
the Atwood number, $At$, is defined as the ratio of the difference of the initial densities of the two phases to their sum, and expressed by: 
\begin{equation}
At = \frac{{\rho_{l,0} - \rho_{h,0}}}{{\rho_{l,0} + \rho_{h,0}}}.
\label{glg:AtwoodNumber}
\end{equation}
According to this definition, a negative (positive) $At$ suggests the propagation of the shock from the heavier (lighter) to the lighter (heavier) fluid. We will focus below on flows parameterised by negative $At$. This study is primarily motivated by the impact of Non-Newtonian effects on RMI in ICF, as exemplified by the work of \citeauthor{Derentowicz_ThermonuclearFusionNeutrons}\cite{Derentowicz_ThermonuclearFusionNeutrons} on thermonuclear fusion neutrons, which operates within the negative $At$ range. 

Equations (\ref{glg:continuity_nonDim})-(\ref{glg:energy_nonDim})
are solved using the open-source software \textit{blastFOAM} \cite{blastfoam}, which has been developed to model highly-compressible flows. \textit{blastFOAM} also employs the volume-of-fluid (VOF) method to capture the interface by advecting the volume fraction of the heavy fluid, $\alpha$, according to
\begin{equation}
\frac{\partial \alpha}{\partial t} +  \boldsymbol{\tilde{\nabla}} \cdot (\alpha \mathbf{\tilde{u}} ) = \alpha \boldsymbol{\tilde{\nabla}} \cdot \mathbf{\tilde{u}},
\end{equation}
where $\alpha=0$ and $\alpha=1$ for cells occupied by the light and heavy fluids, respectively; the interface resides in cells wherein 
$0 < \alpha < 1$. 
Within the VOF framework, $\alpha$ is used to calculate the local density and viscosity as follows: 
\begin{equation}
\tilde{\rho} = \alpha \tilde{\rho}_h+(1-\alpha)\tilde{\rho}_l,
\end{equation}
\begin{equation}
\tilde{\mu} = \alpha\tilde{\mu}_h+(1-\alpha)\tilde{\mu}_l.
\label{eq:mu_equation}
\end{equation}

We employ the stiffened gas equation-of-state to calculate the pressure in each phase $\tilde{p}_i$ and the mixture pressure $\tilde{p}$ in Mie-Gruneisen form \cite{Zheng_2011},
\begin{equation}
\centering
\tilde{p}_i = (\gamma_i - 1)\tilde{\rho}_i\tilde{e}-\gamma_i \tilde{a}_{i},
\label{glg:EOS}
\end{equation}
\begin{equation}
\centering
\tilde{p} = \dfrac{\tilde{\rho}\tilde{e} - \left( \frac{\alpha \gamma_h \tilde{a}_{h}}{\gamma_h-1} + \frac{\left(1-\alpha \right) \gamma_l \tilde{a}_{l}}{\gamma_l-1} \right)}{\frac{\alpha}{\gamma_h-1} + \frac{1-\alpha}{\gamma_l-1}},
\label{glg:pMieGruneisen}
\end{equation}
where $\gamma_i$ and $\tilde{a}_i$ are the compressibility and the reference pressures, respectively, which are (constant) material properties in each phase. 
The non-Newtonian behaviour of the heavier fluid is modelled using the Bird-Carreau-Yasuda (BCY) \cite{BCY} model in which we take into account 
the shear-thinning 
dependence:
\begin{equation}
\centering
\tilde{\mu}_h(\tilde{\dot{\gamma}}) = \tilde{\mu}_{h,\infty}+ \frac{\tilde{\mu}_{h,0} - \tilde{\mu}_{h,\infty}}{[1+(\tilde{\tau} \tilde{\dot{\gamma}})^{2}]^{(1-n)/2}};
\label{glg:BCY_Model}
\end{equation}
here, $\tilde{\mu}_{h,\infty}$ and $\tilde{\mu}_{h,0}$ represent the dimensionless infinite and zero-shear rate viscosities, respectively, $\tilde{\dot{\gamma}}$ is the dimensionless shear rate, the flow index is denoted by $n$, while $\tilde{\tau}$ represents the ratio of the relaxation time, $\tau$, to the flow time scale, $t_S$. The EOS parameters are given in Table 
\ref{tab:Rheological_Paramters} as are the rheological properties used for the two phases. 

We utilize the advantages of \textit{blastFOAM}'s explicit solution approach for conservative variables that offers both high accuracy and computational efficiency. We use a third-order accurate Runge-Kutta time integration method along with the SFCD interpolation scheme and HLLC flux scheme \cite{Toro}. 
We conduct simulations of the RMI within a rectangular domain of size $10\times 1$, where unity is the initial wavelength of RMI that is equal to the width of the domain. The shock is initiated in the heavier fluid, on the left side of the interface, as shown in Figure 1, and propagates towards the interface. 

In scenarios similar to ICF, the shock travels from the heavier to the lighter fluid \cite{chapmann,Derentowicz_ThermonuclearFusionNeutrons}. Given this, our simulation is designed to emulate conditions relevant to ICF thereby closely reflecting the RMI phenomena we are aiming to study. The top and bottom boundaries are set to be symmetric to the shock axis. We use a zero-gradient boundary condition for density, volume fraction, and pressure at both the left and right boundaries. Moreover, a zero-gradient boundary condition is employed for velocity at the left boundary, while a zero velocity is imposed at the right boundary to simulate a reflected shock. 
%
%
%
%
%
The interface at the start of our simulations is modelled as 
$f(y) = \alpha \cos\left(\frac{2\pi y}{\lambda}\right)$ where $y$ is the vertical coordinate, as shown in Figure 1. 
Furthermore, to mitigate the influence of small-scale perturbations originating from the VOF method and the mesh, a diffusion-type layer is introduced \cite{peng_vortex-accelerated_2003}.

\begin{table}
  \caption{Dimensionless stiffened gas equation of state parameters which appear in equation (\ref{glg:EOS}). The initial densities are scaled on $\rho_{h,0}$. Rheological parameters for the light and heavy phases which appear in equations (\ref{glg:BCY_Model}). }
  \label{tab:Rheological_Paramters}
  \begin{ruledtabular}
  \begin{tabular}{ccccccc}
  \hline
      Phase & $\gamma_i$ & $\tilde{a}_i$ & $\mu_{0}$ & $\mu_{\infty}$ & $\tilde{\tau}$  & $n$  \\[3pt]
      \hline
      Light  & 4.0 & 0.74 & 6.2$\times 10^{-9}$ & 0.0 & 0.0 & 1.0  \\
      \hline
      Heavy  & 4.0 & 0.75 & 1.0 &2.5$\times 10^{-3}$& 8505 & 0.55  \\
      \hline
  \end{tabular}
  \end{ruledtabular}
  \end{table}
We investigate the impact of non-Newtonian effects on the RMI and vorticity evolution for Mach numbers in the range of $Ma = [2.5,10]$ and Atwood numbers in the range of $At = [-0.048,-0.66]$, respectively. Our focus is on studying how shear-thinning effects on the RMI and vorticity evolution vary with Mach numbers and Atwood numbers. 
 The negative $At$ values reflect the flow situations we will focus on exclusively in the present work, which are associated with shock propagation from the heavier to the lighter phase, typical of ICF applications. The results of these simulations will be compared with those generated for cases wherein both fluids are either inviscid or Newtonian to highlight the role played by the shear-thinning behaviour in the development of instabilities. A discussion of our numerical results is presented next. 

\begin{figure*}
    \begin{minipage}{0.45\textwidth}
        \includegraphics[width=\textwidth]{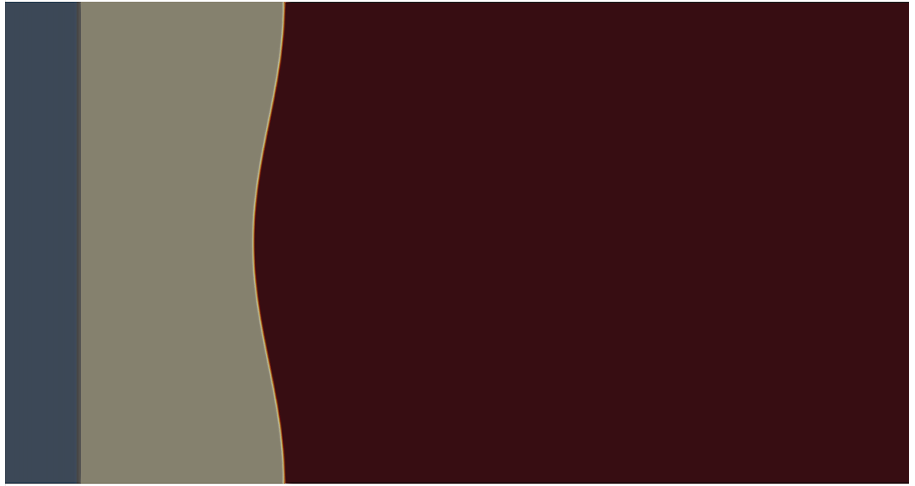}
        \subcaption{}
        \label{fig:stage_a}
    \end{minipage}%
   \hfill
    \begin{minipage}{0.45\textwidth}
        \includegraphics[width=\textwidth]{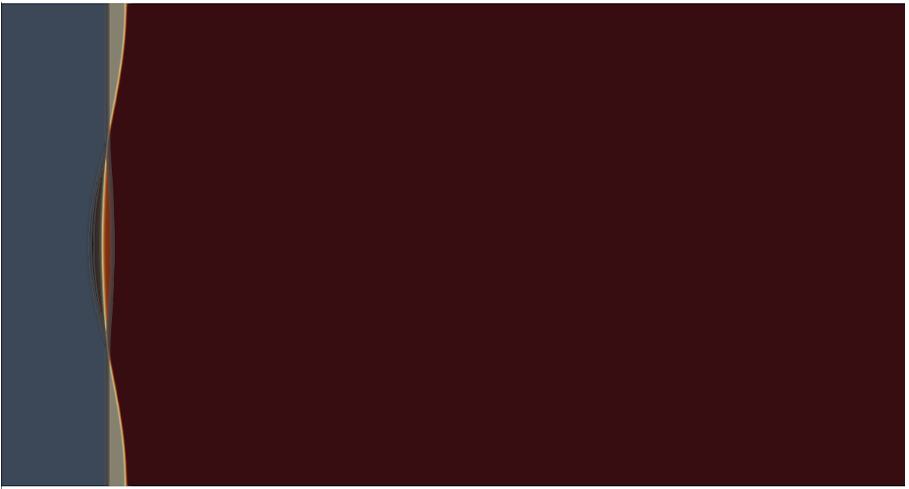}
        \subcaption{}
        \label{fig:stage_b}
   \end{minipage}%
    
    \vspace{0.5cm} 
    
    \begin{minipage}{0.45\textwidth}
        \includegraphics[width=\textwidth]{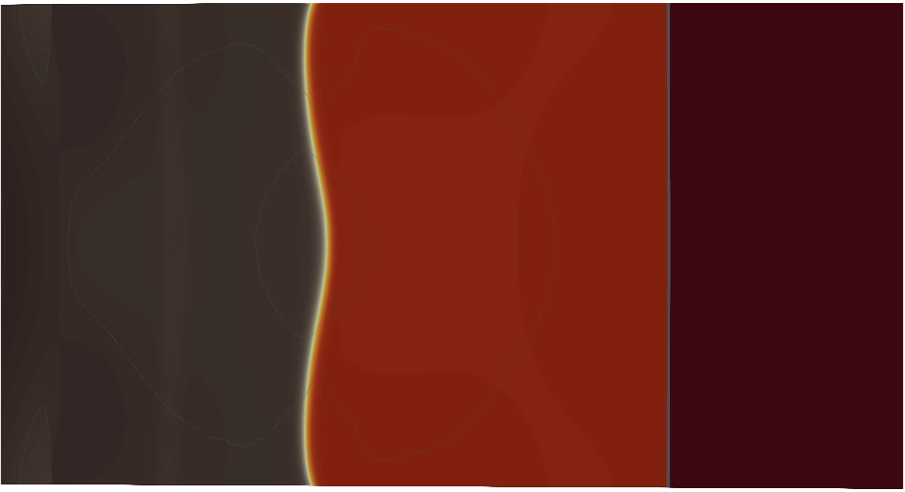}
        \subcaption{}
        \label{fig:stage_c}
    \end{minipage}%
    \hfill
    \begin{minipage}{0.45\textwidth}
        \includegraphics[width=\textwidth]{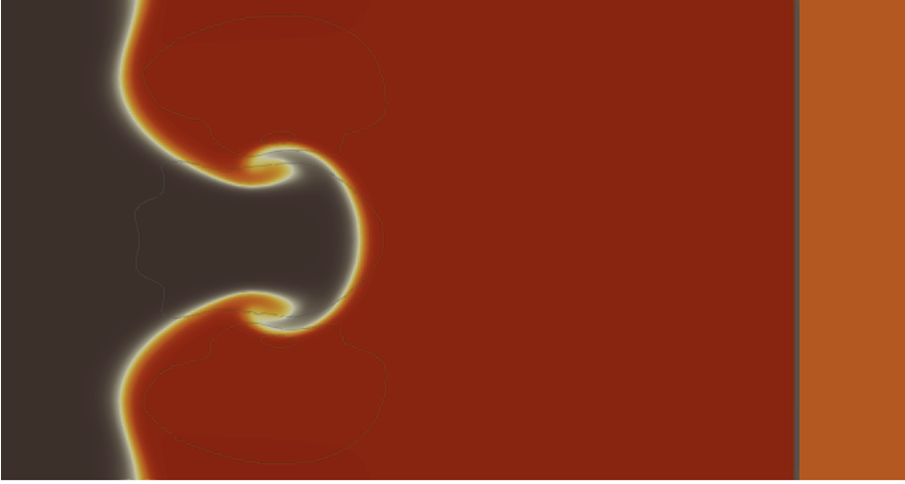}
        \subcaption{}
        \label{fig:stage_d}
    \end{minipage}%
    
    \vspace{0.5cm} 
    
    \begin{minipage}{0.45\textwidth}
        \centering
        \includegraphics[width=\textwidth]{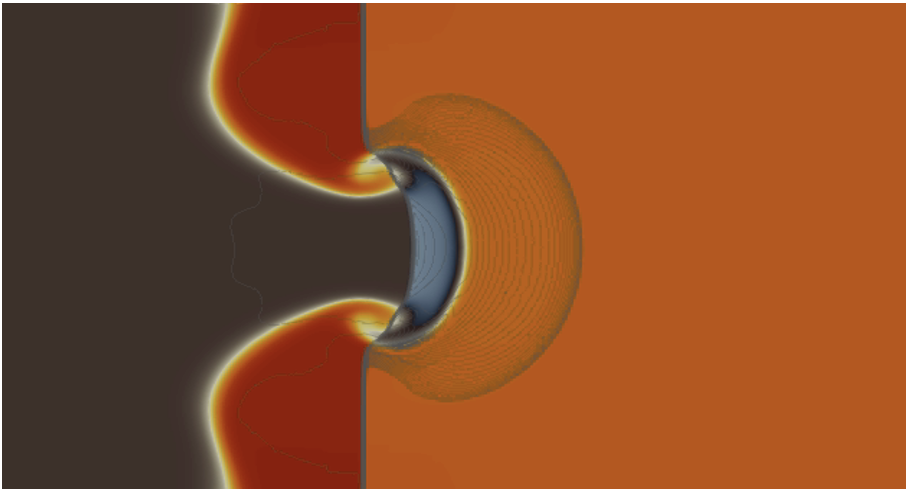}
        \subcaption{}
        \label{fig:stage_e}
    \end{minipage}%
    \hfill
    \begin{minipage}{0.45\textwidth}
        \centering
        \includegraphics[width=\textwidth]{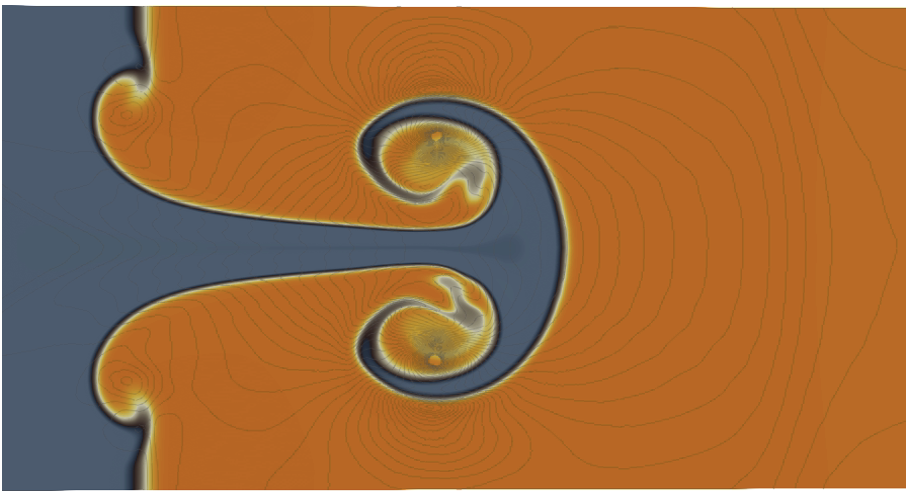}
        \subcaption{}
        \label{fig:stage_f}
    \end{minipage}%
    
    \vspace{0.5cm} 
    
    \begin{center}
        \includegraphics[width=0.7\textwidth]{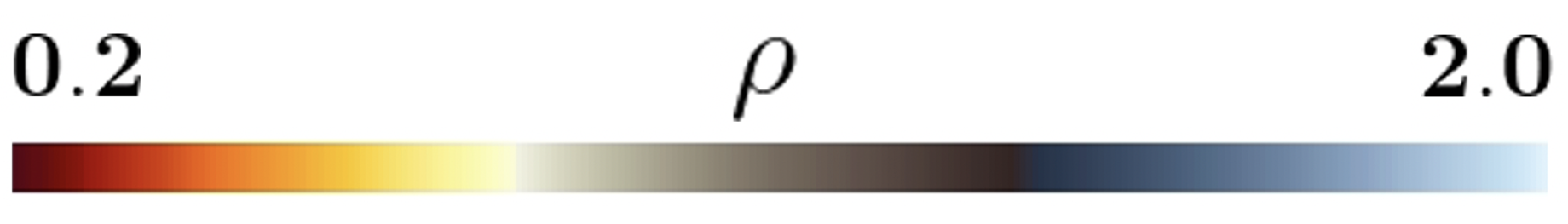}
    \end{center}
    \caption{Projections of the density field in the $x-y$ plane depicting the development of a RMI in the inviscid case, and a subsequent KHI, driven by the left-to-right propagation of a shock wave from the heavier towards the lighter phase for $t=$ 0, 0.06, 0.25, 1.35, 1.42, 1.80 shown in (a)-(f), respectively. The parameter values are $At=-0.66$, $Ma=5$, and $Fr=48571$; the remaining parameters are in Table \ref{tab:Rheological_Paramters}. 
    }
    \label{fig:RMI_stages}
\end{figure*}

\section{Results and Analysis}\label{sec:Results and Analysis}

\subsection{Perturbation growth and vorticity generation: inviscid and BCY fluids}
We begin by describing the development of the RMI at different stages for a situation wherein a shock propagates from the heavier to the lighter fluid (see Figure \ref{fig:stage_a}) characterised by $At = - 0.66$ and $Ma = 5$. The shock impacts the interface, as depicted in Figure \ref{fig:stage_b} which leads to the development of counter-clockwise vorticity and, in turn, the protrusion of the heavier phase into the lighter one (see Figure \ref{fig:stage_c}). The mechanisms giving rise to vorticity generation and damping, with a particular focus on the role of non-Newtonian effects, are discussed below. 
%
%
%
Following the propagation of the shock past the interface, the protrusion becomes elongated and the development of roll-up driven by a KHI is clearly evident, giving rise to a classic mushroom-like structure at the protrusion leading edge, as shown in 
Figure \ref{fig:stage_d}. 
The shock is reflected by 
the right boundary and propagates towards the interface from the lighter fluid side of the domain (see Figure \ref{fig:stage_d}). As shown in Figure \ref{fig:stage_e}, upon impact, the reflected shock is repelled by the interface due to the higher density of the heavier fluid, which induces the formation of shock waves that travel towards the upper, lower, and right boundaries. These waves are then reflected from these boundaries and interact once again with the interface, only to be reflected back, and the process is then repeated. As shown in Figure \ref{fig:stage_f}, the mushroom-like structure becomes larger, with a flattened leading edge; highly pronounced KHI-driven roll-up phenomena are also clearly evident in this figure. 

\begin{figure*}
 \begin{minipage}[b]{\textwidth}
 \includegraphics[scale=0.73]{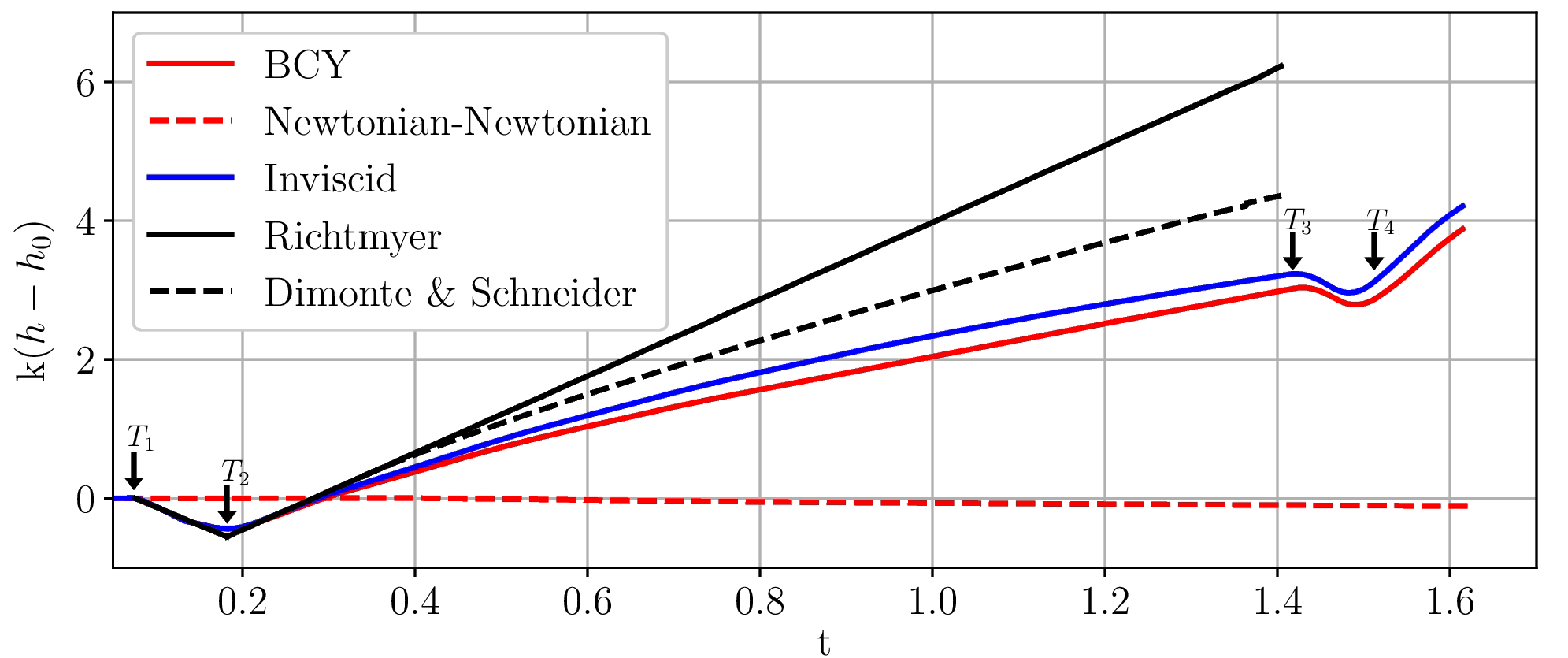}\\
 \subcaption{}
 \end{minipage}
        \begin{minipage}[b]{0.24\textwidth}
\includegraphics[width=\textwidth]{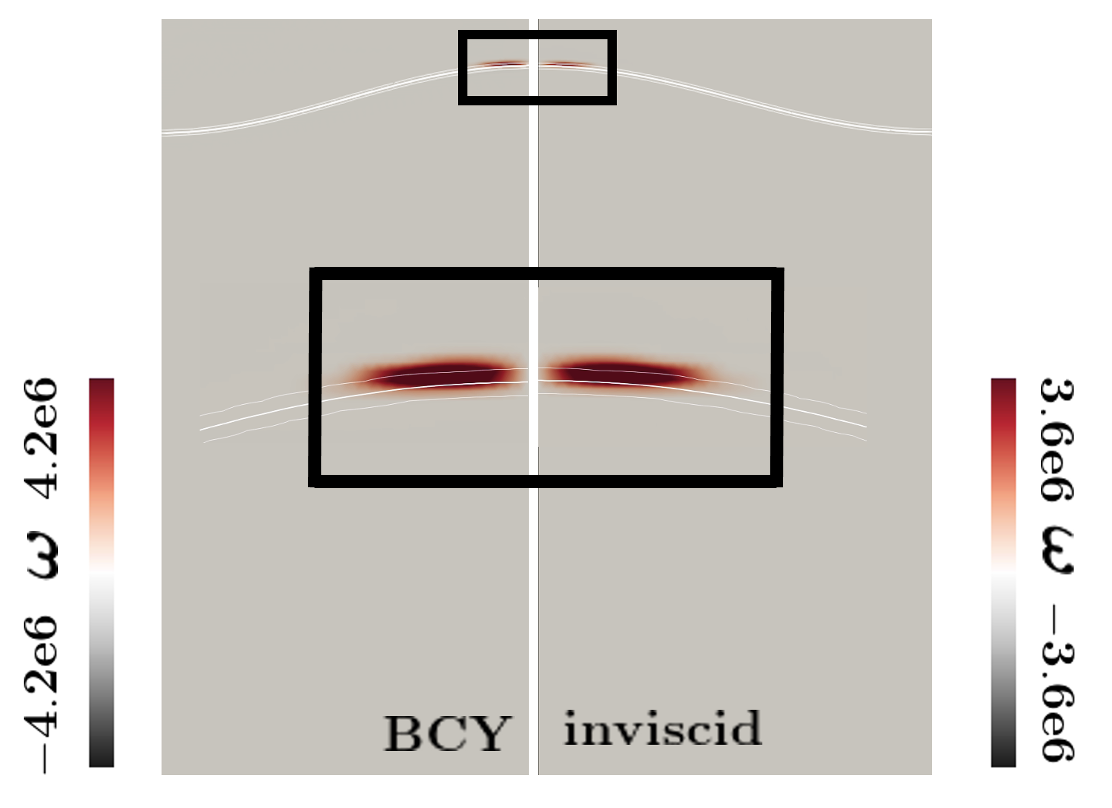}
            \subcaption{}
            \label{fig:mean and std of net14}
        \end{minipage}
        \hfill
        \begin{minipage}[b]{0.24\textwidth}  
\includegraphics[width=\textwidth]{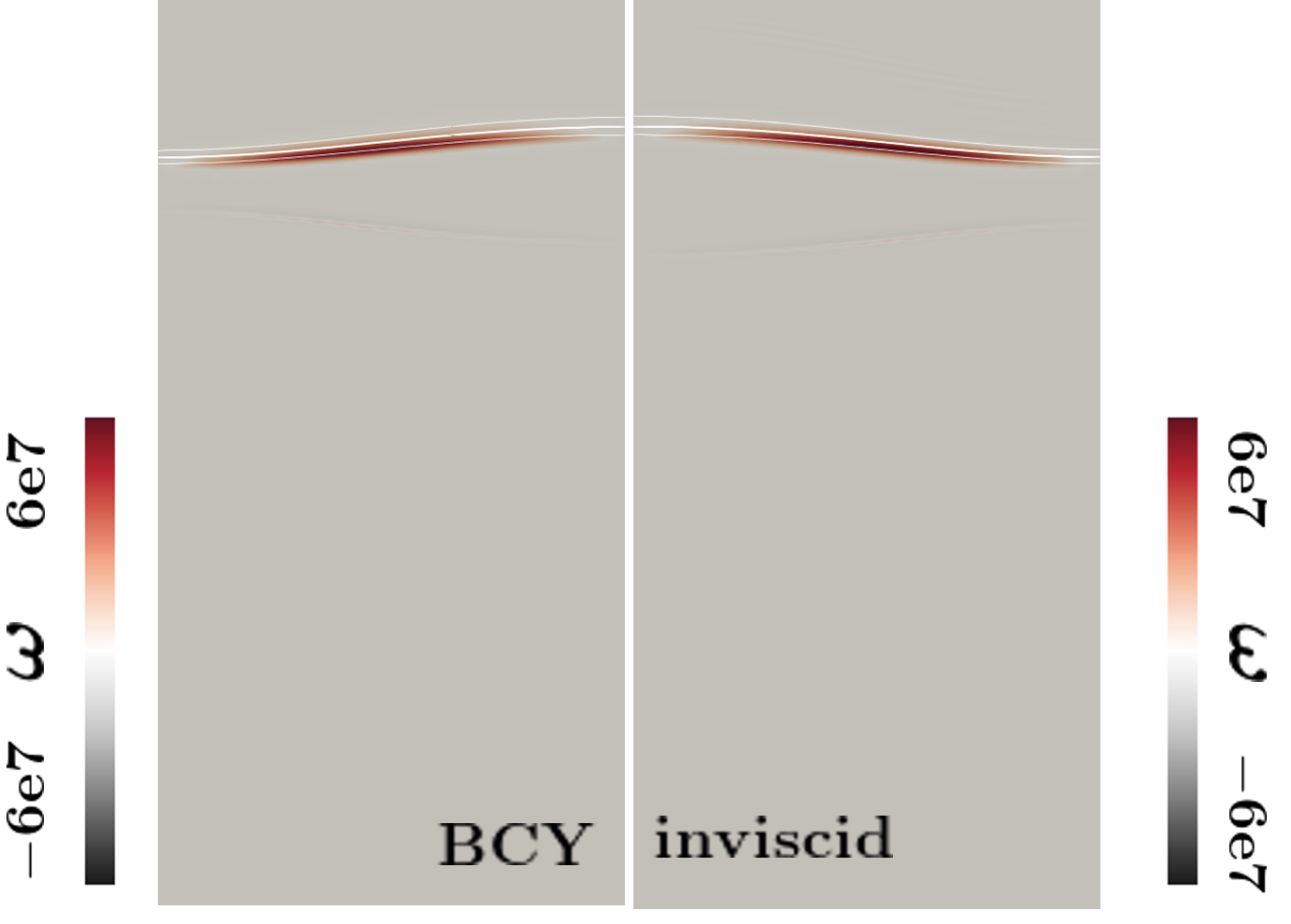}
            \subcaption{}
            \label{fig:mean and std of net24}
        \end{minipage}
        \begin{minipage}[b]{0.24\textwidth}   
\includegraphics[width=\textwidth]{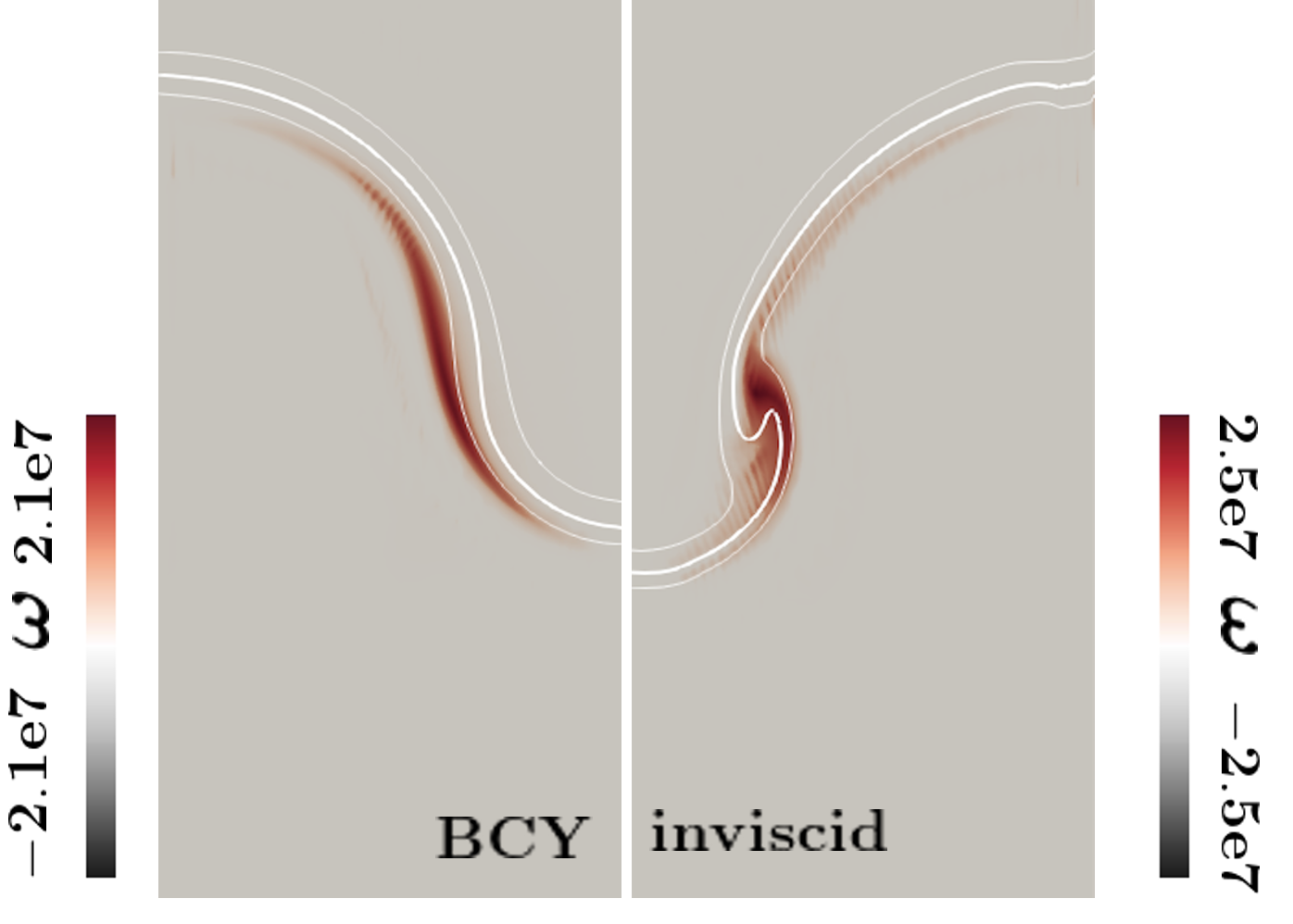}
            \subcaption{}
            \label{fig:mean and std of net34}
        \end{minipage}
        \hfill
        \begin{minipage}[b]{0.24\textwidth}   
\includegraphics[width=\textwidth]{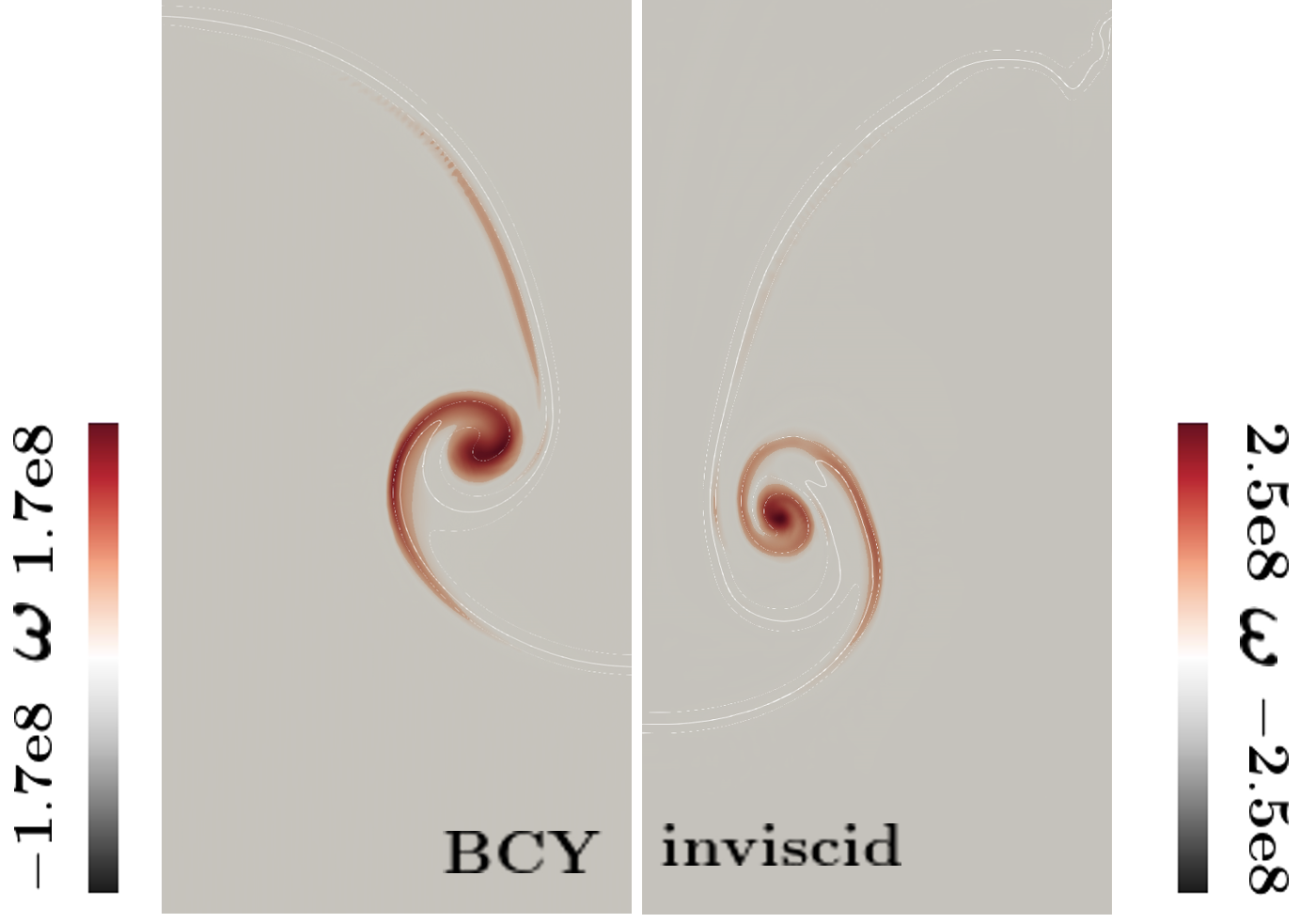}
            \subcaption{}
            \label{fig:mean and std of net44}
        \end{minipage}
        \begin{minipage}[b]{0.24\textwidth}
       \includegraphics[width=\textwidth]{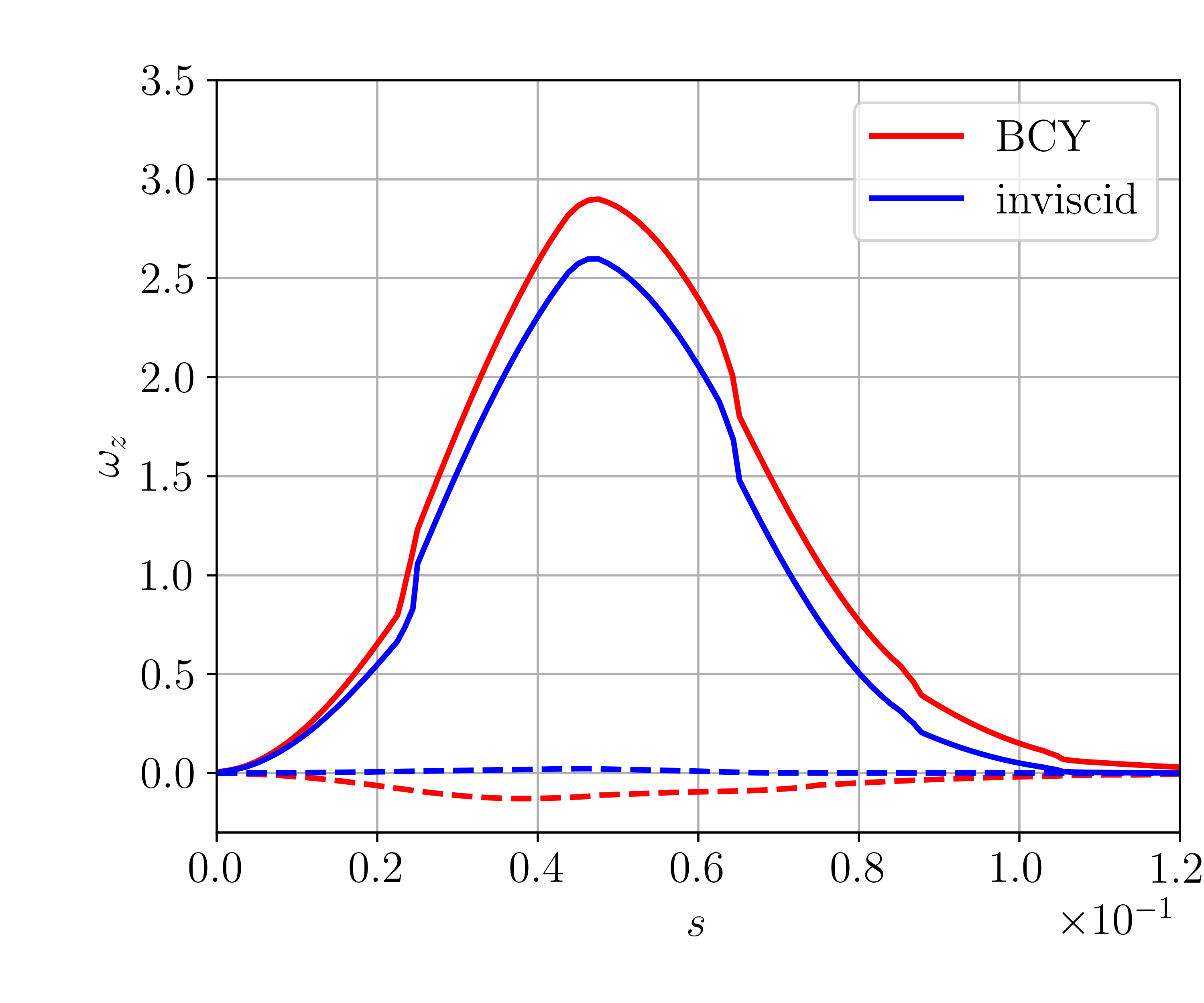}
            \subcaption{}
            \label{fig:mean and std of net14-2}
        \end{minipage}
        \hfill
        \begin{minipage}[b]{0.24\textwidth}  
       \includegraphics[width=\textwidth]{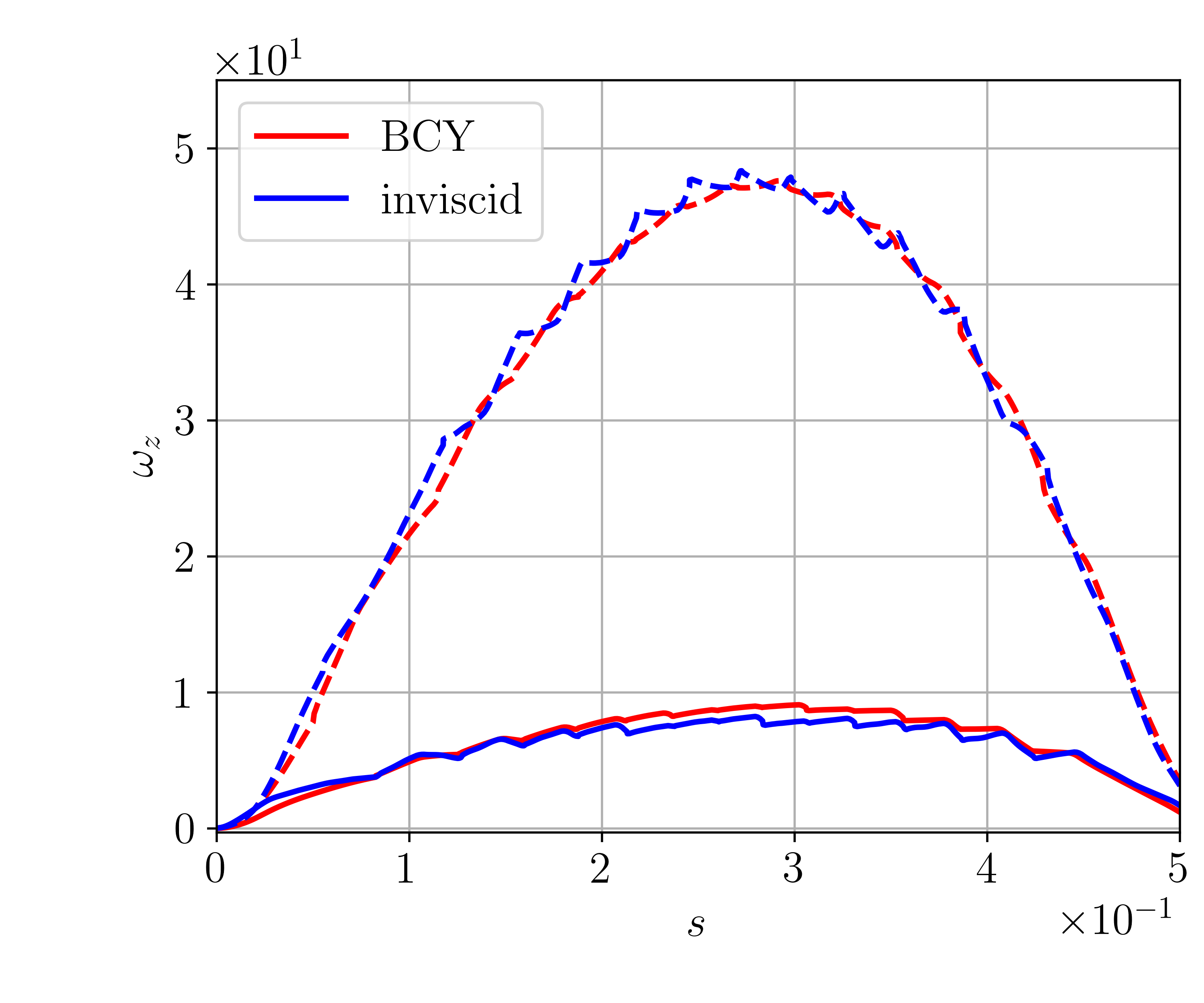}
            \subcaption{}
            \label{fig:mean and std of net24-2}
        \end{minipage}
        \begin{minipage}[b]{0.24\textwidth}   
       \includegraphics[width=\textwidth]{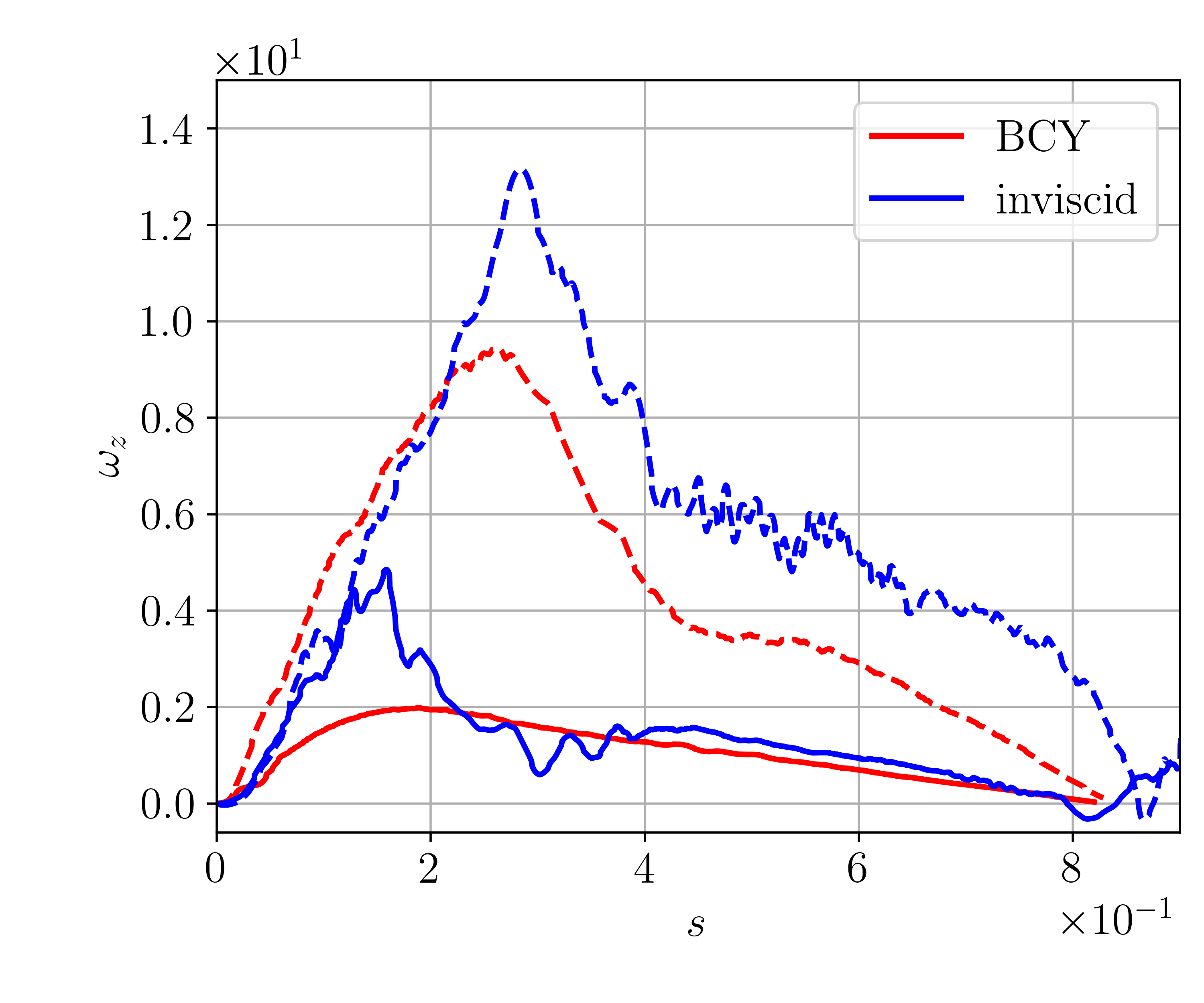}
            \subcaption{}
            \label{fig:mean and std of net34-2}
        \end{minipage}
        \hfill
        \begin{minipage}[b]{0.24\textwidth}   
            \includegraphics[width=\textwidth]{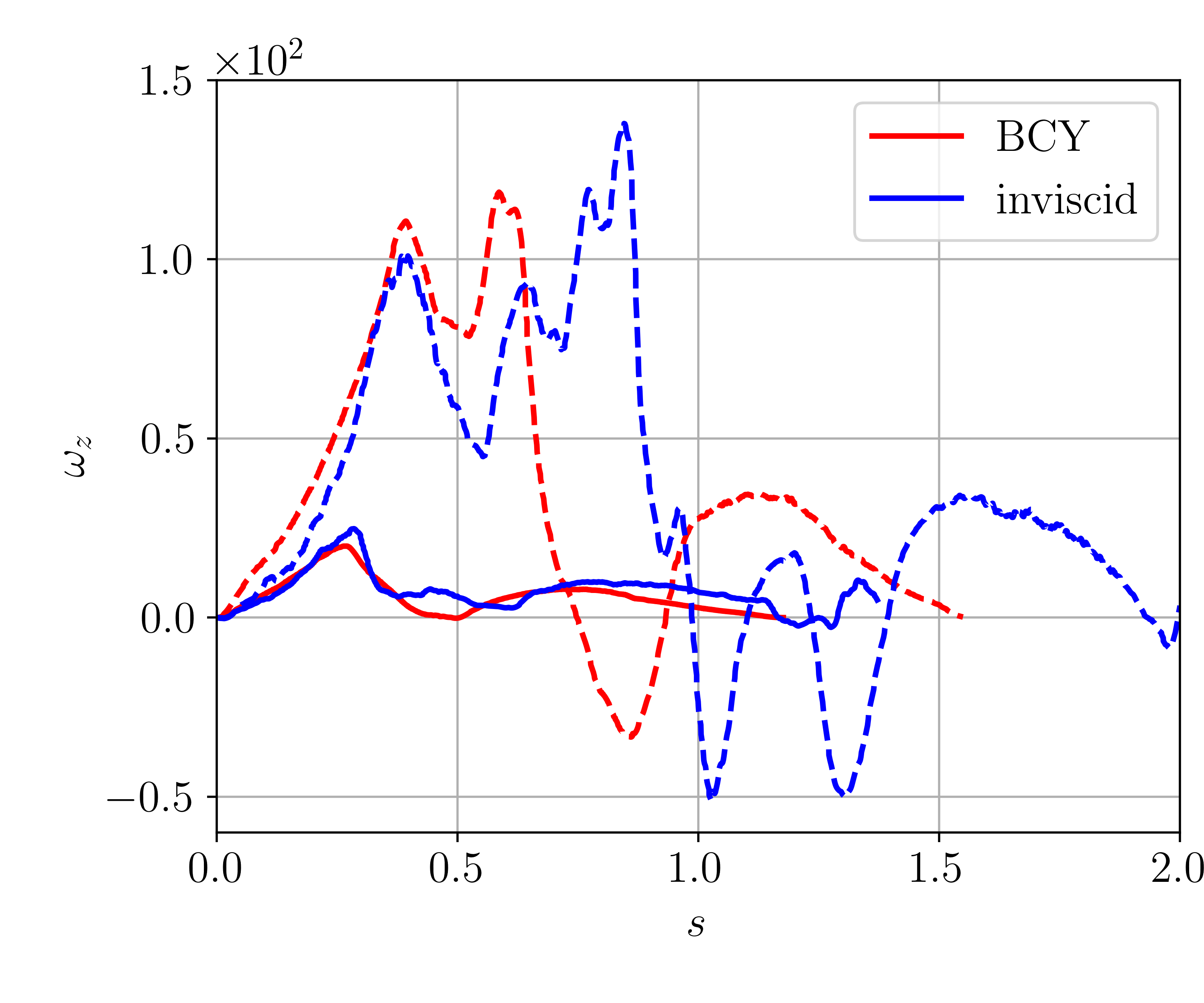}
            \subcaption{}
            \label{fig:mean and std of net44-2}
        \end{minipage}
        \caption{Temporal variation of the perturbation amplitude for $At = -0.66$ and $Ma = 5$, (a), where the black solid and dashed curves represent the predictions of the \citeauthor{richtmyer_taylor_1960}\cite{richtmyer_taylor_1960} and \citeauthor{dimonte_density_2000}\cite{dimonte_density_2000} models, the blue curves represent the inviscid case, while the red dashed and solid lines represent the numerical predictions obtained for the Newtonian-Newtonian and BCY cases, respectively; in the latter case, the light and heavy fluids correspond to Newtonian and BCY fluids, respectively; here, $T_1-T_4$ correspond to $t=$ 0.075, 0.18,1.41, and 1.48, respectively. Snapshots of the $x-y$ projections of the density field at $t=T_1-T_4$ depicting the interface shape superimposed on vorticity contours shown in (b)-(e), respectively; in each panel the predictions associated with the BCY and inviscid cases are shown on the left and right, respectively. Variation of the $z$-component of the vorticity with arc length for the inviscid and viscous cases represented by blue and red lines for $t=T_1-T_4$ shown in (f)-(i), respectively, plotted on the heavy- (with $\alpha=0.9$) and light-side (with $\alpha=0.1$) of the interface using solid and dashed lines, respectively. The rest of the parameters remain unchanged from the previous Figure.} 
        \label{fig:Ma5-At066}
    \end{figure*}

In Figure \ref{fig:Ma5-At066}(a), 
we plot the temporal variation of the perturbation amplitude for three situations: one in which both phases are inviscid; the second in which both phases are Newtonian; and the third which involves a Newtonian light fluid and a heavier non-Newtonian fluid, the properties of which are given in Table \ref{tab:Rheological_Paramters}; the rest of the parameters remain unchanged from those used to generate Figure \ref{fig:RMI_stages}. In the Newtonian-Newtonian case, 
the impact time between the shock and the interface is denoted by $T_1$ and is followed in the other cases by the protrusion of the heavier phase into the lighter one at  $T_2$. 
The results presented in Figure \ref{fig:Ma5-At066}(a) are compared to predictions obtained from the \citeauthor{richtmyer_taylor_1960} model \cite{richtmyer_taylor_1960} given by
\begin{equation}
\frac{d\tilde{h}}{d\tilde{t}} = At \, \tilde{h_0} \, \tilde{k} \, \Delta \tilde{u},
\label{glg:RMI}
\end{equation}
where $\tilde{h_0}$ is the initial amplitude of the perturbation, $\tilde{k}$ is its wavenumber, and $\Delta \tilde{u}$ is the velocity jump in the direction of travel of the shock following its propagation through the interface, at which stage
this model assumes that the flow becomes incompressible. 
The Nova laser experiments  \cite{dimonte_richtmyermeshkov_1996,farley_high_1999,dimonte_richtmyer-meshkov_1993} showed a lower growth rate at high $Ma$ as compared to the predictions by the \citeauthor{richtmyer_taylor_1960} incompressible model \cite{richtmyer_taylor_1960}; the latter model is only valid for the early, linear stage of the RMI, prior to the development of secondary KHI whose vortices decelerate the RMI growth. 

At high $Ma$ numbers strong shocks are formed; for this range of $Ma$, the behaviour of fluid dynamics significantly diverges from simpler, incompressible models. In the case of strong shocks, due to high compression, the transmitted shock recedes slowly and represents an effectively impenetrable boundary, which reduces the growth of the RMI. This phenomenon leads to a complex interaction between the shock waves and the interface perturbations, further complicating the fluid motion and the subsequent development of instabilities. It is therefore unsurprising that the agreement with the \citeauthor{richtmyer_taylor_1960} model \cite{richtmyer_taylor_1960} deteriorates after early times giving way to significantly larger growth rate estimates than those generated numerically for both the inviscid and viscous cases.

We also compare the numerical predictions for the perturbation amplitude with those obtained from the work of \citeauthor{dimonte_density_2000}\cite{dimonte_density_2000} who introduced a late-stage model which accounts for the drag effects on the growth of RMI. 
The  evolution of the amplitude $\tilde{h}$ of the single-mode RMI is described in this model by the following equation:
%
\begin{equation}
\frac{d^2\tilde{h}}{d\tilde{t}^2} = -c_d \left(\frac{d\tilde{h}}{d\tilde{t}}\right)\left|\frac{d\tilde{h}}{d\tilde{t}}\right| \frac{1}{\tilde{h}}.
\label{glg:DS_RMI_Model}
\end{equation}
In Eq. (\ref{glg:DS_RMI_Model}), $c_d= \frac{24}{Re}$ represents a drag coefficient for simple laminar flow around a sphere.
A comparison of the \citeauthor{dimonte_density_2000} model \cite{dimonte_density_2000} predictions and those from the inviscid and viscous numerical models demonstrate an improvement over those obtained from the work of \citeauthor{richtmyer_taylor_1960}\cite{richtmyer_taylor_1960}.

We show $x-y$ projections of the interface shape superimposed on vorticity contour plots in Figure \ref{fig:Ma5-At066}(b)-(e) for $t=0.075,0.18, 1.41, 1.48$, respectively, for the inviscid and viscous cases; the variation of the $z$-component of the vorticity, $w_z$, along the arc length is also shown for the same times. It is seen clearly that vorticity is generated in the regions immediately adjacent to the interface, and this is closely correlated to the interfacial deformation associated with the RMI and the KHI-related roll-up phenomena; the development of the latter, whose onset is at $T_3$, leads to a deceleration of the perturbation amplitude. Furthermore, the dominant contribution is due to the heavier phase during the earliest stages of the flow; following the crossing of the shock into the lighter phase, the latter becomes the dominant contributor to vorticity generation. 

\begin{figure*}
 \begin{minipage}[b]{\textwidth}
 \includegraphics[scale=0.73]{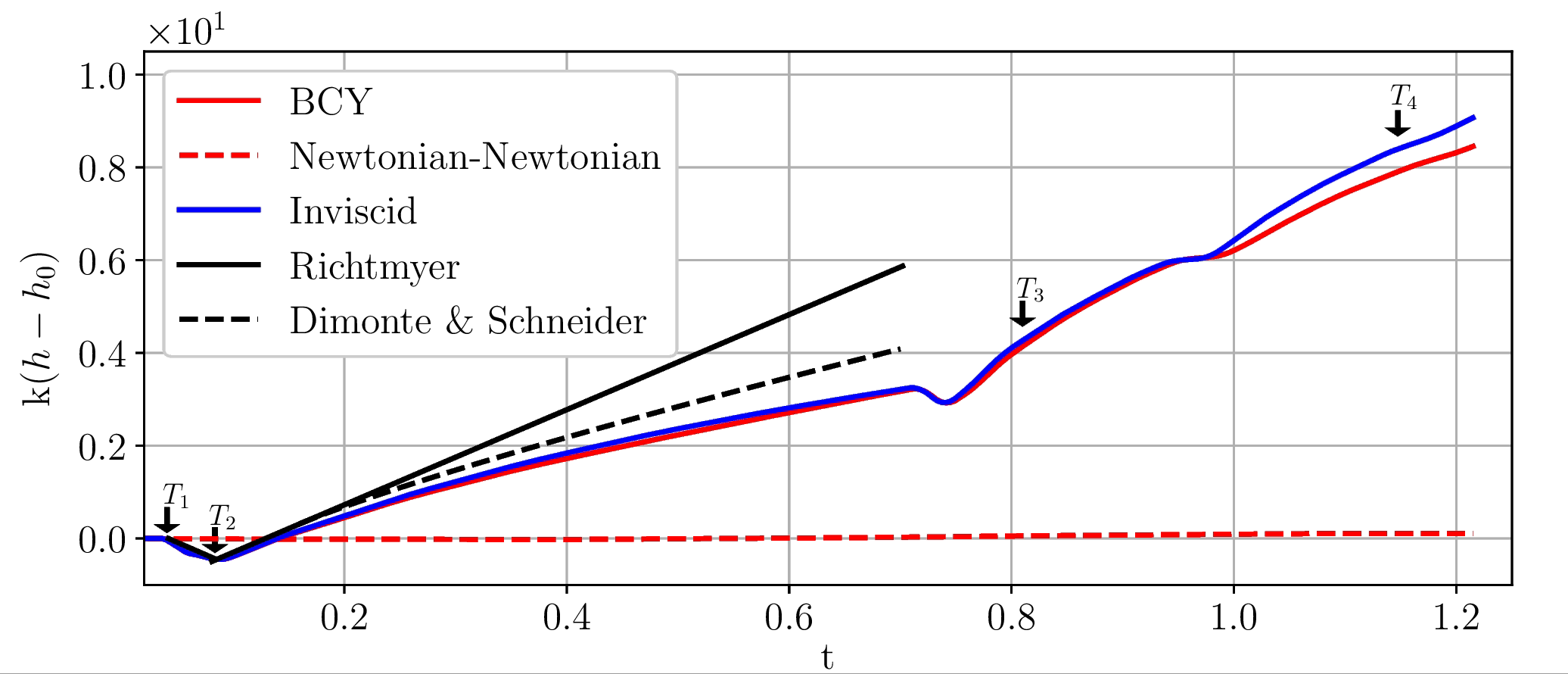}\\
 \subcaption{}
 \end{minipage}
        \begin{minipage}[b]{0.22\textwidth}
    \includegraphics[width=\textwidth]{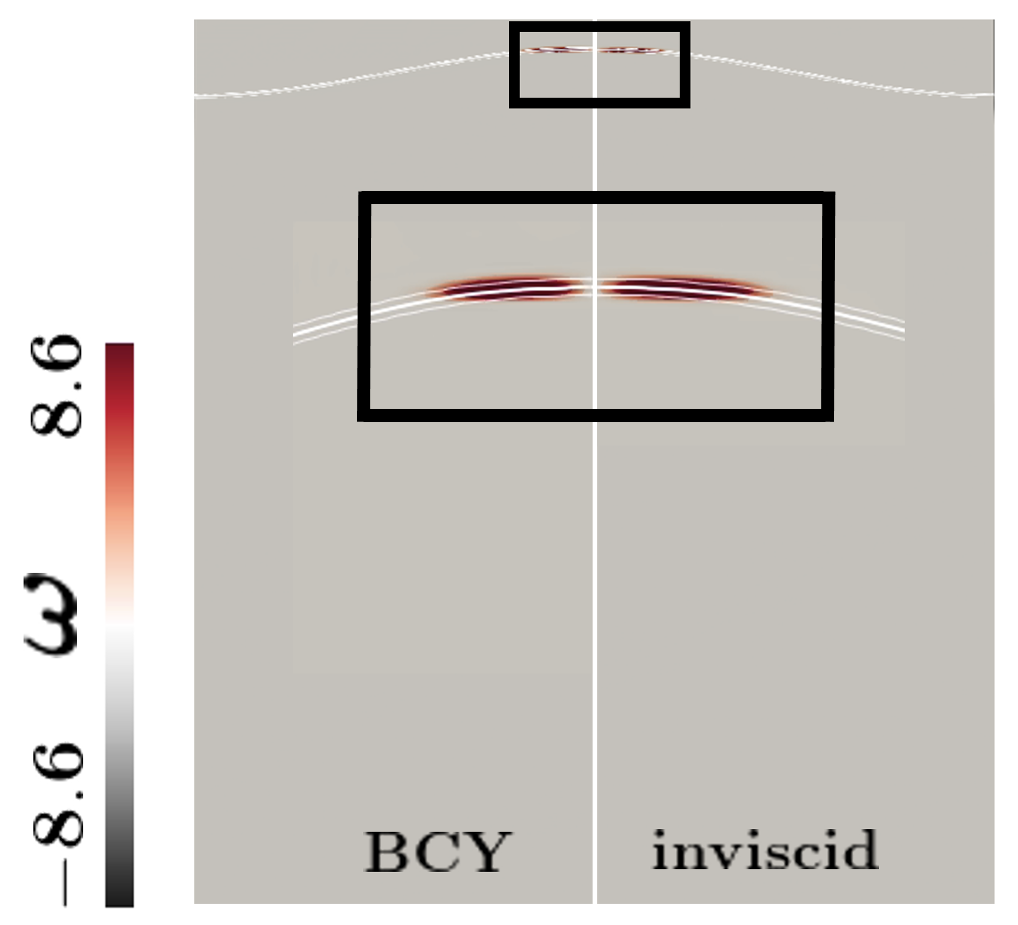}
            \subcaption{}
            \label{fig:mean and std of net14-Mach10}
        \end{minipage}
        \hfill
       \begin{minipage}[b]{0.22\textwidth}  
\includegraphics[width=\textwidth]{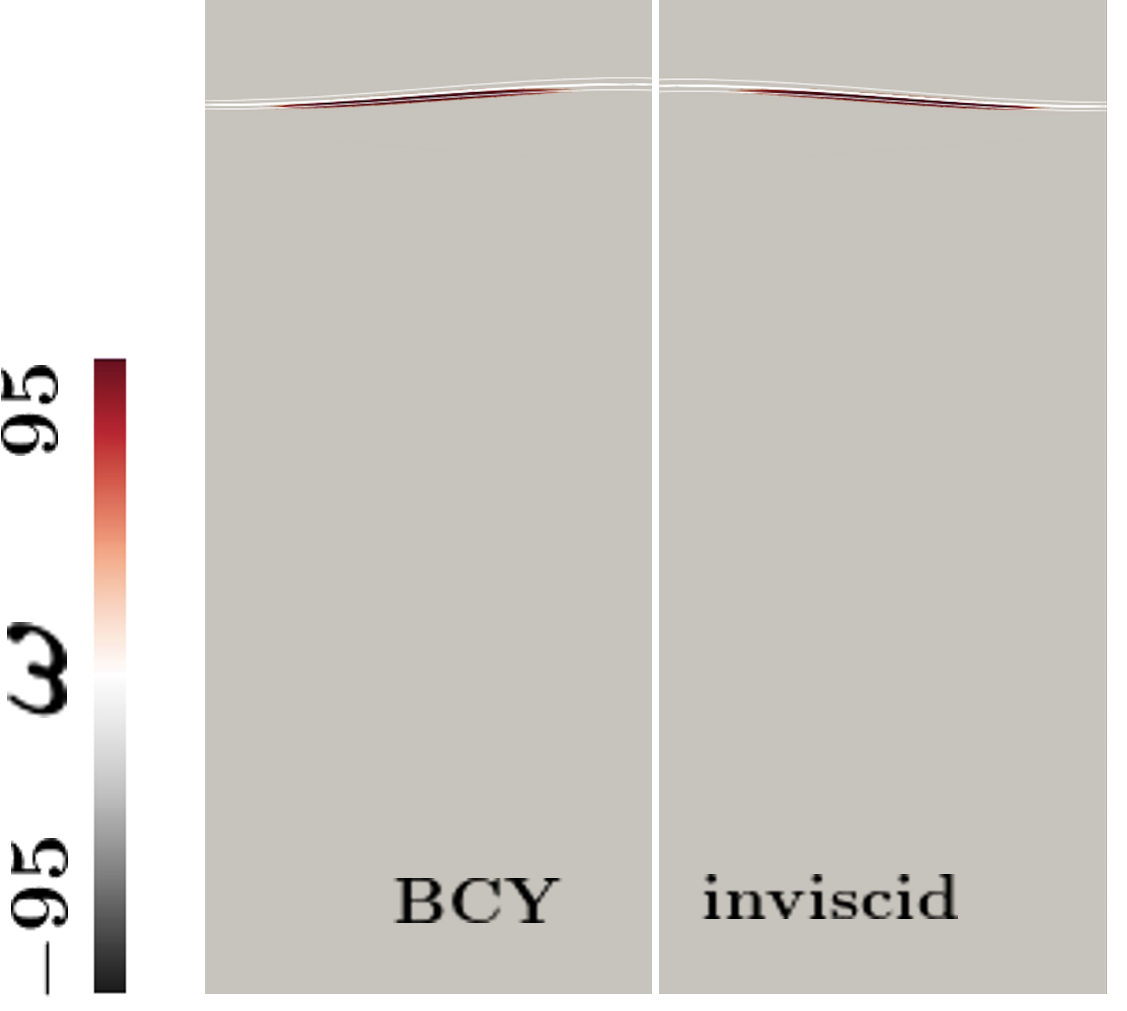}
            \subcaption{}
            \label{fig:mean and std of net24-Mach10}
        \end{minipage}
        \begin{minipage}[b]{0.22\textwidth}   
\includegraphics[width=\textwidth]{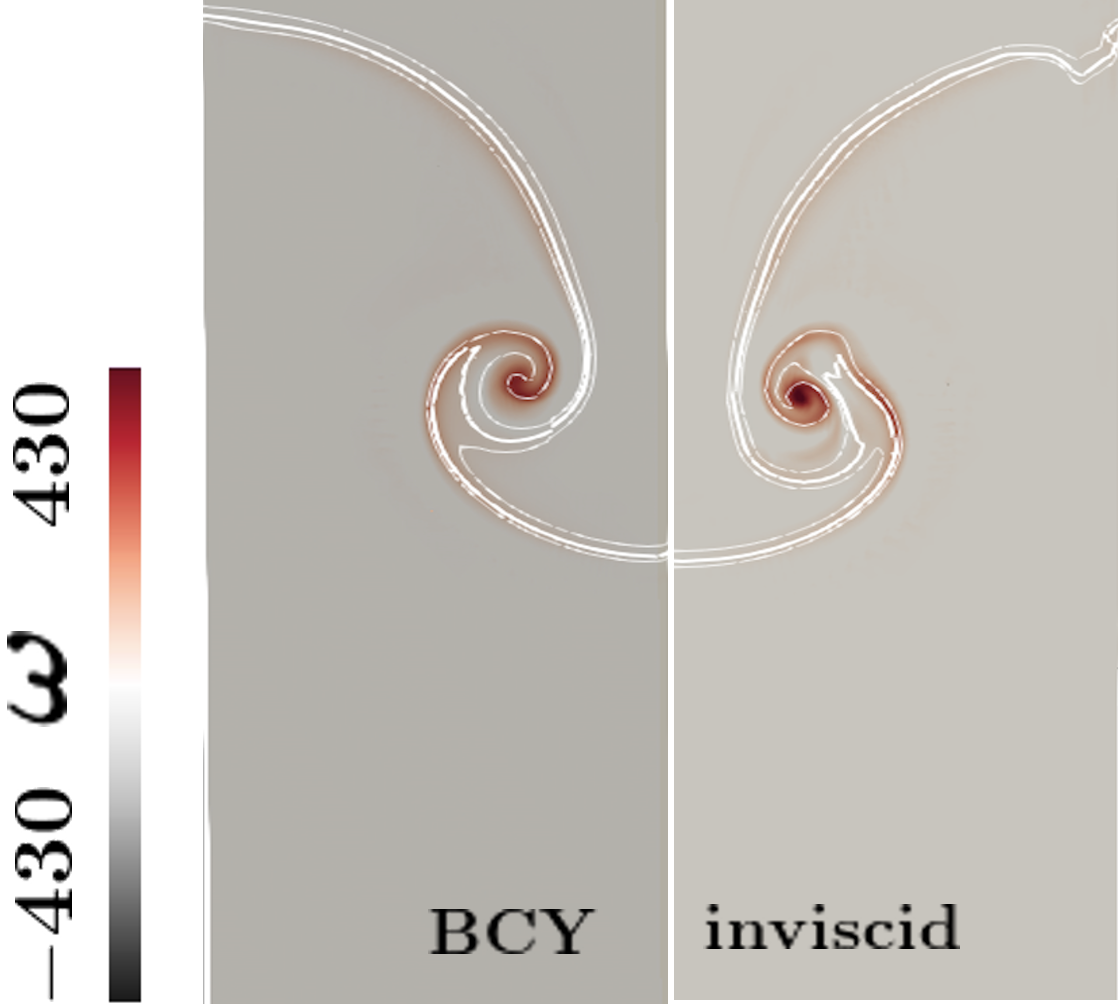}
            \subcaption{}
            \label{fig:mean and std of net34-Mach10}
        \end{minipage}
        \hfill
        \begin{minipage}[b]{0.215\textwidth}   
      \includegraphics[width=\textwidth]{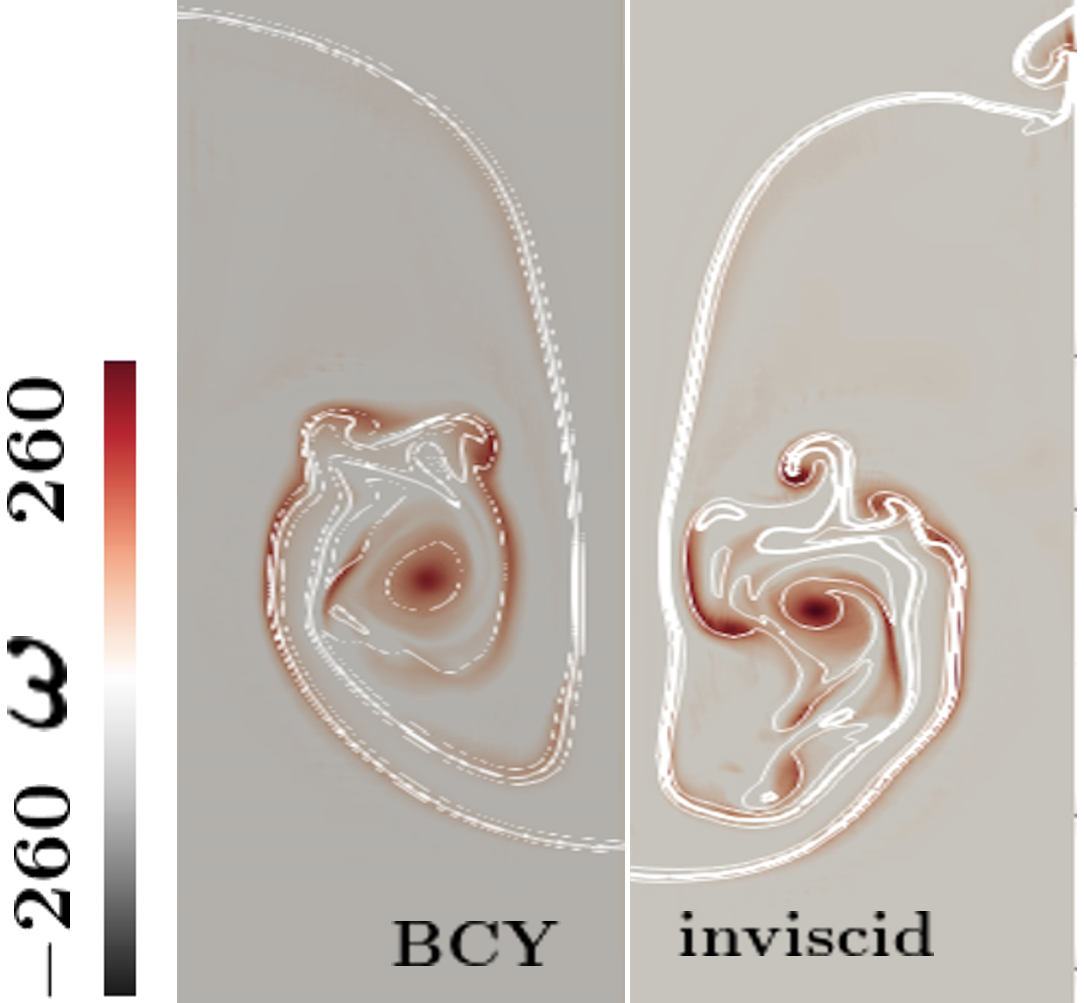}
            \subcaption{}
            \label{fig:mean and std of net44-Mach10}
        \end{minipage}
        \begin{minipage}[b]{0.24\textwidth}
      \includegraphics[width=\textwidth]{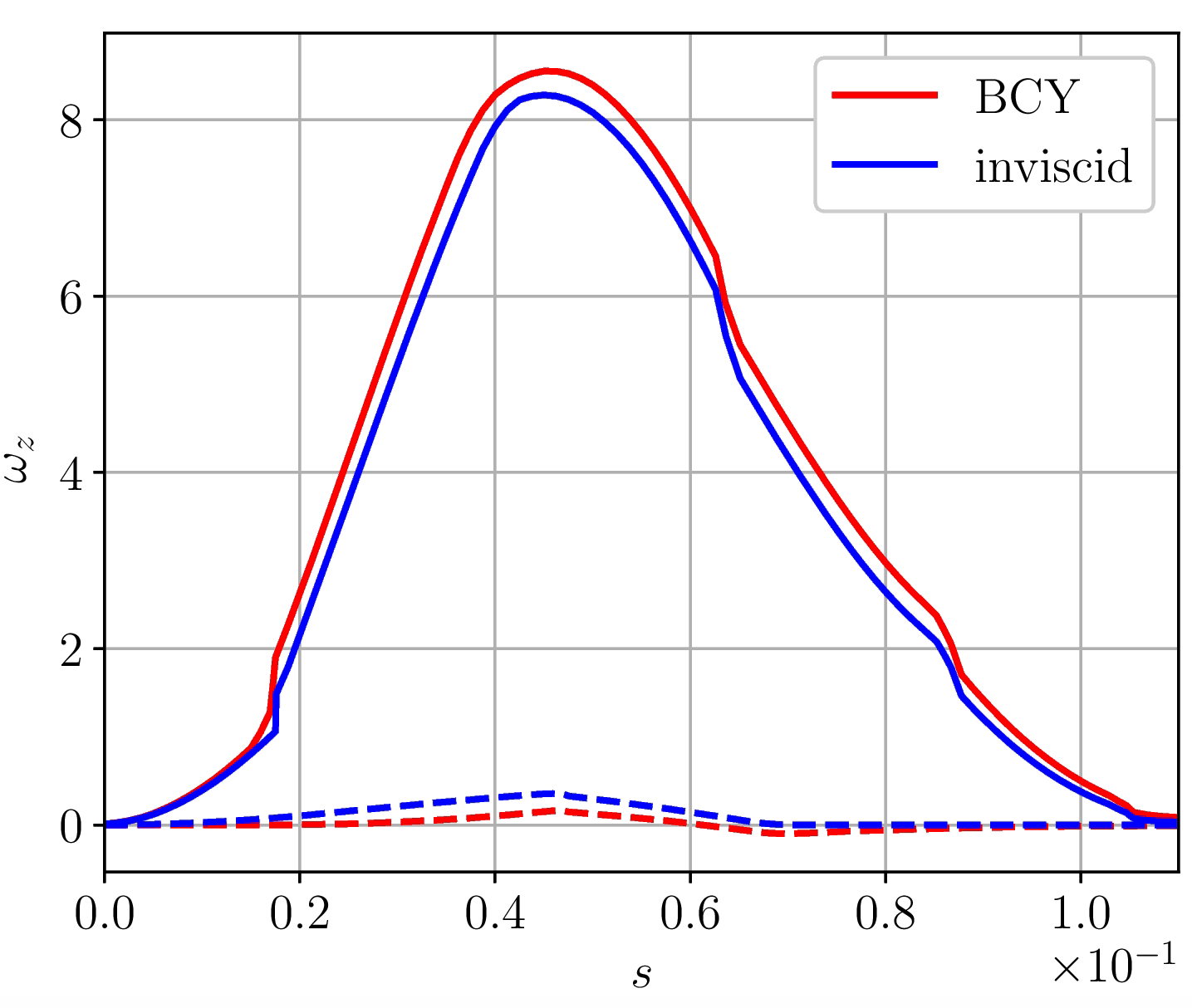}
            \subcaption{}
            \label{fig:mean and std of net14-Mach10-2}
       \end{minipage}
        \hfill
        \begin{minipage}[b]{0.24\textwidth}  
      \includegraphics[width=\textwidth]{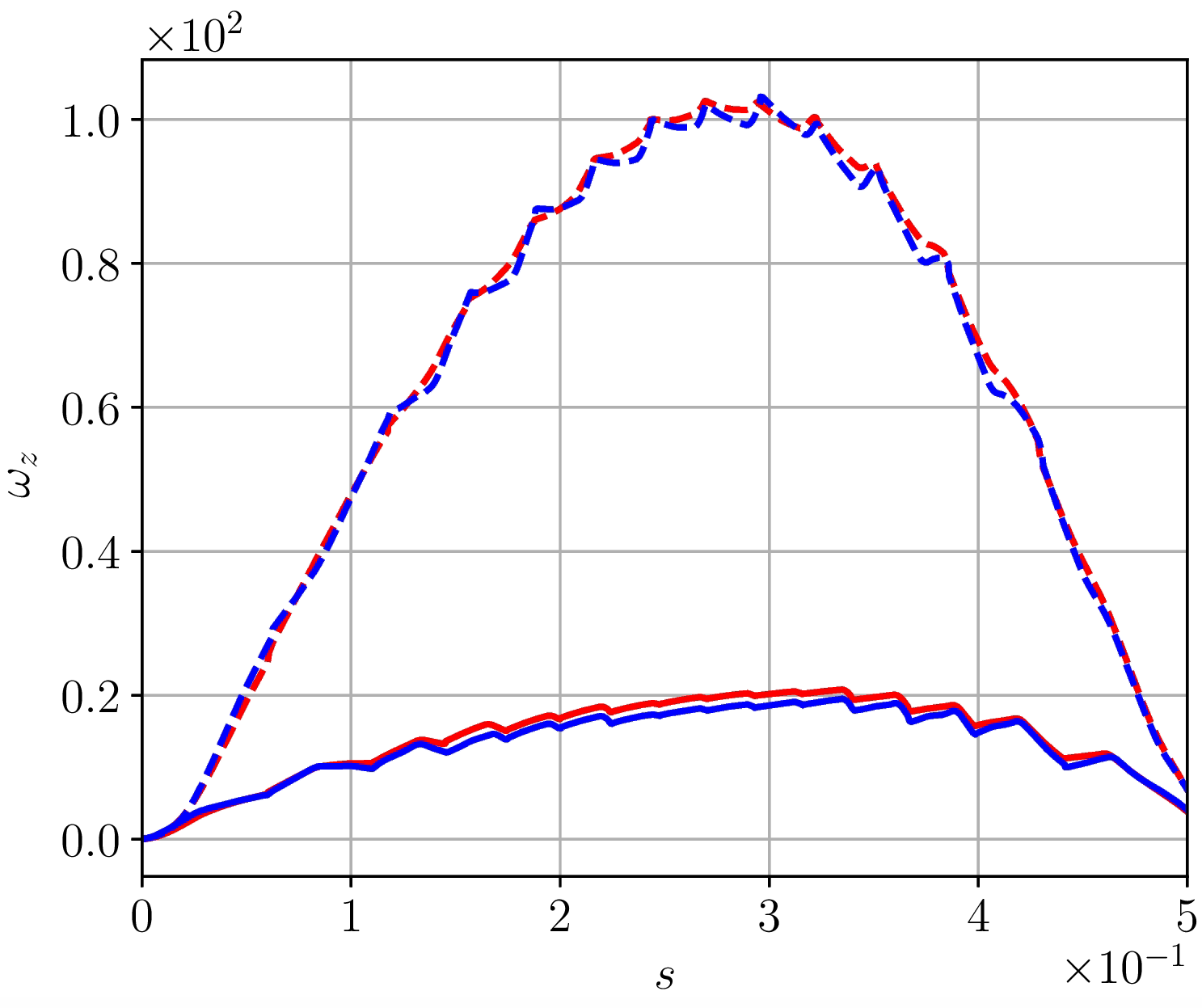}
            \subcaption{}
            \label{fig:mean and std of net24-Mach10-2}
        \end{minipage}
        \begin{minipage}[b]{0.24\textwidth}   
      \includegraphics[width=\textwidth]{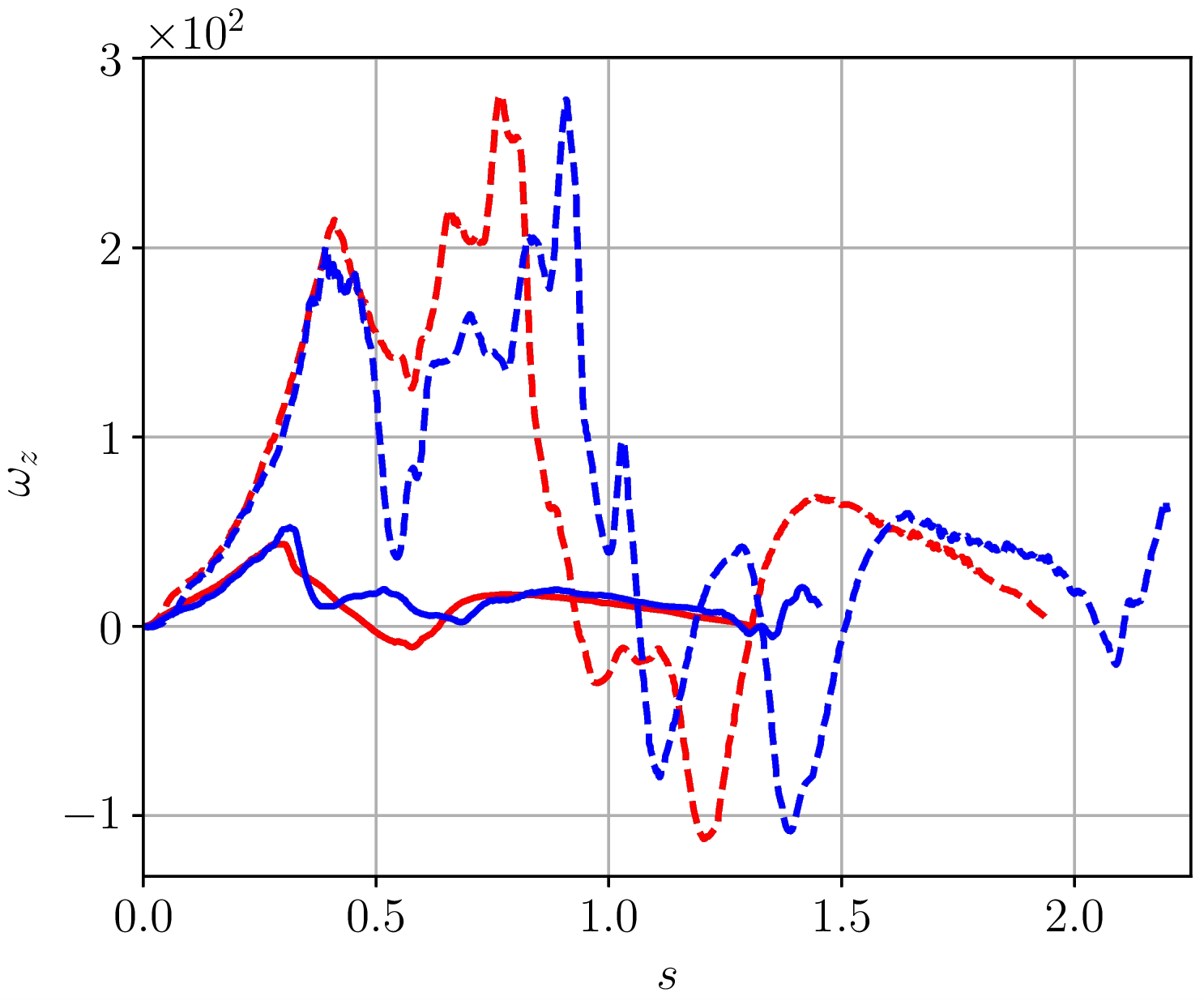}
            \subcaption{}    
           \label{fig:mean and std of net34-Mach10-2}
        \end{minipage}
        \hfill
        \begin{minipage}[b]{0.24\textwidth}   
            \includegraphics[width=\textwidth]{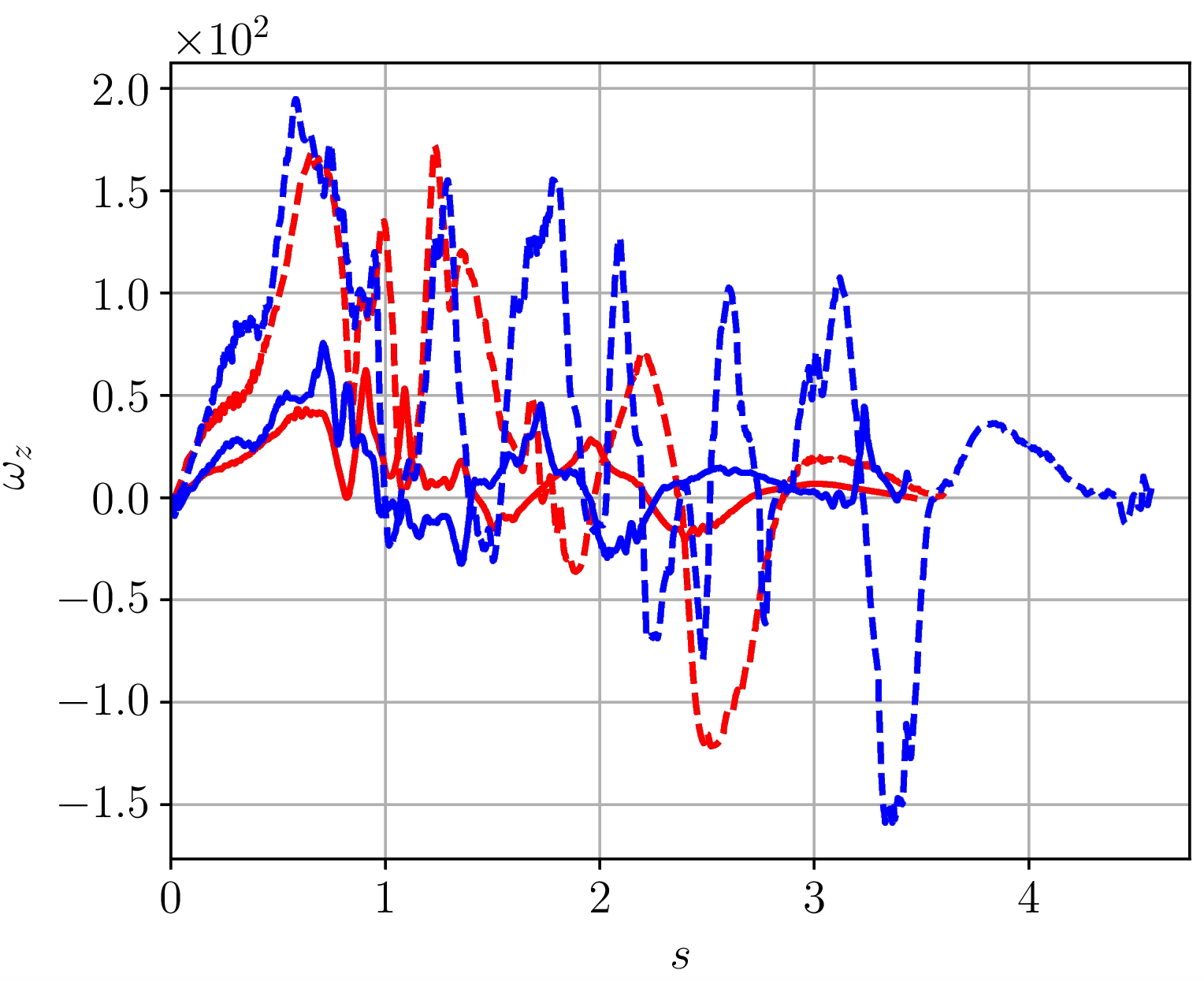}
            \subcaption{}
            \label{fig:Ma10-At066}
        \end{minipage}
        \caption{Temporal variation of the perturbation amplitude for $At = -0.66$ and $Ma = 10$, (a), where the black solid and dashed lines represent the predictions of the \citeauthor{richtmyer_taylor_1960}\cite{richtmyer_taylor_1960} and \citeauthor{dimonte_density_2000}\cite{dimonte_density_2000} models, the blue curves represent the inviscid case, while the red dashed and solid lines represent the numerical predictions obtained for the Newtonian-Newtonian and BCY cases, respectively; in the latter case, the light and heavy fluids correspond to Newtonian and BCY fluids, respectively; here, $T_1-T_4$ correspond to $t=$ 0.037, 0.085, 0.815, and 1.175, respectively. Snapshots of the $x-y$ projections of the density field at $t=T_1-T_4$ depicting the interface shape superimposed on vorticity contours shown in (b)-(e), respectively; in each panel the predictions associated with the BCY and inviscid cases are shown on the left and right, respectively. Variation of the $z$-component of the vorticity with arc length for the inviscid and viscous cases represented by blue and red lines for $t=T_1-T_4$ shown in (f)-(i), respectively, plotted on the heavy- (with $\alpha=0.9$) and light-side (with $\alpha=0.1$) of the interface using solid and dashed lines, respectively. The rest of the parameters remain unchanged from the previous Figure.} 
        \label{fig:Ma10_final-At066}
    \end{figure*} 
    
The divergence in the evolving dynamics between the inviscid and viscous cases becomes more accentuated with increasing time as can be confirmed upon inspection of Figures \ref{fig:Ma5-At066}(d), (e), (h) and (i); it is also seen that the vortical structures are more pronounced for the inviscid case, for which the interface becomes more stretched as revealed by the expanded arc length range associated with this case (see Figure \ref{fig:Ma5-At066}(i)). Lastly, the impact of the reflected shock on the interface at $t=T_4$ leads to further development of the RMI and KHI and acceleration of perturbation growth. 
In Figure \ref{fig:Ma10_final-At066}, we plot the analogous variables to those shown in Figure \ref{fig:Ma5-At066} except $Ma=10$, and the rest of the parameters remain unchanged. 
A comparison of the results presented in these figures reveals similar trends in terms of perturbation amplitude growth and the close association between the vorticity in the vicinity of the interface and the development of the RMI and KHI roll-up phenomena. An exception is provided by the Newtonian-Newtonian case where the strong influence of viscosity prevents any increase in perturbation amplitude, as can be seen from Figures \ref{fig:Ma5-At066}(a) and \ref{fig:Ma10_final-At066}(a). 
At the higher $Ma$, it is seen clearly that not only are these phenomena significantly more pronounced, the difference in vorticity generated between the inviscid and BCY cases is lower than that associated with the 
$Ma=5$ case. This is due to the higher shear rates at $Ma=10$ and therefore lower viscosity in the heavy phase in comparison to that at $Ma=5$. In contrast, the Newtonian-Newtonian cases exhibits comparatively very little perturbation growth, even at the elevated $Ma$. 

Following the passage of the incident shock, viscous damping associated with the residual viscosity becomes dominant and acts to suppress the RMI and KHI roll-up. This is seen to be more effective for the lower $Ma$ studied due to the comparatively smaller shear rates and therefore larger residual viscosity. 
There is also evidence for perturbation amplitude deceleration due to the development of KHI roll-up in the $Ma=10$ case, as can be seen in Figure \ref{fig:Ma10_final-At066}(a). 

Furthermore, the impact of the first reflected shock at $t=T_3$ on the KHI and RMI growth rate is similar in both the viscous and inviscid cases, as shown in Figure \ref{fig:Ma10_final-At066}(d). The impact of the second reflected shock at $t=T_4$ leads to the development of small-scale roll-up structures. 
These small-scale structures lead to additional vorticity generation, the so-called vortex-accelerated secondary baroclinic vorticity, which has been investigated in detail by Peng et al. \citep{peng_vortex-accelerated_2003}. 
This 
mechanism involves the interaction of vortices generated by RMI and KHI, causing a further enhancement of baroclinic vorticity deposition, which contributes to an increase in the growth rate of RMI and a more developed KHI. In our simulations, we can observe the effects of vortex-accelerated secondary baroclinic vorticity on the RMI and KHI: the RMI growth rate is higher, and KHI is further developed in the inviscid case. In the BCY case, however, the span of the RMI, which is a low shear rate region, is characterised by a relatively high viscosity
that leads to a lower growth rate of the RMI after the second reflected shock. 

\begin{figure}
    \includegraphics[width=0.26\textwidth]{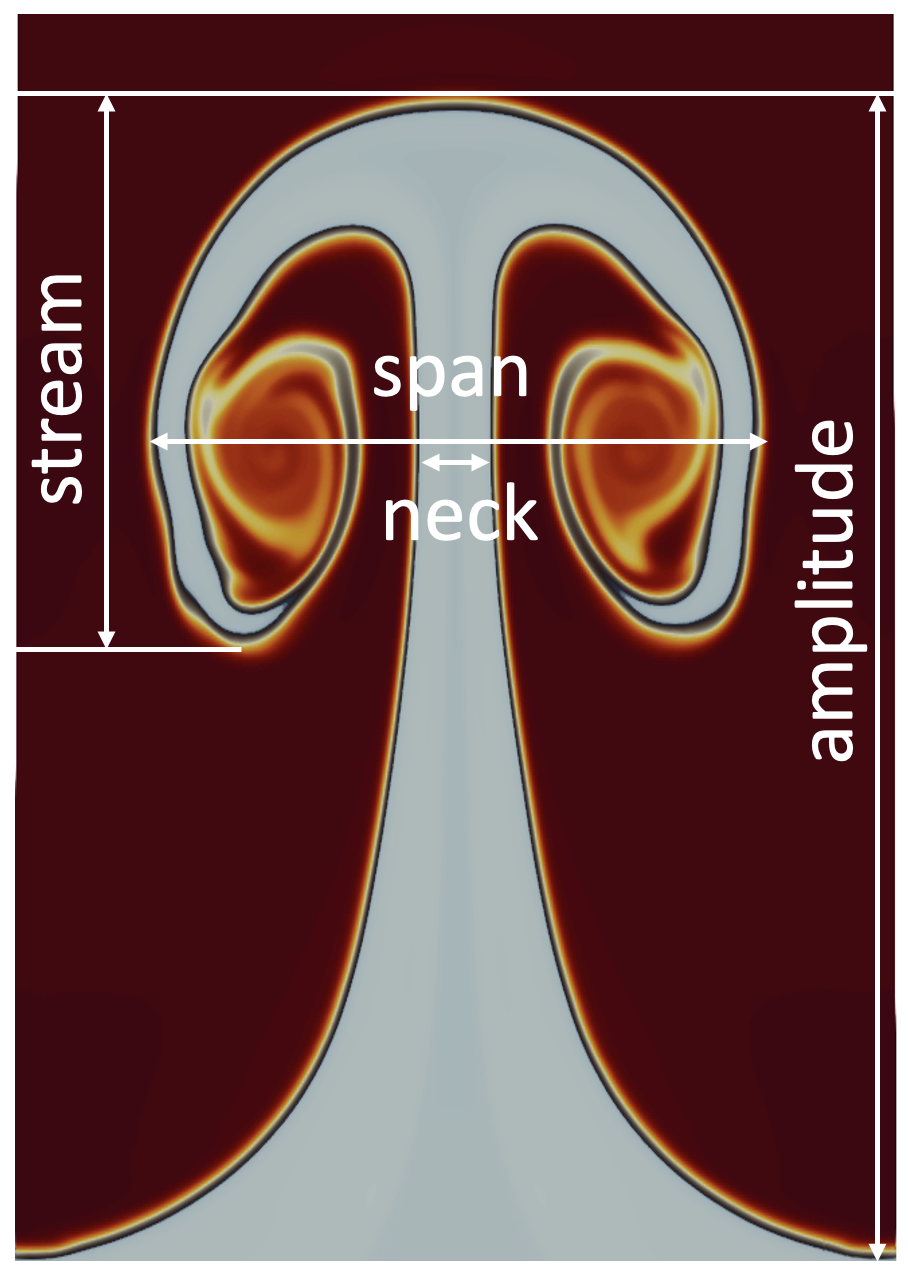}
    \caption{Schematic of a fully-developed mushroom-like structure formed following the passage of a shock travelling from a heavy phase towards a lighter one through the separating interface. The features stream, span, neck and amplitude \cite{peng_vortex-accelerated_2003} characterise the structure are shown together with their defined dimensions.}
    \label{fig:RMI_Sketch}
\end{figure}

\begin{figure}
\vspace{-5mm}
  \begin{minipage}{0.41\textwidth}
    \centering
    \includegraphics[width=\linewidth]{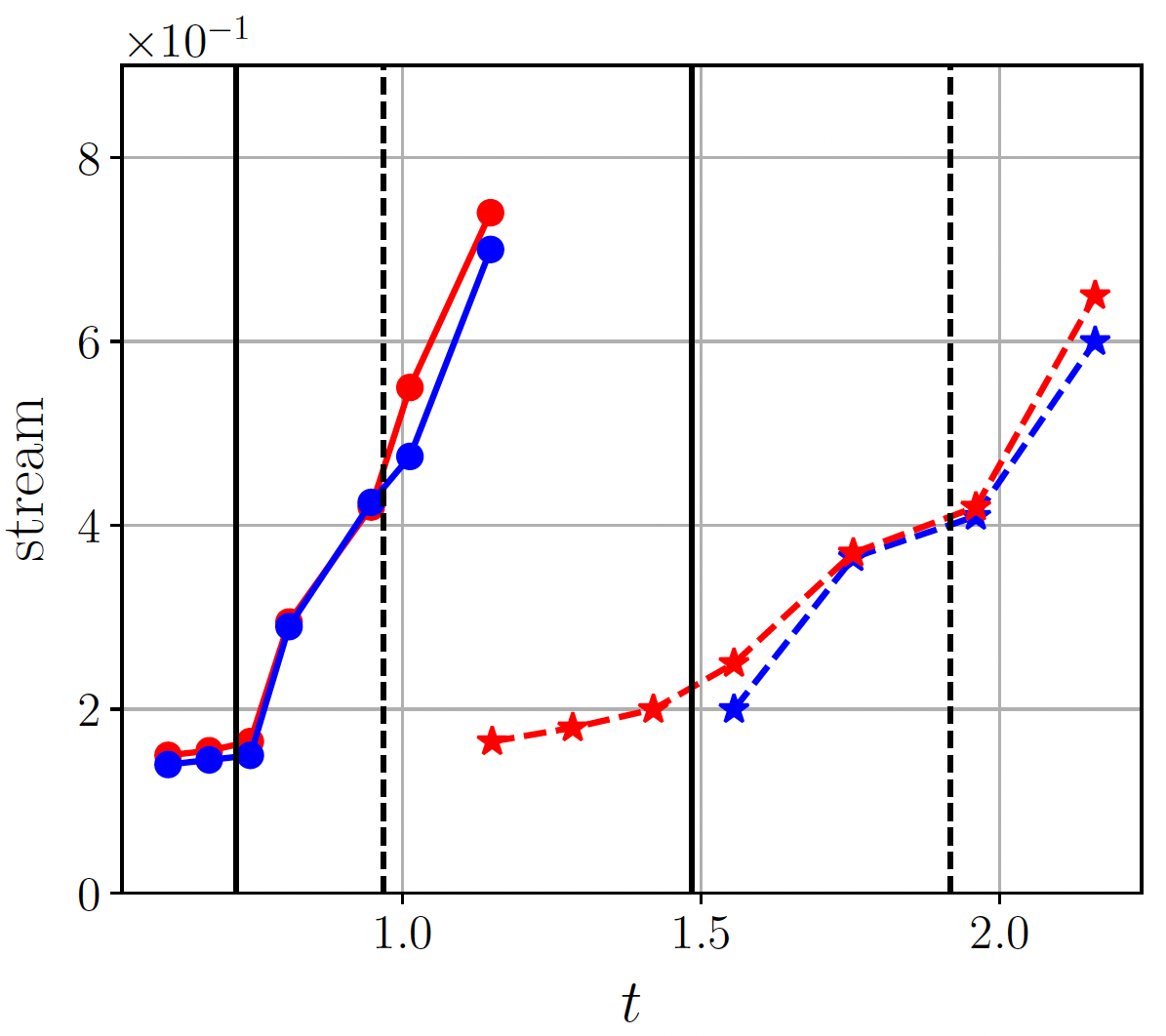}
    \subcaption{}
    \label{subfig:your_label1}
    \vspace{-2mm}
  \end{minipage}
    \begin{minipage}{0.41\textwidth}
    \centering
    \includegraphics[width=\linewidth]{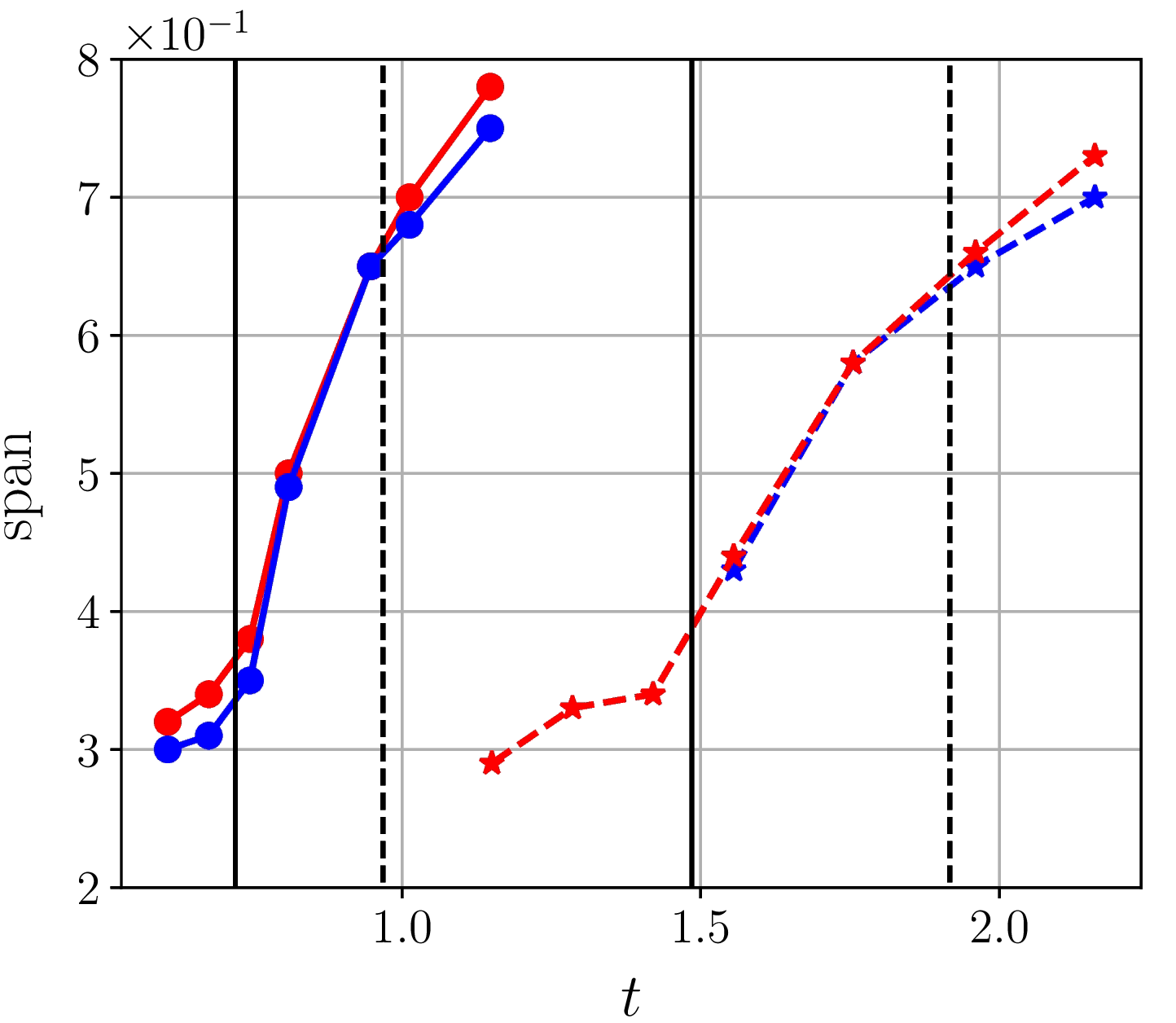}
    \subcaption{}
    \label{subfig:your_label2}
    \vspace{-2mm}
  \end{minipage}
  \begin{minipage}{0.41\textwidth}
    \centering
    \includegraphics[width=\textwidth]{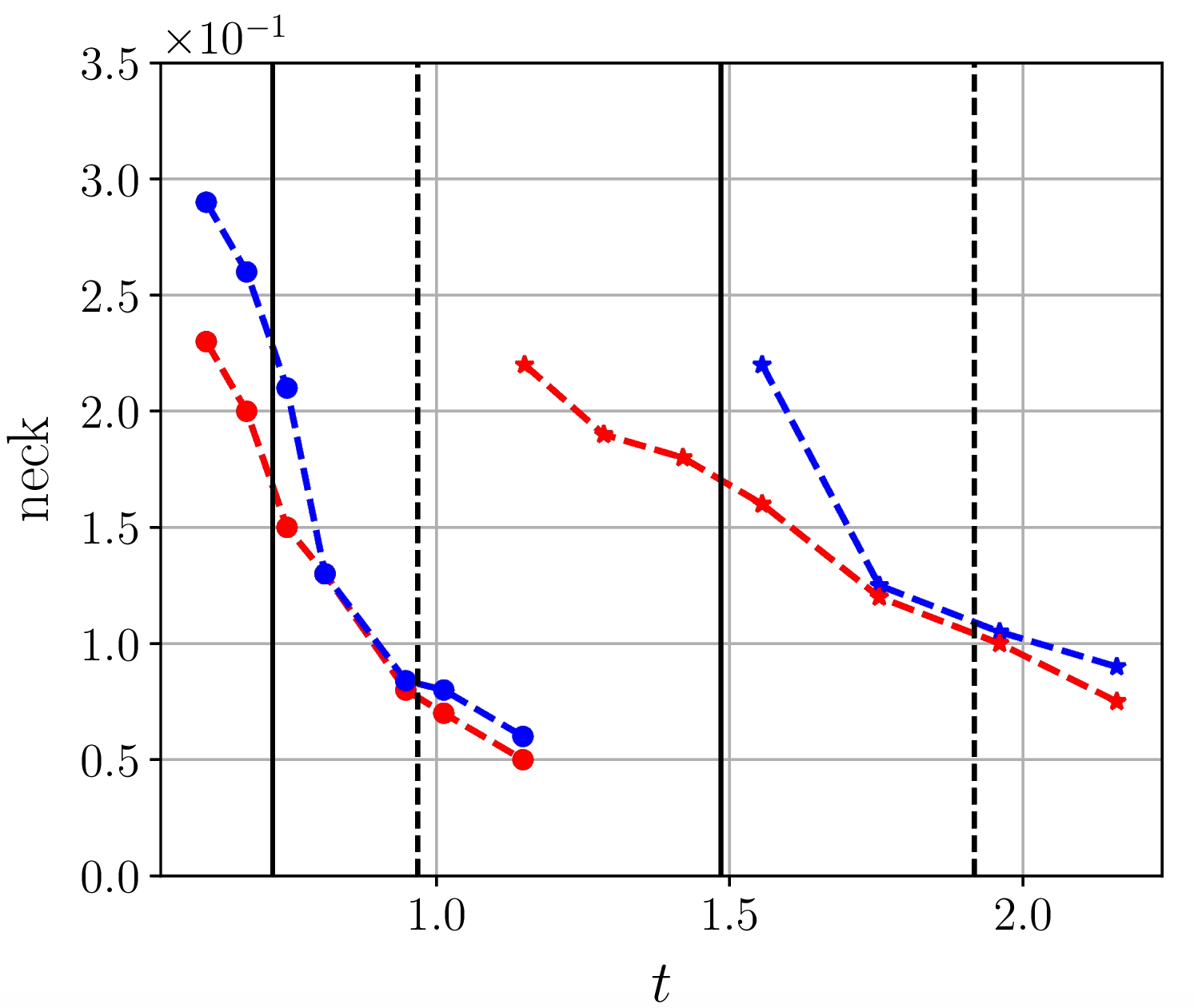}
    \subcaption{}
    \label{subfig:your_label3}
    \vspace{-2mm}
  \end{minipage}
  \caption{Temporal evolution of the length of the stream, span, and neck (see Figure \ref{fig:RMI_Sketch} for the definitions of these features) $At = - 0.66$ shown in (a)-(c), respectively. In each panel, the red and blue solid lines designate the inviscid and BCY cases for $Ma=10$, and their dashed counterparts the $Ma=5$ case, respectively. The solid and dashed vertical lines at $t=$ 0.71 and $t=$ 0.95, and $t=1.41$ and $t=1.89$ highlight the times at which the interface is impacted by the first and second reflected shocks for $Ma=10$ and $Ma=5$, respectively. The rest of the parameters are unchanged from previous Figures.}
  \label{fig:RMI_features}
\end{figure}

We now explore the effects of system parameters on the features of the mushroom-like structure discussed above, and show in Figure \ref{fig:RMI_Sketch}
an annotated schematic of a fully-developed such structure; in this figure, we highlight the dimensions of the various features: the `span', `stream', and `neck'. 
In Figure \ref{fig:RMI_features} we demonstrate the influence of the non-Newtonian effects on these features. As shown in Figure  \ref{fig:RMI_features}(a), there is very little difference in stream development between the inviscid and BCY cases after the first reflected shock for $Ma=10$. 
The second reflected shock leads to the development of small-scale structures, and the occurrence of these structures induces vortex-accelerated vorticity. This additional vorticity in the inviscid case results in a higher stream in comparison to the BCY case wherein the higher viscosity due to the comparatively lower shear rates dampens the small-scale structures, leading to a lower stream. 

For $Ma=5$, the stream does not develop until the first reflected shock for the BCY case in contrast to the inviscid case. The second reflected shock leads to the development of small-scale structures and vortex-accelerated vorticity, resulting in elevation of the BCY stream, which remains below that of the inviscid case due to larger viscous damping in the BCY case. 
Similar behaviour is observed in the development of the span and neck for both $Ma=5$ and $Ma=10$, as shown in Figures \ref{fig:RMI_features}(b) and (c), respectively. 

\begin{figure}
  \hspace{0cm}
  \includegraphics[scale=0.65]{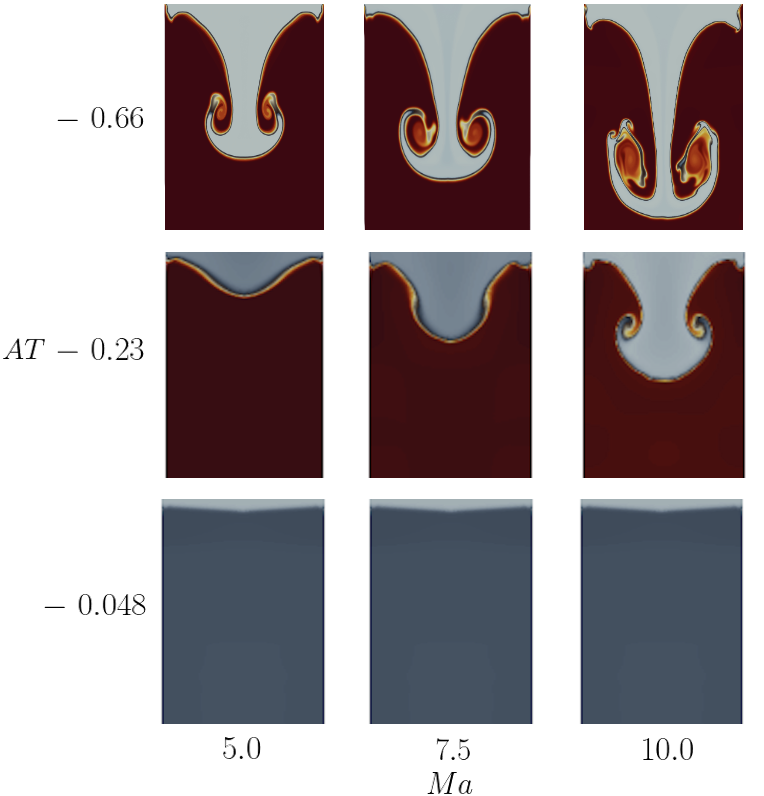}
  \caption{Fully-developed $x-y$ projections of the density field for the inviscid case in the $(At,Ma)$ space for $At=(-0.048,-0.230,-0.66)$ and $Ma=(5,7.5,10)$; the rest of the parameters remain unchanged from previous Figures.}
  \label{fig:inviscid_map}
\end{figure}

\begin{figure}
 \includegraphics[scale=0.33]{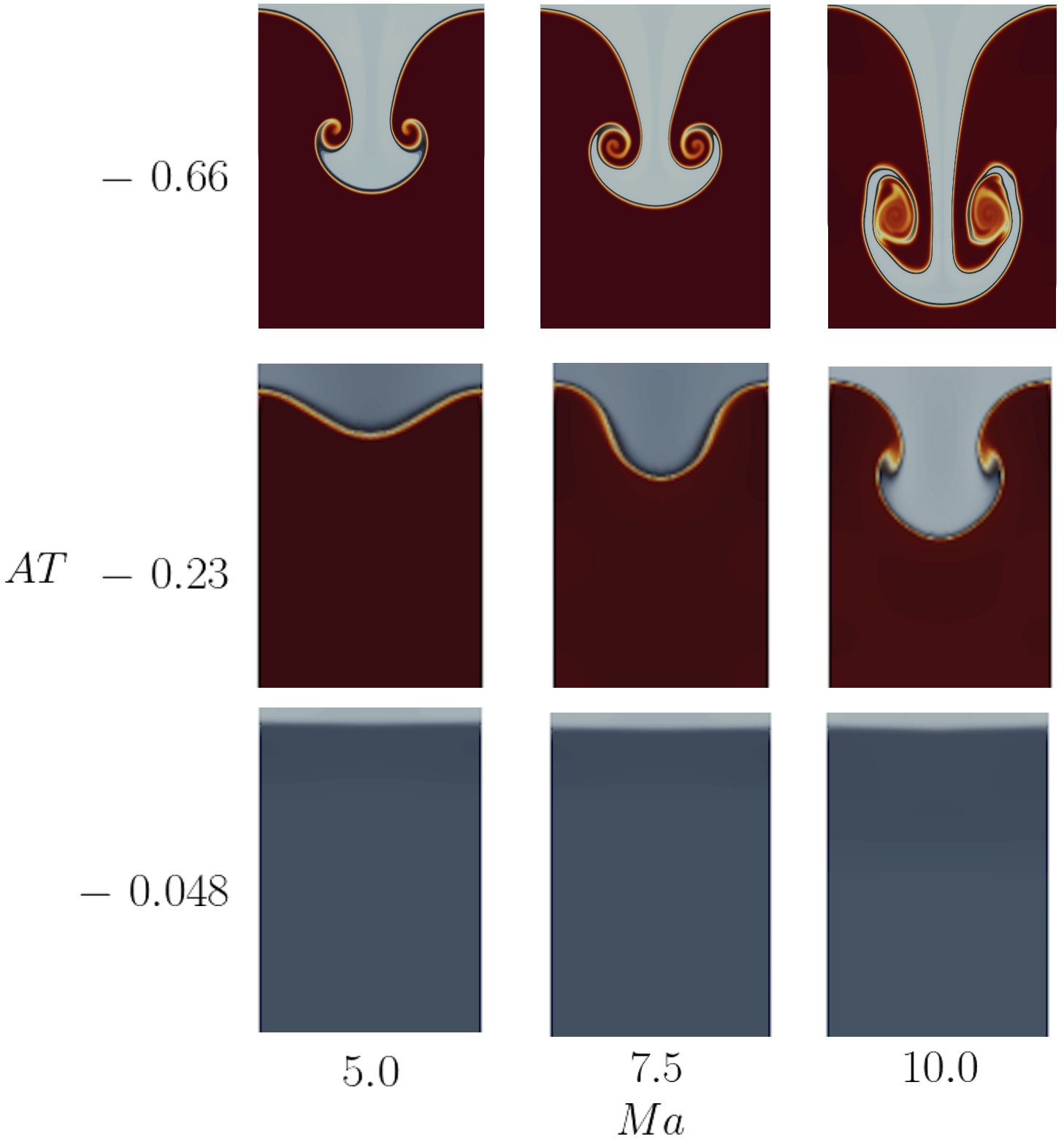}
 \caption{Fully-developed $x-y$ projections of the density field for the case wherein the lighter and heavier phases correspond to Newtonian and BCY fluids, respectively, in the $(At,Ma)$ space  for $At=(-0.048,-0.230,-0.66)$ and $Ma=(5,7.5,10)$; the rest of the parameters remain unchanged from previous Figures.}
  \label{fig:bcy_map}
\end{figure}

We have also carried out a parametric study of the effects of $At$ and $Ma$ on the development of RMI and associated phenomena. In Figures \ref{fig:inviscid_map} and \ref{fig:bcy_map}, we show the fully-developed $x-y$ projections of the density field for the inviscid and BCY cases, respectively, generated for $At=(-0.048,-0.230,-0.66)$ and $Ma=(5,7.5,10)$.
Inspection of the results presented in Figure \ref{fig:inviscid_map} demonstrates that an increase in $Ma$ and/or $At$ is destabilising. 
At $Ma=5$ and low $At$, characterised by $At=(-0.048,-0.23)$, the generation of vorticity is insufficient for the formation of the characteristic mushroom-like structure  associated with the RMI. Furthermore, at $At=-0.048$, i.e., when the fluid densities are similar in magnitude, and $Ma=(5,7.5,10)$, minimal vorticity generation and a relatively flat interface are observed with no RMI development. 

Figure \ref{fig:bcy_map} highlights the combined effects of $At$, $Ma$, and the non-Newtonian rheology in the heavy phase on RMI development. Inspection of this Figure reveals that viscous damping decelerates the development of the RMI in comparison to the inviscid case shown in Figure \ref{fig:inviscid_map} for all Atwood and Mach number values investigated. Notably, however, the flow features exhibited by both cases are qualitatively similar due to the shear-induced viscosity reduction in the BCY fluid. This is in marked contrast to the Newtonian-Newtonian case (not shown) which is significantly more stable and is accompanied by virtually no perturbation growth. 

\begin{figure}
\centering
  \includegraphics[keepaspectratio=true,scale=0.4]{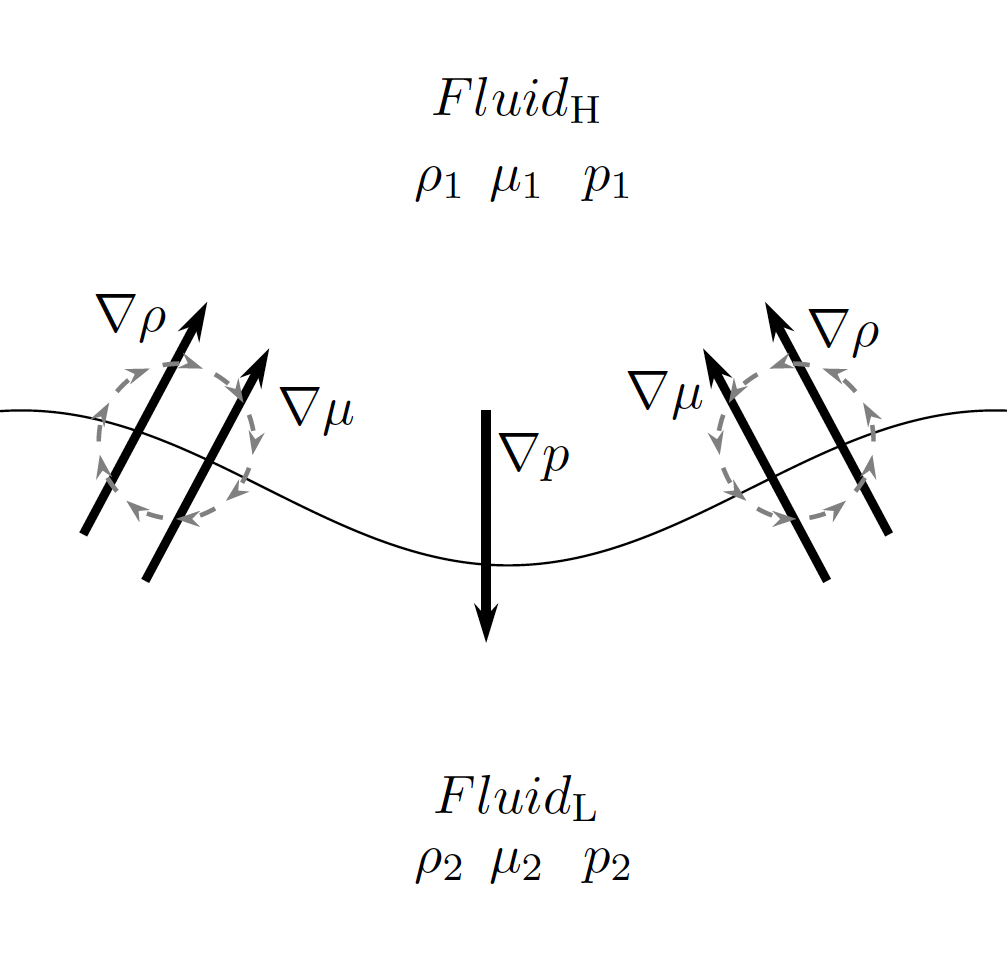}
\caption{Orientation of the pressure, density, and viscosity gradients following shock impact and protrusion of the heavier phase into the lighter one. }\label{orientation_of_gradient}
\end{figure}
%
\subsection{Mechanisms for vorticity evolution}
In the previous sections, we highlighted the significant role of vorticity in the dynamics associated with the KHI and RTI roll-up processes. It was particularly noted that vorticity generation is enhanced with increasing Atwood and Mach numbers. The non-Newtonian effects can slow down the mixing rate within the span of the KHI. In this region, the non-Newtonian effects influence the small-scale structures that play an important role in mixing. For example, in general ICF scenarios such as those studied by \citeauthor{Derentowicz_ThermonuclearFusionNeutrons}\cite{Derentowicz_ThermonuclearFusionNeutrons}, shear-thinning effects can influence the development of small-scale structures, which in turn can slow down the mixing rate of hotter and colder fluids.

In this section, we identify the mechanisms that lead to vorticity generation and damping, paying particular attention to the role of non-Newtonian rheology. In the vicinity of an interface between two immiscible fluids of different densities, it is well known that the presence of a pressure gradient as a result of hydrostatic pressure or a shock leads to vorticity generation  in RTIs and RMIs, respectively. However, when considering generalised Newtonian fluids, not only should the viscous dissipation be taken into account in the vorticity balance, but also the viscosity gradients. Taking the curl of Eq. (\ref{glg:momentum_nonDim}), the two-dimensional vorticity equation can be written in the one-fluid formulation framework, expressed by Eqs. (\ref{glg:continuity_nonDim})-(\ref{glg:energy_nonDim}) (details are in Appendix \ref{appendix}):
\begin{align}
\dfrac{D \boldsymbol{\omega}}{D t}  = & \ 
   \boldsymbol{\omega} \dfrac{Ma^2}{{\rho}} \dfrac{D {p}}{D {t}} 
   + \dfrac{1}{Re} \nabla^2 \boldsymbol{\omega}  
   + \dfrac{1}{Re} \dfrac{1}{\rho^2} \boldsymbol{\omega} \nabla \rho  \cdot \nabla \mu 
  \nonumber \\ & 
  + \dfrac{1}{\rho^2} \nabla \rho \times \left[\nabla p 
  + \dfrac{4 \, Ma^2}{3 \, Re} \mu \nabla \left(\dfrac{1}{\rho} \dfrac{D p}{D t}\right) \right] \nonumber \\ &
  + \dfrac{1}{\rho^2} \nabla \rho \times \left[
   \dfrac{1}{Re} \mu \nabla \times \boldsymbol{\omega} 
  -  \dfrac{2}{Re} \nabla \mu \cdot \nabla \mathbf{u} 
  -  \dfrac{2 \, Ma^2}{3 \, \rho \, Re} \dfrac{D p}{D t} \nabla \mu \right] 
  \nonumber \\  & 
 - \dfrac{1}{Re} \dfrac{1}{\rho} \nabla \mu  \times \left[ \dfrac{2 \, Ma^2}{3} \nabla \left( \dfrac{1}{\rho}\dfrac{D p}{D t}\right) 
 +\nabla \times \boldsymbol{\omega}\right] \nonumber \\
  & 
 + \dfrac{1}{Re} \dfrac{1}{\rho}  \nabla \times \left[ 2 \nabla \mu \cdot \nabla \mathbf{u} + \nabla \mu \times \boldsymbol{\omega} \right].  
 \label{eq:vorticity_balance}
\end{align}
Note that in the simultaneous limits $Ma \rightarrow 0$ and $Re \rightarrow \infty$, this equation reduces to the greatly simplified form of the vorticity equation associated with the inviscid case \cite{Batchelor2000} 
\begin{equation}
    \frac{D\boldsymbol{\omega}}{Dt}=\frac{1}{\rho^2}\nabla\rho \times \nabla p,
    \label{glg:baroclinic_vorti}
    \end{equation}
which implies that the baroclinic term on the right-hand-side of this equation
is responsible for vorticity generation resulting from misalignments of density and pressure at the interface. 
We illustrate in Figure \ref{orientation_of_gradient} how the major gradients such as $\nabla \rho$, $\nabla \mu$, and $\nabla p$ are oriented after the passage of the shock and the creation of the protrusion of the heavier phase into the lighter one. 
This protrusion leads to large gradients in pressure, density, and viscosity across the interface, transitioning from high values in the heavier fluid to lower ones in the lighter fluid. The misalignment of $\nabla p$ and $\nabla \rho$ leads to a destabilising baroclinic mechanism, which is present in the inviscid case, that drives the RMI. For finite $Re$, other effects compete with this mechanism. The presence of non-Newtonian rheology in one of the phases gives rise to additional mechanisms related to $\nabla\mu$.

\begin{table}
\centering
\caption{A breakdown of the rate of change of vorticity into the terms that appear on the right-hand-side of equation (\ref{eq:vorticity_balance}). The terms are characterised as `Stabilising' or `Destabilising' depending on their time-averaged contribution to this equation, and are ranked according to the magnitude of their respective contributions, e.g, `S1' and `D1' correspond to the terms with the largest stabilising and destabilising contributions, respectively. Note: Some terms, like Term 5,6,9 and 10 have mixed effects (both stabilising and destabilising at different times), but are classified by their dominant time-averaged effect.}
\begin{tabular}{c|c|c}
\hline
Term 1 &  $\omega \dfrac{Ma^2}{{\rho}} \dfrac{D {p}}{D {t}}$ 
    & Stabilising  (S5)
\\
\hline
Term 2 & $\frac{1}{Re} \nabla^2 \omega$ 
  & Stabilising (S3)\\
\hline
Term 3 &  $\dfrac{1}{Re} \dfrac{1}{\rho^2} \omega \nabla \rho  \cdot \nabla \mu$ 
& Destabilising (D3)\\
\hline
Term 4 &  $\dfrac{1}{\rho^2} \nabla \rho \times \nabla p$ 
& Destabilising (D2)\\
\hline
Term 5 &  $\dfrac{1}{\rho^2} \nabla \rho \times \dfrac{4 \, Ma^2}{3 \, Re} \mu \nabla \left(\dfrac{1}{\rho} \dfrac{D p}{D t}\right)$ 
& Destabilising (D4) \\
\hline
Term 6 &  $\dfrac{1}{\rho^2} \nabla \rho \times \dfrac{1}{Re} \mu \nabla \times \omega$ 
& Stabilising (S7) \\
\hline
Term 7 &  $-\dfrac{1}{\rho^2} \nabla \rho \times \dfrac{2}{Re} \nabla \mu \cdot \nabla \mathbf{u}$
& Stabilising (S2) \\
\hline
Term 8 & $ - \dfrac{1}{\rho^2} \nabla \rho \times \dfrac{2 \, Ma^2}{3 \, \rho \, Re} \dfrac{D p}{D t} \nabla \mu $
& Destabilising (D5) \\
\hline
Term 9 &  $-\dfrac{1}{Re} \dfrac{1}{\rho} \nabla \mu  \times \dfrac{2 \, Ma^2}{3} \nabla \left( \dfrac{1}{\rho}\dfrac{D p}{D t}\right)$ 
& Stabilising (S6) \\
\hline
Term 10 &  $ \dfrac{1}{Re} \dfrac{1}{\rho}  \nabla \times 2 \nabla \mu \cdot \nabla \mathbf{u}$ 
& Stabilising (S4) \\
\hline
Term 11 &  $- \dfrac{1}{Re} \dfrac{1}{\rho} \nabla \mu  \times\nabla \times \omega$  
& Destabilising (D1) \\
\hline
Term 12 &  $\dfrac{1}{Re} \dfrac{1}{\rho}  \nabla \times 2 \nabla \mu \cdot \nabla \mathbf{u}$ 
& Stabilising (S1)\\
\hline
\label{tab:vorticity_terms}
\end{tabular}
\end{table}

\begin{figure*}
    \centering
    \begin{minipage}{.49\textwidth}
        \centering
        \includegraphics[width=\linewidth]{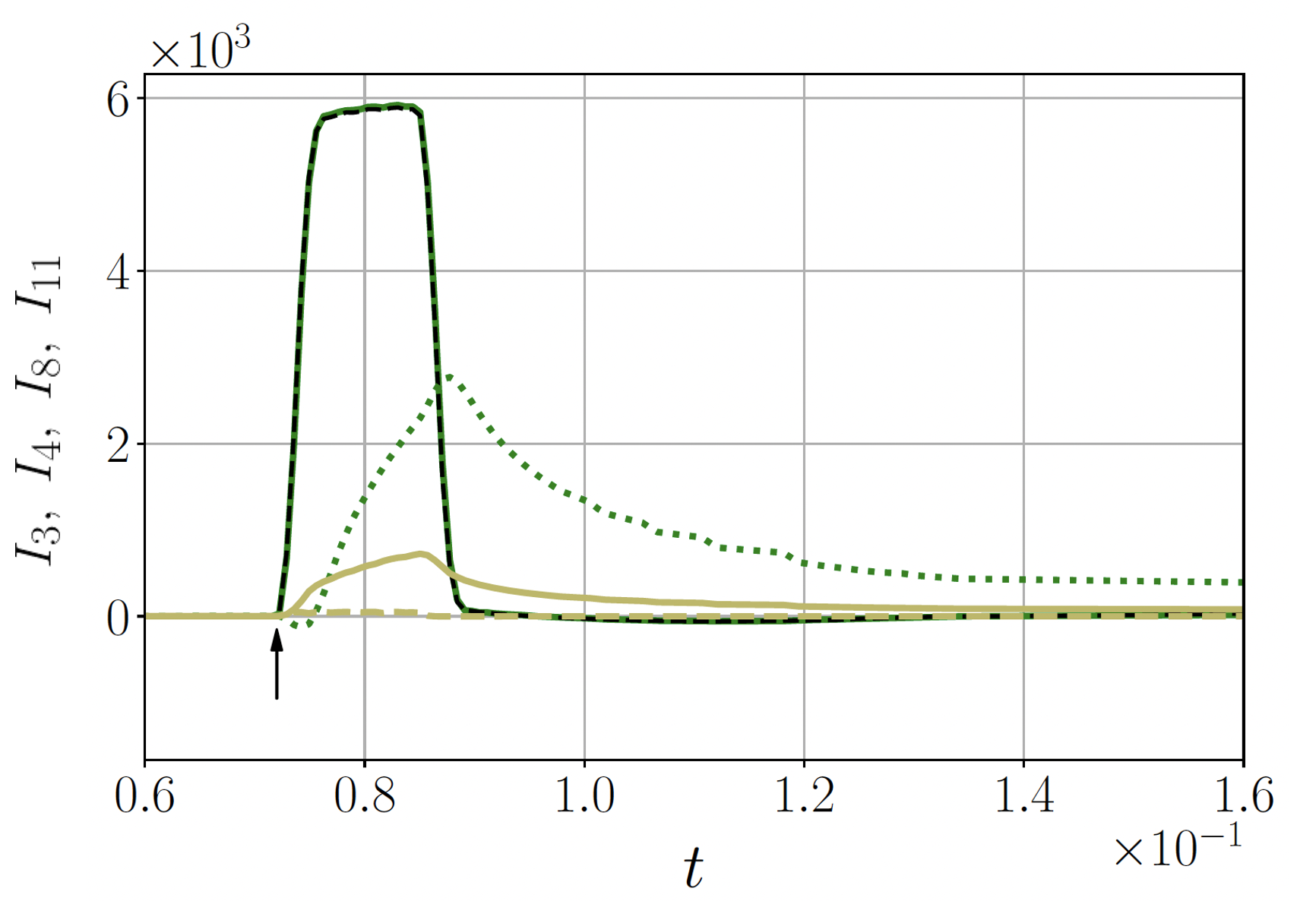}
        \subcaption{}
    \end{minipage}%
    \begin{minipage}{.485\textwidth}
        \centering
        \includegraphics[width=\linewidth]{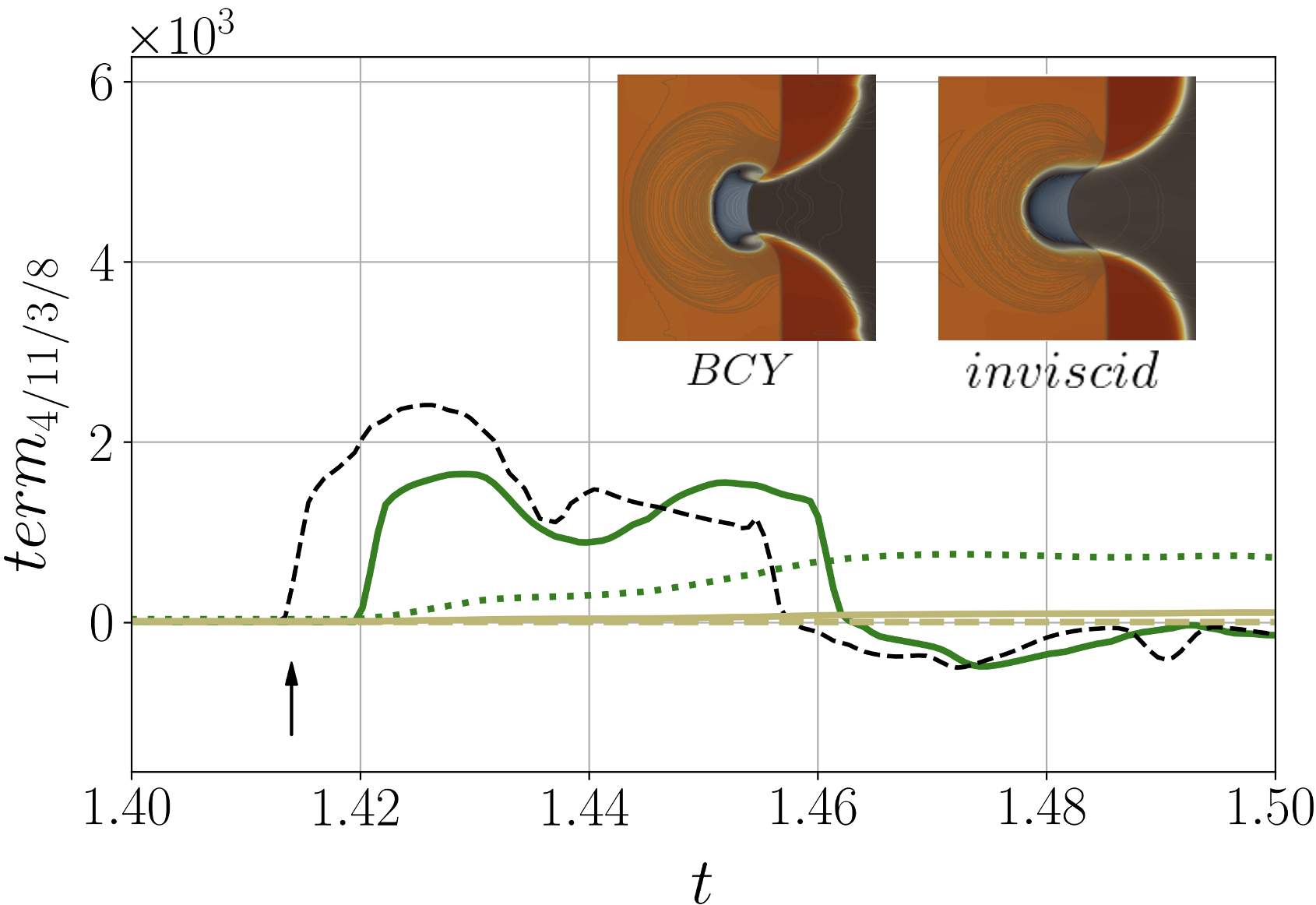}
        \subcaption{}
    \end{minipage}%
    \caption{Temporal evolution of the spatial integrals of Terms 3, 4, 8, and 11, $I_3$, $I_4$, $I_8$, and $I_{11}$ shown between $t=0.06-0.16$ and $t=1.4-1.5$ in (a) and (b), respectively, highlighting the effect of impact by the primary shock at $t\approx 0.7$ in (a), and reflected shock at $t\approx 1.415$ in (b), for \(At=-0.66\) and \(Ma=5\). The green solid and dotted lines represent $I_4$ and $I_{11}$, the solid and dashed khaki lines represent $I_3$ and $I_8$ for there case wherein the heavier and lighter phases correspond to BCY and Newtonian fluids, respectively. The dashed black line represents $I_4$ for the case wherein both fluids are treated as inviscid. The insets in panel (b) represent $x-y$ projections of the density field for the inviscid and BCY cases following the impact of the reflected shock. All other parameters mirror those from preceding Figures.}
    \label{fig:destabilisingterms}
\end{figure*}

\begin{figure*}
    \centering
    \begin{minipage}{.49\textwidth}
        \centering
        \includegraphics[width=\linewidth]{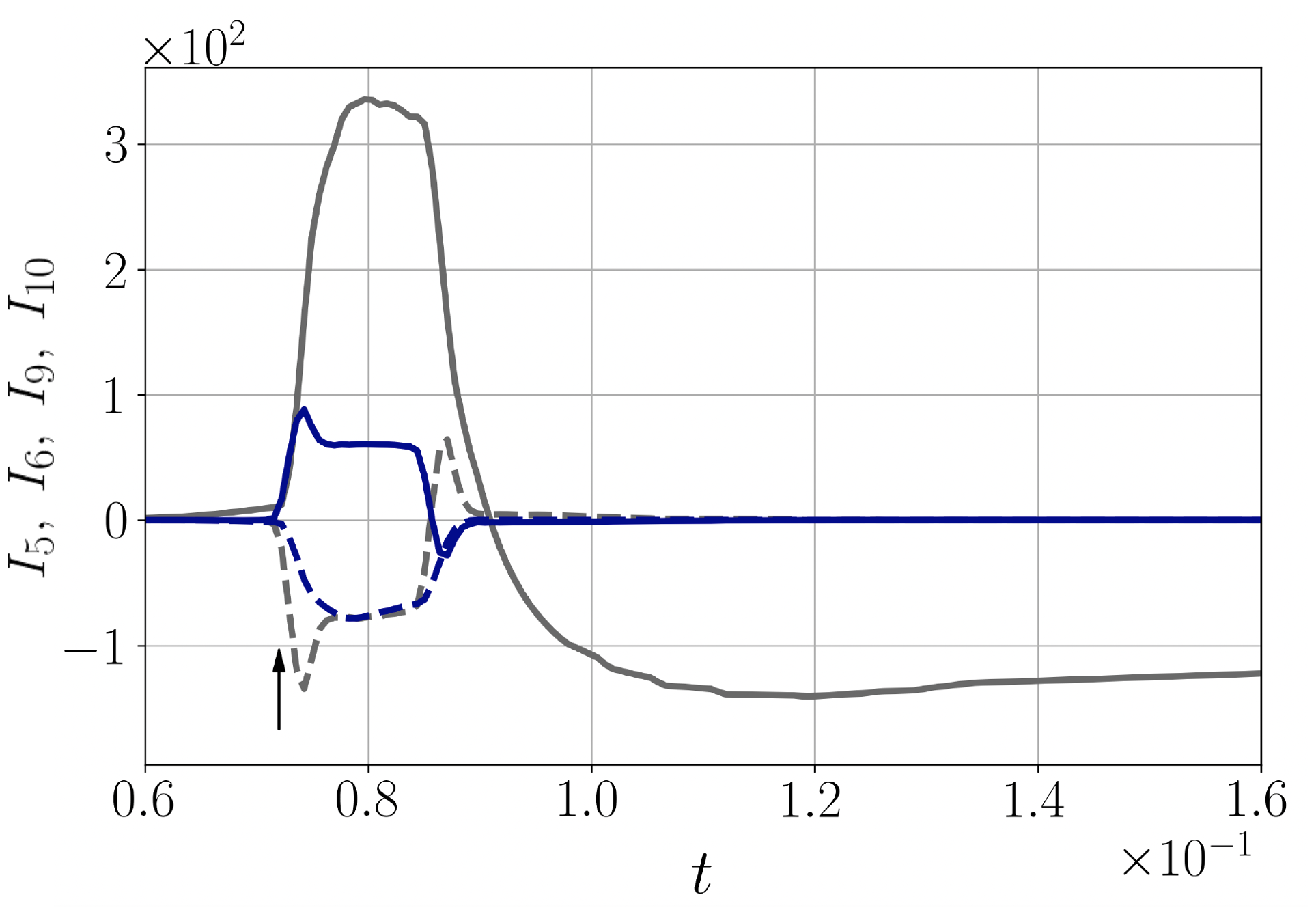}
        \subcaption{}
    \end{minipage}%
    \begin{minipage}{.49\textwidth}
        \centering
        \includegraphics[width=\linewidth]{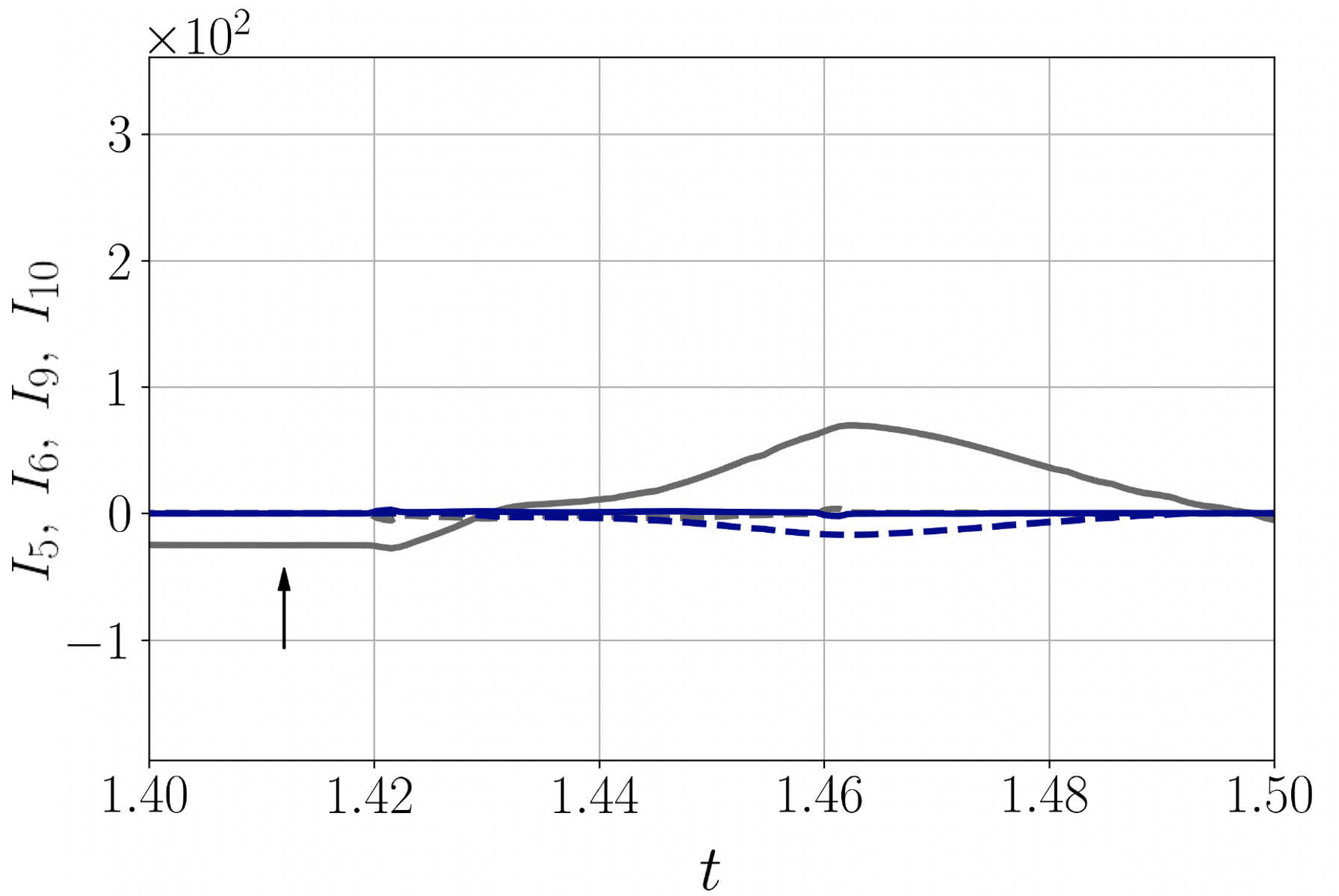}
        \subcaption{}
    \end{minipage}%
    \caption{Temporal evolution of the spatial integrals of Terms 5, 6, 9, and 10, $I_5$, $I_6$, $I_9$, and $I_{10}$ shown between $t=0.06-0.16$ and $t=1.4-1.5$ in (a) and (b), respectively, highlighting the effect of impact by the primary shock at $t\approx 0.7$ in (a), and reflected shock at $t\approx 1.415$ in (b), for \(At=-0.66\) and \(Ma=5\). The blue solid and dashed lines represent $I_5$ and $I_6$, and the gray solid and dashed lines represent $I_{10}$ and $I_9$, respectively. All other parameters mirror those from preceding Figures.}
    \label{fig:destabilising_stabilising}
\end{figure*}

\begin{figure*}
    \centering
    \begin{minipage}{.49\textwidth}
        \centering
        \includegraphics[width=\linewidth]{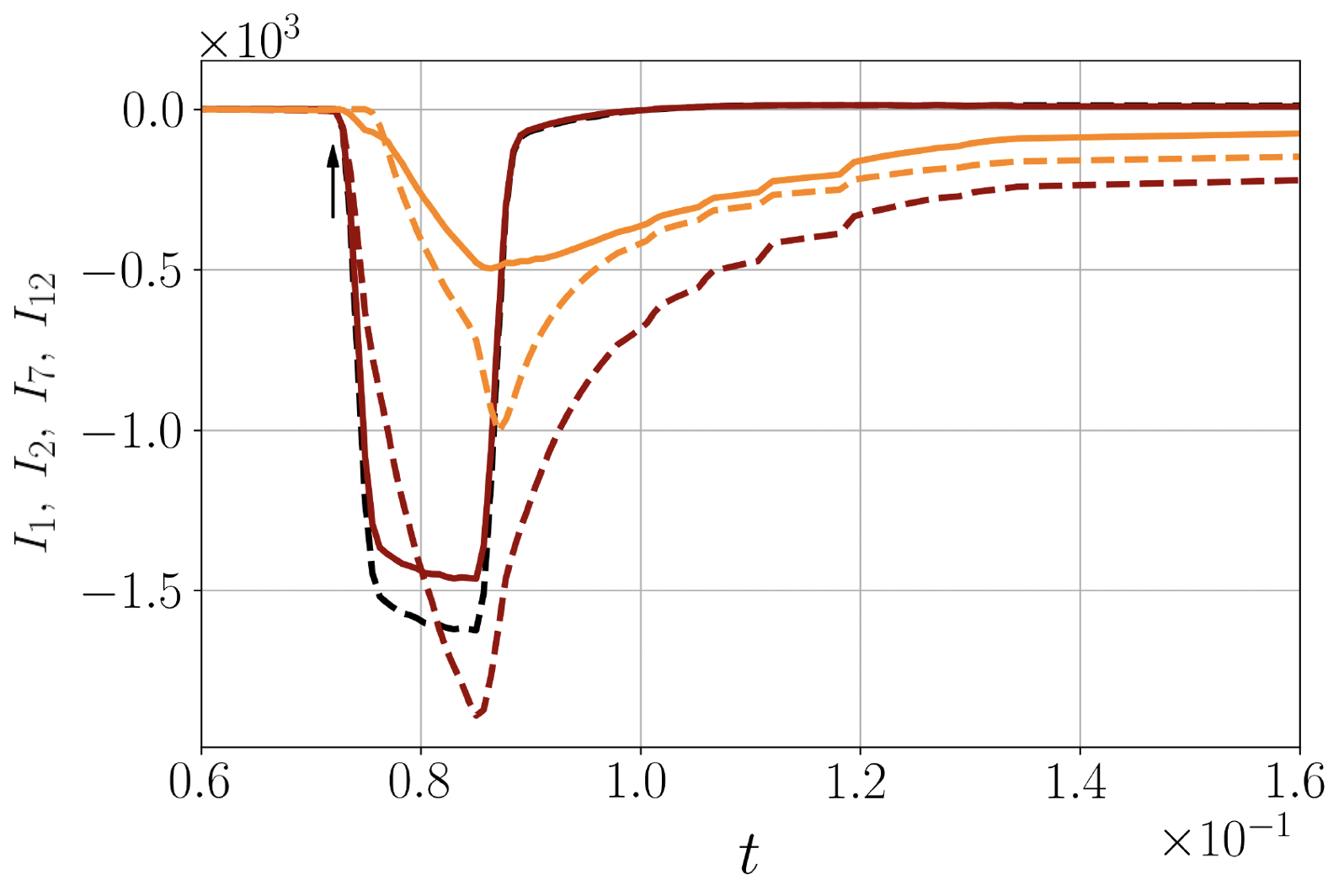}
        \subcaption{}
    \end{minipage}%
    \begin{minipage}{.49\textwidth}
        \centering
        \includegraphics[width=\linewidth]{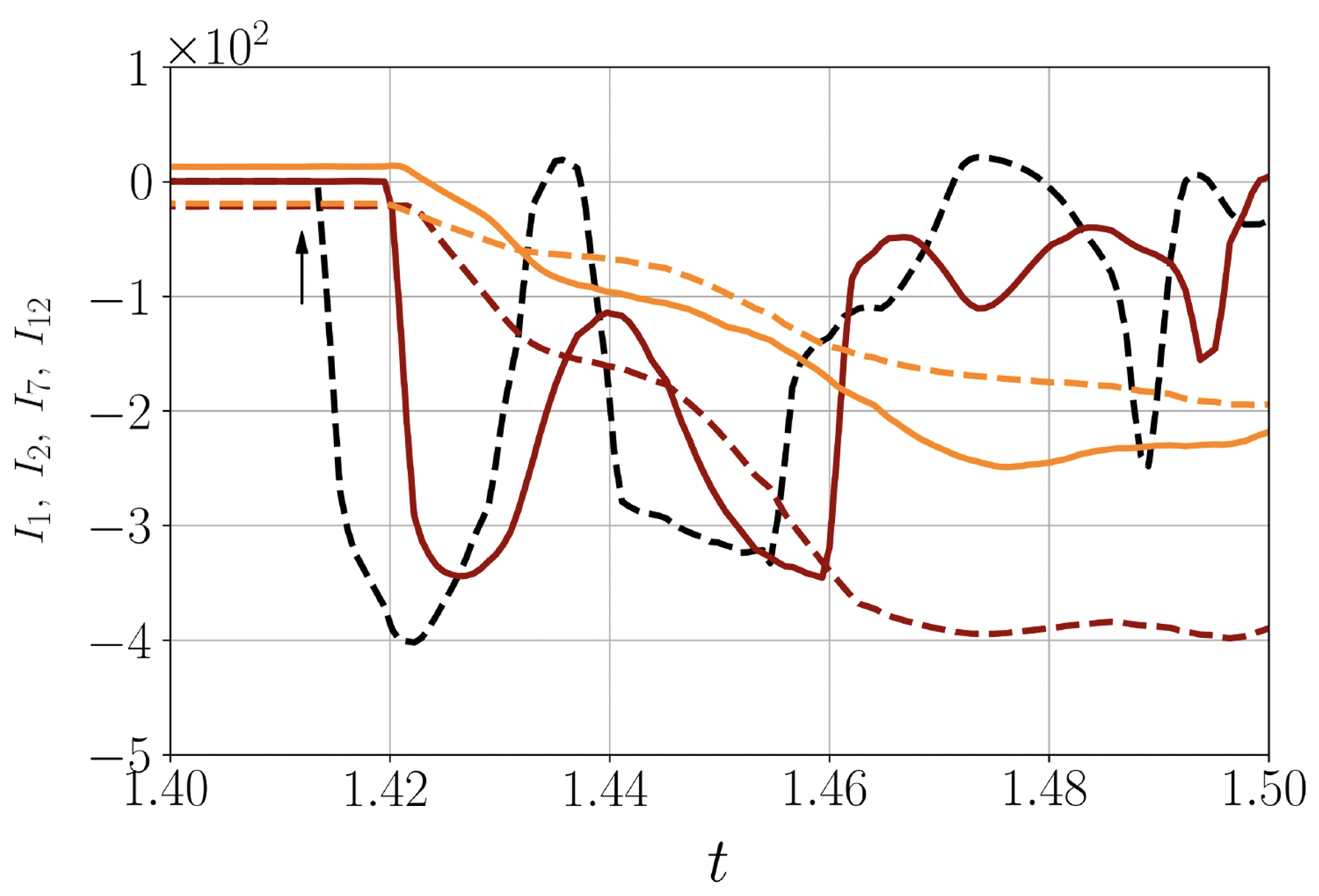}
        \subcaption{}
    \end{minipage}%
    \caption{Temporal evolution of the spatial integrals of Terms 1, 2, 7, and 12, $I_1$, $I_2$, $I_7$, and $I_{12}$ shown between $t=0.06-0.16$ and $t=1.4-1.5$ in (a) and (b), respectively, highlighting the effect of impact by the primary shock at $t\approx 0.7$ in (a), and reflected shock at $t\approx 1.415$ in (b), for \(At=-0.66\) and \(Ma=5\). The red solid and dashed lines represent $I_1$ and $I_{12}$, and the orange solid and dashed lines represent $I_2$ and $I_7$, respectively. The black dashed lines represent $I_1$ for the inviscid case. All other parameters mirror those from preceding Figures.}
    \label{fig:stabilising_terms}
\end{figure*}

We list the various terms in Table \ref{tab:vorticity_terms}, Terms 1-12  that appear in the dimensionless vorticity Eq. (\ref{eq:vorticity_balance}) derived from the dimensionless momentum equations; time is non-dimensionalised as specified in the earlier section. These dimensionless terms are integrated over half of the dimensionless wavelength for volume fractions  $\alpha=0.1$ and  $\alpha=0.9$ yielding integrals $I_1-I_{12}$, respectively, and  the temporal evolution of these integrals is then plotted in Figures \ref{fig:destabilisingterms},   \ref{fig:destabilising_stabilising}, and \ref{fig:stabilising_terms}.  These Figures provide an indication of the instantaneous contributions of $I_1-I_{12}$, and their associated terms and mechanisms, to the stability of the flow following impact by the primary shock and reflected shocks during the the early and latter stages of the flow, as shown in panels (a) and (b) of Figures \ref{fig:destabilisingterms}-\ref{fig:stabilising_terms}, respectively.  
We also obtain time-averages of $I_1-I_{12}$, which highlight the overall contributions of Terms 1-12 to the system stability over the duration of the flow investigated, and this is also indicated in Table \ref{tab:vorticity_terms}. We now examine the contribution of each term. 

Inspection of Figure \ref{fig:destabilisingterms}(a) reveals that the primary shock impact leads to a rapid rise in $I_3$, $I_4$, and $I_{11}$ which exhibit positive values highlighting the destabilising contributions of Terms 3, 4, and 11; in contrast, $I_8$ is comparatively smaller in magnitude though it is also destabilising albeit marginally. It is also seen that $I_4$ associated with the inviscid case is virtually indistinguishable from that of the BCY case.
The largest instantaneous contributor to instability is Term 4, which is related to the baroclinic mechanism discussed above, and this is followed by 
Term 11, represented by $- \frac{1}{Re} \frac{1}{\rho} \nabla \mu \times \nabla \times \omega$, a distinct source of vorticity generation specifically driven by the gradient in viscosity. 
This, in turn, is followed by the contribution from Term 3, which is represented by $\frac{1}{Re} \frac{1}{\rho^2} \boldsymbol{\omega} \nabla \rho \cdot \nabla \mu$, and emerges when the vorticity interacts with regions of sharp density and viscosity variations. Lastly, Term 8, represented by $-\frac{1}{\rho^2} \nabla \rho \times \frac{2 Ma^2}{3 \rho Re} \frac{Dp}{Dt} \nabla \mu$, is marginally destabilising and signifies the interactions between the density and viscosity gradients, the temporal variation in pressure, and the Mach number, which provides a measure of the flow compressibility. 

The ordering of the terms remains unchanged following the impact of the reflected shock, though, as shown in Figure \ref{fig:destabilisingterms}(b), the magnitudes of $I_3$, $I_4$, $I_8$, and $I_{11}$ are smaller than their early-times counterparts. 
We also note that whereas the contribution of the baroclinic term is short-lived, as can be discerned from Figure \ref{fig:destabilisingterms}(a), that of Term 11 is relatively longer-lasting, and the time-averages of $I_3$, $I_4$, $I_8$, and $I_{11}$ indicate that   
%
%
the largest contributor to instability is associated with Term 11 (see Table \ref{tab:vorticity_terms}), which exceeds Term 4 and further highlights the role of viscosity gradients, related to the non-Newtonian rheology of the heavier fluid, in driving instability. Lastly, in the insets in Fig. \ref{fig:destabilisingterms}, we show $x-y$ projections of the density field associated with the inviscid and BCY cases following the impact of the reflected shock; inspection of these Figures reveals that the interface is unstable due to the shock impact in the BCY case, albeit not to the extent of the instability observed in the absence of viscous effects.      

In a similar vain to Figure \ref{fig:destabilisingterms}, we show in Figure \ref{fig:destabilising_stabilising} the temporal evolution of the spatial integrals of Terms 5, 6, 9, and 10, $I_5$, $I_6$, $I_9$, and $I_{10}$, respectively, over the same time duration and parameter values. It is clearly seen that Terms 5 and 6, which are represented by  
$\frac{1}{\rho^2} \times \frac{4 Ma^2}{3 Re} \mu \nabla (\frac{1}{\rho}\frac{Dp}{Dt})$ and $\frac{1}{\rho^2} \nabla \rho \times \frac{1}{Re} \mu \nabla \times \omega$, are destabilising and stabilising, respectively, following the impact of the primary shock, as shown in Figure \ref{fig:destabilising_stabilising}(a); the impact of the reflected shock appears to have virtually no effect on $I_5$, while $I_6$ is slightly negative. This suggests that Terms 5 and 6 have a weakly stabilising influence on the late-stage dynamics.
Although Term 9, represented by $-\frac{1}{Re} \frac{1}{\rho} \nabla \mu \times \frac{2 Ma^2
}{3} \nabla (\frac{1}{\rho}\frac{Dp}{Dt})$, makes a relatively modest, and mostly stabilising, contribution to the dynamics, Term 10, represented by 
$ \dfrac{1}{Re} \dfrac{1}{\rho}  \nabla \times 2 \nabla \mu \cdot \nabla \mathbf{u}$ is destabilising immediately following primary shock impact and strongly stabilising thereafter as manifested by the rise in $I_{10}$.
This is because the shock impact gives rise to sharp velocity and viscosity gradients whose coupling through the Term 10 mechanism induces vorticity generation. The reflected shock impact also leads to an increase in $I_{10}$. The overall contribution to the flow, however, is destabilising due to Term 5, and stabilising due to Terms 6, 9, and 10. In the case of the latter, it appears that its stabilising long-time influence outweighs its destabilising effect during the shock impact episodes at the early- and late-stage dynamics.  

\begin{figure*}
  \begin{minipage}{0.43\textwidth}
    \centering
    \includegraphics[width=\linewidth]{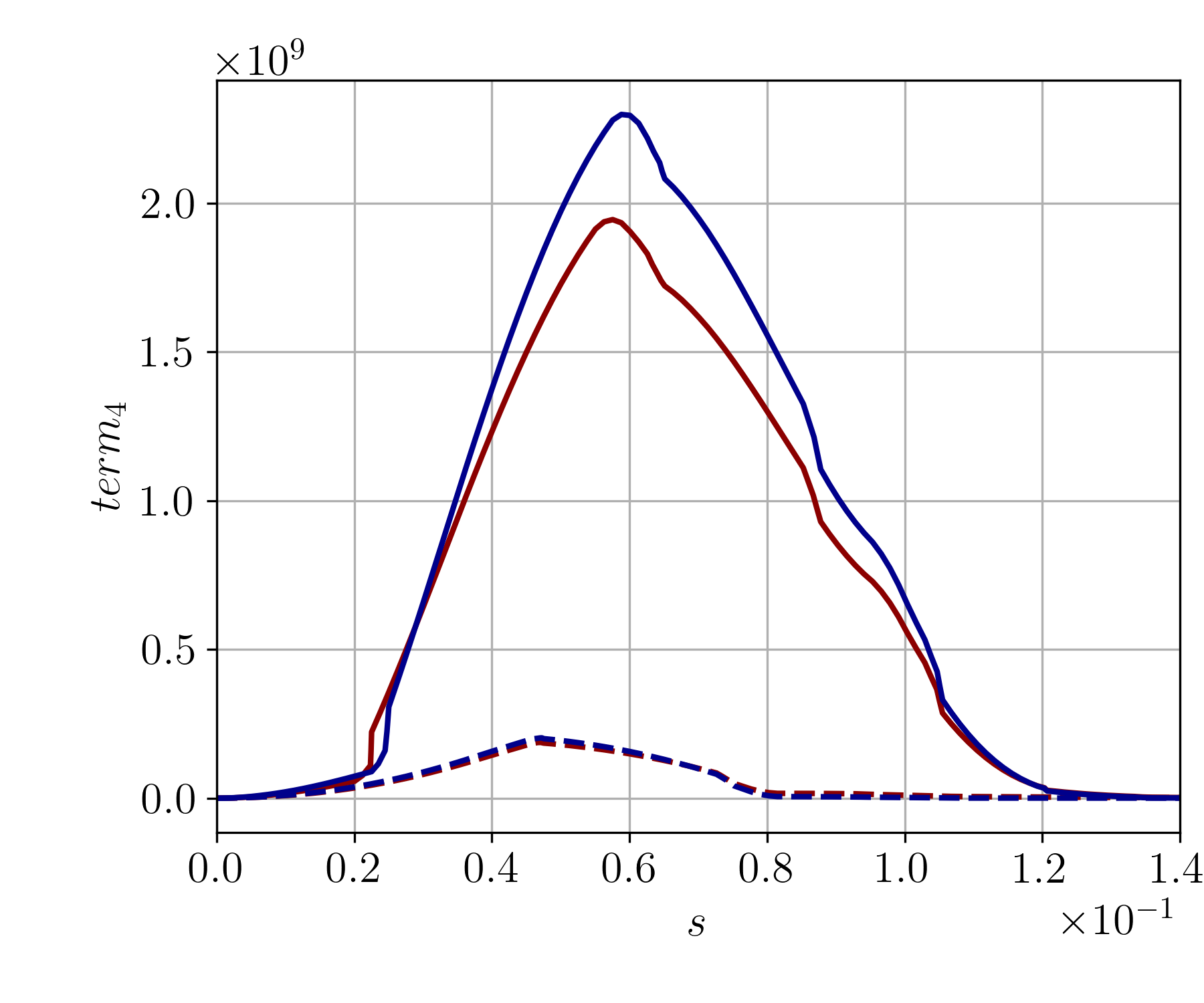}
    \subcaption{}
    \label{subfig:label1}
  \end{minipage}
  \begin{minipage}{0.43\textwidth}
    \centering
    \includegraphics[width=\linewidth]{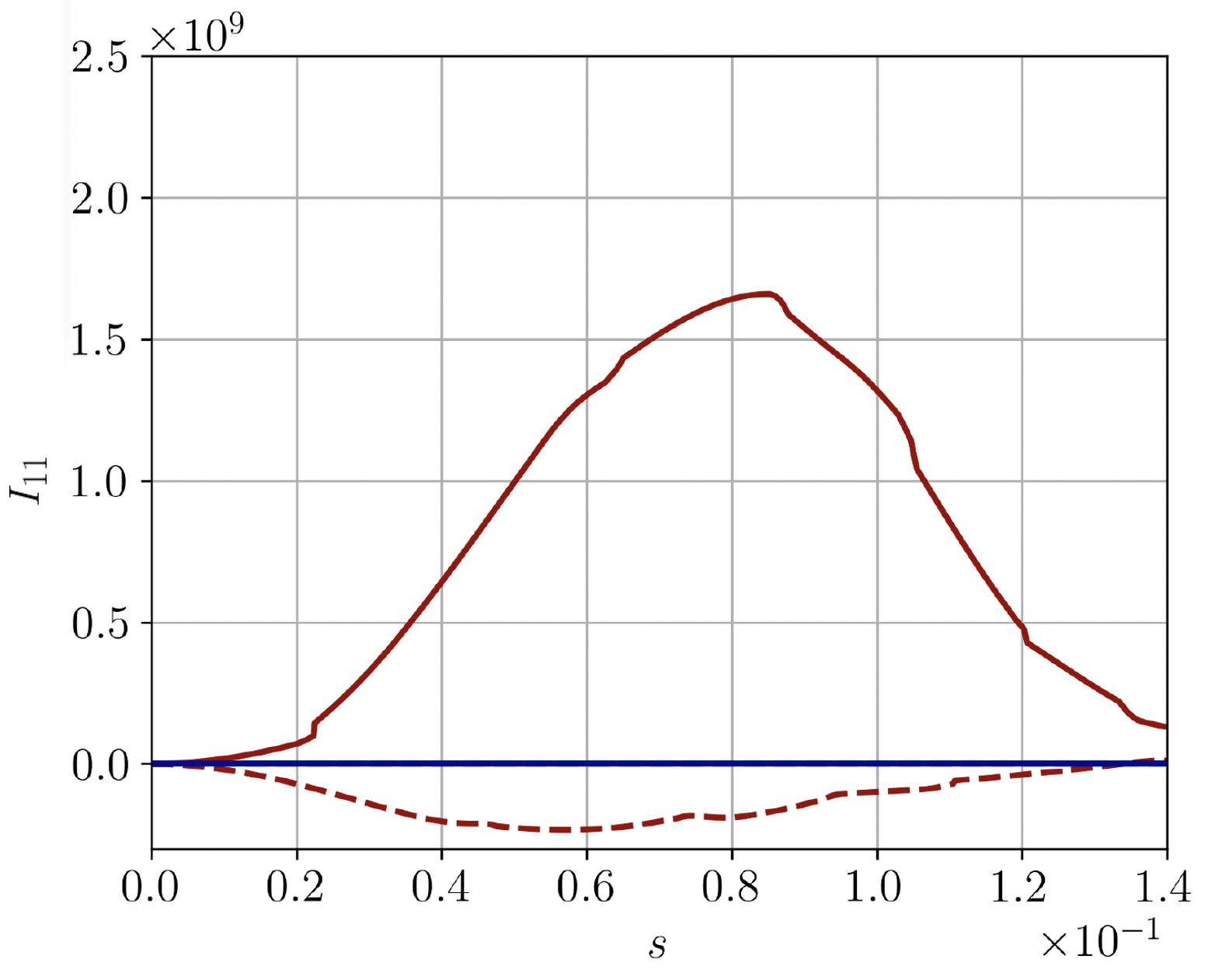}
    \subcaption{}
    \label{subfig:label2}
  \end{minipage}

  \begin{minipage}{0.43\textwidth}
    \centering
    \includegraphics[width=\linewidth]{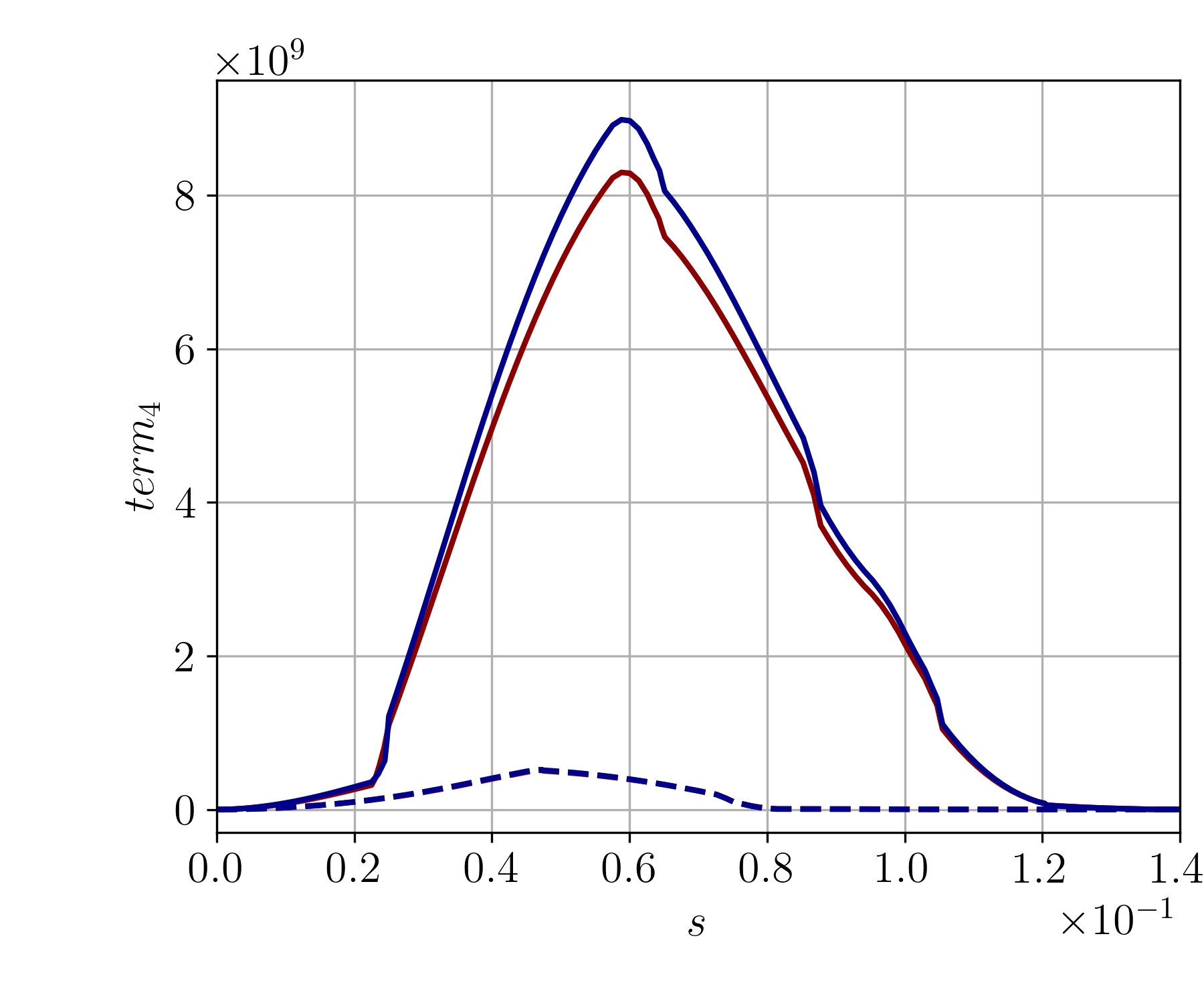}
    \subcaption{}
    \label{subfig:label3}
  \end{minipage}
  \begin{minipage}{0.43\textwidth}
    \centering
    \includegraphics[width=\linewidth]{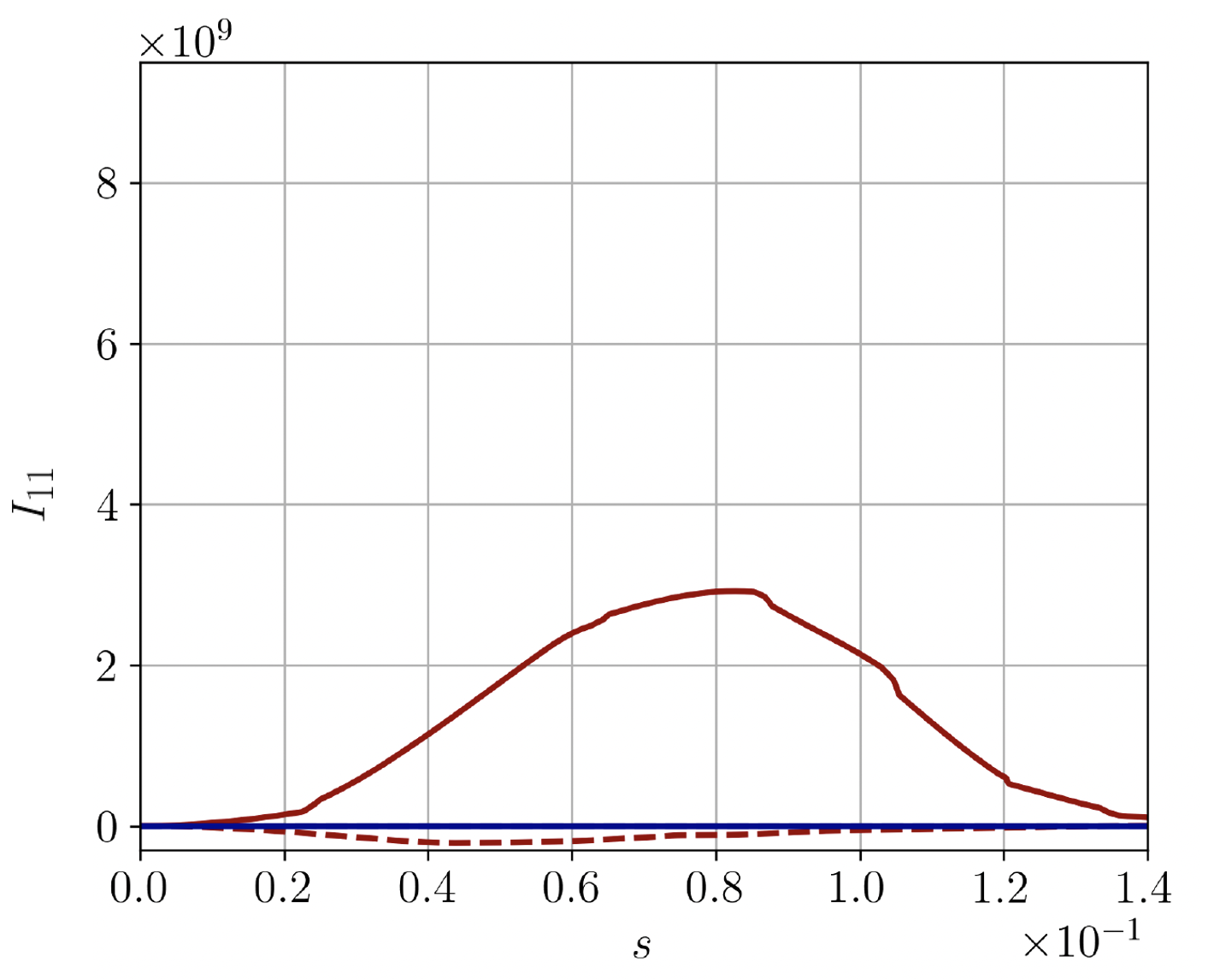}
    \subcaption{}
    \label{subfig:label4}
  \end{minipage}

  \begin{minipage}{0.43\textwidth}
    \centering
    \includegraphics[width=\linewidth]{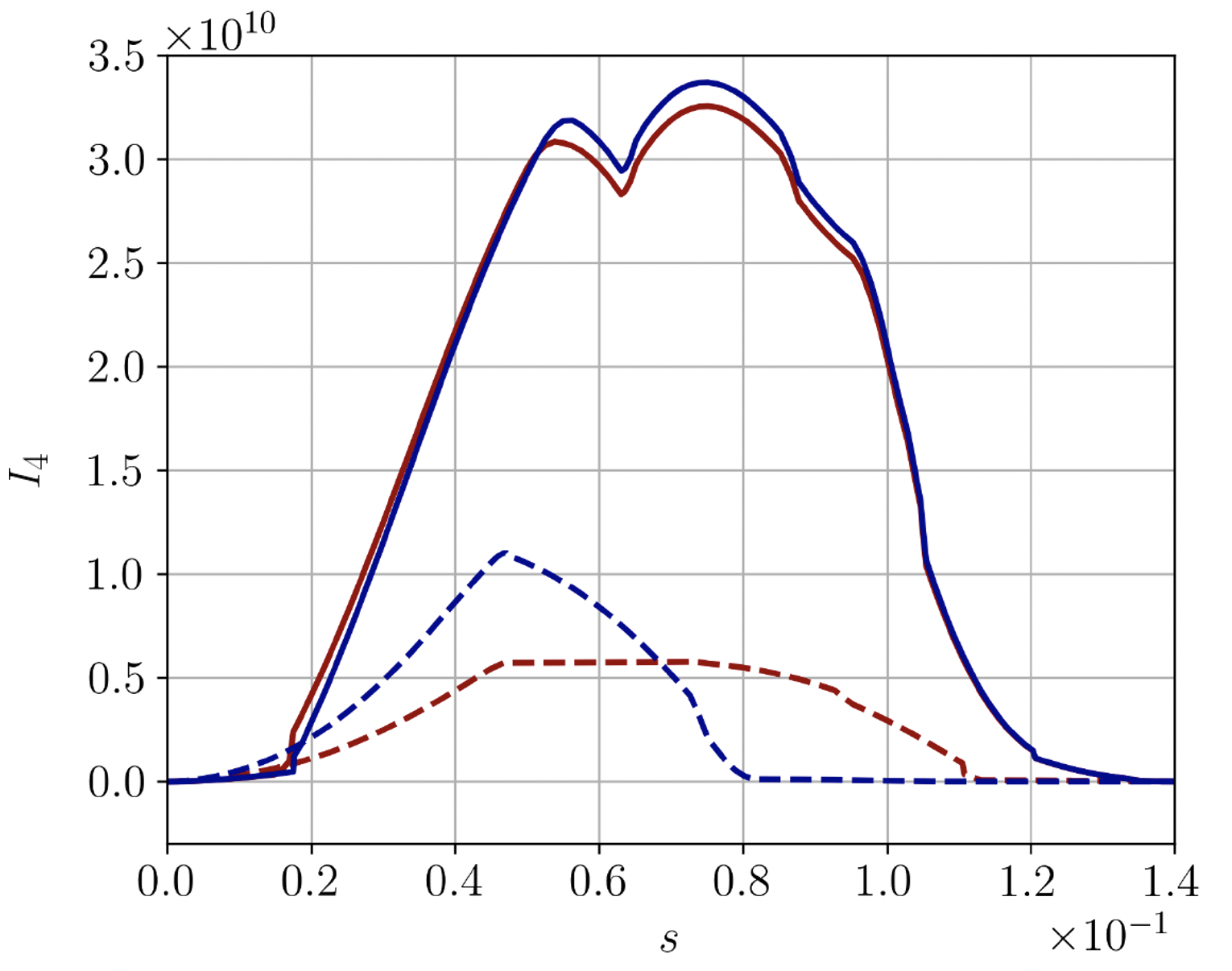}
    \subcaption{}
    \label{subfig:label5}
  \end{minipage}
  \begin{minipage}{0.43\textwidth}
    \centering
    \includegraphics[width=\linewidth]{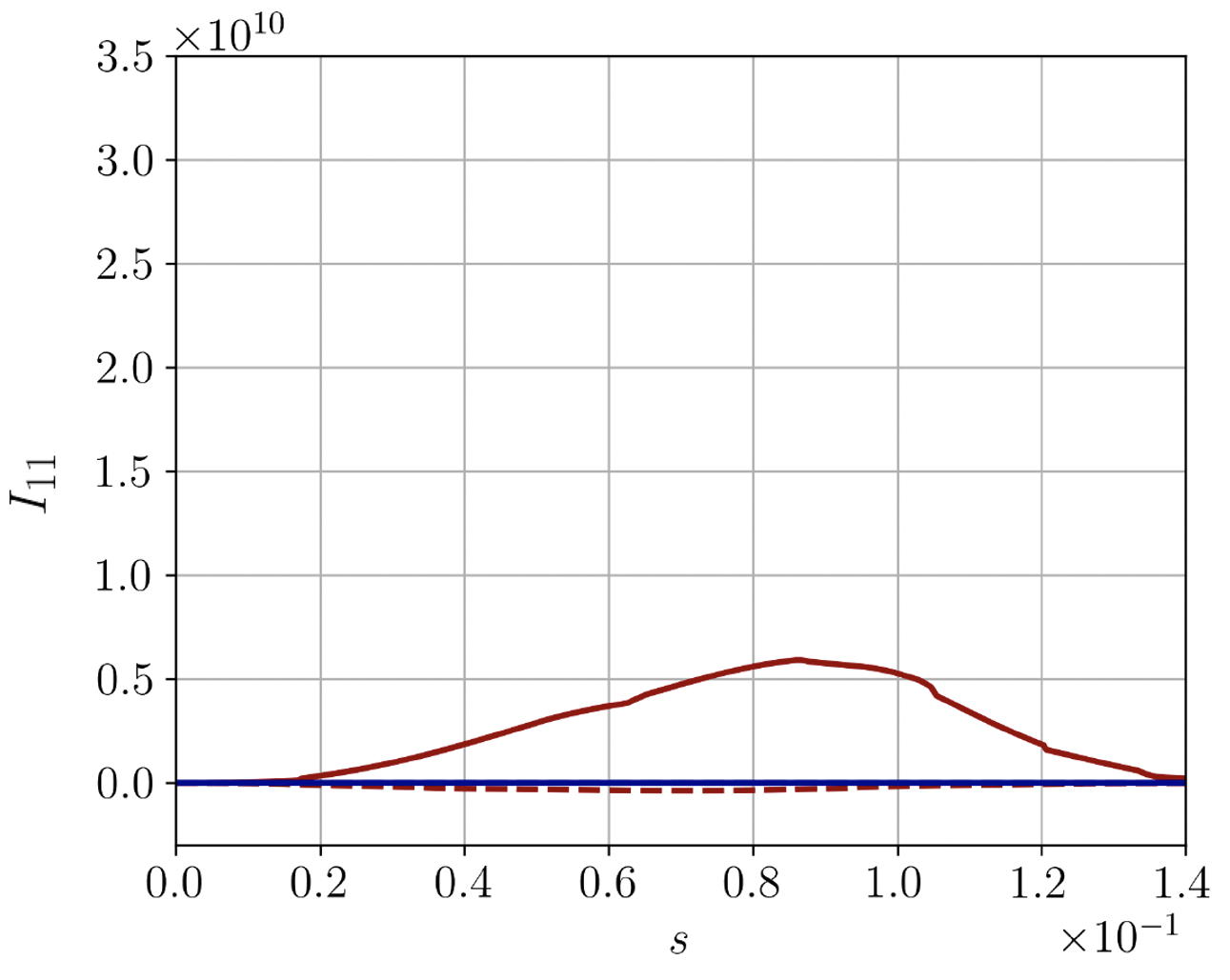}
    \subcaption{}
    \label{subfig:rt}
  \end{minipage}

  \caption{Variation of the arc length integrals of terms 4 and 11, $I^{(s)}_4$ and $I^{(s)}_{11}$, respectively, when the primary shock impacts the interface with $At$ = -0.66 for $Ma=2.5$, $Ma=5$, and $Ma=10$ shown in (a) and (b), (c) and (d), and (e) and (f), respectively. The definition of Terms 4 and 11 is provided in Table \ref{tab:vorticity_terms}. The blue and red lines represent the inviscid case and that wherein the heavier and lighter phases correspond to BCY and Newtonian fluids, respectively. The solid and dashed lines represent the results associated with $\alpha=0.9$ and $\alpha=0.1$, respectively. }
  \label{fig:Vortiicty_arc_lenght_various_Mach}
\end{figure*}

In Figure \ref{fig:stabilising_terms}, we assess the contributions due to Terms 1, 2, 7, and 12.
Term 1, represented by $\boldsymbol{\omega} \frac{Ma^2}{\rho} \frac{ Dp}{Dt}$, is associated with stretching and tilting of vorticity in a flow field, while Term 2, 
$\frac{1}{Re} \, \nabla^2 \boldsymbol{\omega}$, reflects the damping effect of viscosity on vorticity by converting the kinetic energy of the vortical motion into internal energy. When the vorticity and velocity divergences are aligned, the stretching can lead to damping. 
Damping is also provided by Term 7, $\frac{1}{\rho^2}  \nabla \rho \times \frac{2}{Re} \nabla \mu \cdot \nabla \mathbf{u}$, and Term 12, $\frac{1}{Re} \frac{1}{\rho} \nabla \times \nabla \mu \times \boldsymbol{\omega}$, which are related to retardation of the secondary, KHI that can arise in the nonlinear phase of RMI evolution. By hindering the development of these secondary instabilities, viscosity can indirectly damp the overall vorticity in the flow by damping the small-scale structures that form due to the interaction of multiple reflected shocks with the interface in the nonlinear stages of the flow. Confirmation of the damping, vorticity-attenuating effect of Terms 1, 2, 7, and 12 is provided by the temporal evolution of the spatial integrals of these terms, which are shown in Figure \ref{fig:stabilising_terms} as well as their respective time-averages as highlighted in Table \ref{tab:vorticity_terms}. 
In Figure \ref{fig:Vortiicty_arc_lenght_various_Mach}, we show the temporal variation of the major vorticity-generating terms, Terms 4 and 11, integrated along half of the arc length, $I^{(s)}_4$ and $I^{(s)}_{11}$, respectively. This analysis is performed when the shock wave interacts with the interface at Mach numbers 2.5, 5, and 10, with $At=-0.66$, both in the inviscid case and in situations where the denser and lighter phases are represented by BCY and Newtonian fluids, respectively;  results were generated for $\alpha=0.1$ and $\alpha=0.9$. It is seen that the magnitude of Term 4 associated with  inviscid flow is larger than that for the viscous case, as expected, for all $Ma$ considered, and higher in the heavier than the lighter fluid. It can also be observed that although both $I^{(s)}_4$ and $I^{(s)}_{11}$ increase with $Ma$, the ratio of the magnitudes of $I^{(s)}_4$ to $I^{(s)}_{11}$ 
increases. This indicates that the baroclinic mechanism driving vorticity growth gains in significance with increasing $Ma$ relative to the one associated with viscosity gradients arising from non-Newtonian effects in the heavier phase. 


\section{Conclusions}
We investigated the effects of non-Newtonian rheology on the development of a Richtmyer-Meshkov instability (RMI) in flows characterised by negative Atwood numbers wherein a shock is initiated in the heavier, non-Newtonian fluid, and travels towards the lighter, Newtonian one. 
In order to resolve the shock and interfacial dynamics, a one-fluid formulation for the mass, momentum, and energy equations for a compressible flow was used.  A shear-thinning rheology was considered for the heavier phase, and a stiffened gas equation-of-state was used to relate the pressure to the density. 

A volume-of-fluid approach was utilised to capture the interface, and the open-source software \textit{blastFOAM} was employed to obtain numerical solutions of the governing equations for a range of Atwood and Mach numbers and a fixed set of rheological parameters for the heavier phase. We also derived a two-dimensional vorticity transport equation, and using a budget analysis of this equation, we identified terms that generate and dampen vorticity, which provide valuable insights into the interplay between non-Newtonian effects, vorticity generation, and RMI growth in various regimes.

Our results demonstrated that the incident shock creates a viscosity gradient along the interface, which generates additional vorticity due to non-Newtonian effects compared to the inviscid case, which was used as a benchmark. This additional vorticity generation is more pronounced at moderate Mach numbers $Ma$ ($< Ma=5$) than at high Mach numbers ($Ma=10$) because lower shear rates in the former $Ma$ range result in a higher residual viscosity gradient. 

After the passage of the shock, the damping effects of viscosity become dominant, leading to a reduction in the vorticity. The reflected shock then generates additional vorticity due to the non-Newtonian viscosity gradient. For the low-to-moderate $Ma$ range, the damping effects of viscosity hinder the development of secondary instabilities, such as the Kelvin-Helmholtz instability (KHI), leading to lower RMI growth. This observation might seem counterintuitive. One might intuitively assume that suppressing secondary instabilities would lead to enhanced growth of RMI, given that such instabilities act as a mechanism to limit or saturate growth. However the dampening effects of viscosity are not only present in nonlinear phase, where the viscosity effects dampen the onset of KHI at low-moderate Mach numbers. The dampening effects of viscosity also play a crucial role in the linear phase. At high $Ma$, the residual viscosity is low due to the very high shear rates associated with high-$Ma$ flows, resulting in low rates of vorticity generation associated with non-Newtonian effects. Furthermore, viscous effects in these situations damp out the small-scale structure, which could have given rise to vortex-accelerated vorticity deposition that would have enhanced the RMI and associated mixing.

Although neglected in the present work, incorporating temperature dependence into the viscosity model aligns the growth rate more closely with the inviscid scenario, attributed to the substantial reduction of viscosity at elevated temperatures. The significance of viscosity becomes especially apparent within the regime span, marked by the presence of small-scale structures. These structures are instrumental in augmenting the generation of baroclinic vorticity through vortex acceleration, a phenomenon thoroughly investigated by Peng et al. \cite{peng_vortex-accelerated_2003}. A reduced occurrence of these small-scale structures could lead to reduced mixing, presenting a potential area for further exploration as an extension of the present work.

\begin{acknowledgments}
We wish to acknowledge financial support from First Light Fusion for a PhD studentship for us, and the Engineering and Physical Sciences Research Council, UK, through the PREMIERE Programme Grant PREMIERE (grant number, EP/T000414/1). We also acknowledge use of the High Performance Computing facilities provided by the Research Computing Service (RCS) of Imperial College London for the computing time.
\end{acknowledgments}

\section*{Author Declarations}
\subsection*{Conflict of Interest}
The authors have conflicts to disclose. 

\subsection*{Author Contributions}
\textbf{Usman Rana:} Conceptualization (equal); Formal analysis (equal); Writing - review \& editing (equal); 
\textbf{Thomas Abadie:} Conceptualization (equal); Formal analysis (equal); Writing - review \& editing (equal); 
\textbf{David Chapman:} Funding acquisition (equal); Writing - review \& editing (equal); 
\textbf{Nathan Joiner:} Funding acquisition (equal); Writing - review \& editing (equal); 
\textbf{Omar Matar:} Funding acquisition (equal); Conceptualization (equal); Writing - review \& editing (equal).

\section*{Data Availability}
The data that support the findings of this study are available from the corresponding author upon reasonable request.

\appendix

\section{Two-dimensional vorticity equation for compressible, non-Newtonian fluids}
\label{appendix}
Starting from the momentum equation, which may be expressed as follows
\begin{equation}
\dfrac{\partial \mathbf{u}}{\partial t} + \left( \mathbf{u} \cdot \nabla \right) \mathbf{u} = - \dfrac{1}{\rho} \nabla p + \dfrac{1}{\rho} \mathbf{\nabla} \cdot \left(\mu \left[\mathbf{\nabla u} + \left(\mathbf{\nabla u}\right)^{T} - \frac{2}{3}\left(\mathbf{\nabla \cdot u}\right)\mathbf{I}\right]\right),
\label{eq:app_momentum}
\end{equation}
we can re-write the viscous term as follows:
\begin{align}
& \
\mathbf{\nabla} \cdot \left(\mu \left[\mathbf{\nabla u} + \left(\mathbf{\nabla u}\right)^{T} - \frac{2}{3}\left(\mathbf{\nabla \cdot u}\right)\mathbf{I}\right]\right) = 
\mu \mathbf{\nabla} \cdot \left( \left[\mathbf{\nabla u} + \left(\mathbf{\nabla u}\right)^{T}\right]\right) \nonumber \\
& - \frac{2}{3}\mu\mathbf{\nabla}\cdot\left(\left[\left(\mathbf{\nabla \cdot u}\right)\mathbf{I}\right]\right)  + \nabla \mu \cdot \left( \left[\mathbf{\nabla u} + \left(\mathbf{\nabla u}\right)^{T} - \frac{2}{3}\left(\mathbf{\nabla \cdot u}\right)\mathbf{I}\right]\right).
\label{eq:app_viscous_stress}
\end{align}

We use the following identities
\begin{align}
\nabla \cdot \left( \nabla \mathbf{u} \right) & = \nabla \left( \nabla \cdot \mathbf{u} \right) - \nabla \times \left( \nabla \times \mathbf{u} \right), \\
\nabla \cdot \left( \nabla \mathbf{u} \right)^T & = \nabla \left( \nabla \cdot \mathbf{u} \right), \\
\nabla \cdot \left(\left( \nabla \cdot \mathbf{u} \right) \mathbf{I} \right) & = \nabla \left( \nabla \cdot \mathbf{u} \right), 
\end{align}
to write the first two terms on the right-hand-side of Eq. (\ref{eq:app_viscous_stress}) as
\begin{align}
\mu \mathbf{\nabla} \cdot \left( \left[\mathbf{\nabla u} + \left(\mathbf{\nabla u}\right)^{T} - \frac{2}{3}\left(\mathbf{\nabla \cdot u}\right)\mathbf{I}\right]\right)  = 
\mu \left( \dfrac{4}{3} \nabla \left( \nabla \cdot \mathbf{u}\right) - \nabla \times \boldsymbol{\omega} \right),
\label{eq:intermediate_eq1}
\end{align}
where $\boldsymbol{\omega}=\nabla\times\mathbf{u}$ is the vorticity.
In addition, we can also write the following relation
\begin{align}
\nabla \mu \cdot \nabla \mathbf{u}^T & = \nabla \mu \cdot \nabla \mathbf{u} + \nabla \mu \times \left( \nabla \times \mathbf{u} \right),
\end{align}
so that the third term on the right-hand-side of Eq. (\ref{eq:app_viscous_stress}) can be written as follows:
\begin{align}
\nabla \mu \cdot \left( \left[\mathbf{\nabla u} + \left(\mathbf{\nabla u}\right)^{T} - \frac{2}{3}\left(\mathbf{\nabla \cdot u}\right)\mathbf{I}\right]\right)  =
& \
2 \nabla \mu \cdot \nabla \mathbf{u} + \nabla \mu \times \boldsymbol{\omega} \nonumber \\ & - \dfrac{2}{3} \left( \nabla \cdot \mathbf{u} \right) \nabla \mu.
\label{eq:intermediate_eq2}
\end{align}

Insertion of Eqs. (\ref{eq:intermediate_eq1}) and (\ref{eq:intermediate_eq2}) into Eq. (\ref{eq:app_momentum}), and taking the curl of the resultant equation, leads to the following equation for the vorticity:
\begin{align}
\dfrac{D \boldsymbol{\omega}}{D t}  = & \ \boldsymbol{\omega} \cdot \nabla \mathbf{u} - \boldsymbol{\omega} \left(\nabla \cdot \mathbf{u} \right) 
  -\dfrac{1}{\rho^2} \left(\nabla \rho \times \nabla p \right) \nonumber \nonumber \\
 & + \nabla \times \left[\dfrac{\mu}{\rho} \left( \dfrac{4}{3} \nabla \left( \nabla \cdot \mathbf{u}\right) - \nabla \times \boldsymbol{\omega} \right) \right] \nonumber \\
 & + \nabla \times \left[ \dfrac{1}{\rho} \left( 2 \nabla \mu \cdot \nabla \mathbf{u} + \nabla \mu \times \boldsymbol{\omega} - \dfrac{2}{3} \left( \nabla \cdot \mathbf{u} \right) \nabla \mu \right) \right].
 \label{eq:app_vorticity}
\end{align}

For two-dimensional (2D) flows, $\boldsymbol{\omega} \cdot \nabla \mathbf{u} = 0$, and Eq. (\ref{eq:app_vorticity}) can be expressed as follows:   
\begin{align}
\dfrac{D \boldsymbol{\omega}}{D t}  = & \ - \boldsymbol{\omega} \left(\nabla \cdot \mathbf{u} \right) 
  -\dfrac{1}{\rho^2} \left(\nabla \rho \times \nabla p \right) + \dfrac{\mu}{\rho} \nabla^2 \boldsymbol{\omega} \nonumber \\
 & + \dfrac{1}{\rho} \nabla \mu  \times \left[\left( \dfrac{4}{3} \nabla \left( \nabla \cdot \mathbf{u}\right) - \nabla \times \boldsymbol{\omega} \right) \right] \nonumber \\
 & - \dfrac{\mu}{\rho^2} \nabla \rho  \times \left[\left( \dfrac{4}{3} \nabla \left( \nabla \cdot \mathbf{u}\right) - \nabla \times \boldsymbol{\omega} \right) \right] \nonumber \\
 & + \dfrac{1}{\rho}  \nabla \times \left[ \left( 2 \nabla \mu \cdot \nabla \mathbf{u} + \nabla \mu \times \boldsymbol{\omega} - \dfrac{2}{3} \left( \nabla \cdot \mathbf{u} \right) \nabla \mu \right) \right] \nonumber \\
 & - \dfrac{1}{\rho^2} \nabla \rho  \times \left[ \left( 2 \nabla \mu \cdot \nabla \mathbf{u} + \nabla \mu \times \boldsymbol{\omega} - \dfrac{2}{3} \left( \nabla \cdot \mathbf{u} \right) \nabla \mu \right) \right].
\end{align}
Using $\nabla \times \left[ \left(\nabla \cdot \mathbf{u} \right) \nabla \mu \right] = \nabla \mu \times \nabla \left( \nabla \cdot \mathbf{u}\right)$ and $\boldsymbol{\omega} \cdot \nabla = 0$ (valid in 2D), this equation can be slightly rearranged as follows:
%
\begin{align}
\dfrac{D \boldsymbol{\omega}}{D t}  = & \ 
  - \boldsymbol{\omega} \left(\nabla \cdot \mathbf{u} \right) + \dfrac{\mu}{\rho} \nabla^2 \boldsymbol{\omega}  + \dfrac{1}{\rho^2} \boldsymbol{\omega} \nabla \rho  \cdot \nabla \mu 
  \nonumber \\ & 
  + \dfrac{1}{\rho^2} \nabla \rho \times \left[\nabla p - \mu \dfrac{4}{3} \nabla \left( \nabla \cdot \mathbf{u}\right) + \mu \nabla \times \boldsymbol{\omega}\right] \nonumber \\ &
  - \frac{1}{\rho^2} \nabla{\rho}\times\left[2 \nabla \mu \cdot \nabla \mathbf{u} + \dfrac{2}{3}\left( \nabla \cdot \mathbf{u} \right) \nabla \mu \right] 
  \nonumber \\  & 
 +\dfrac{1}{\rho} \nabla \mu  \times \left[ \dfrac{2}{3} \nabla \left( \nabla \cdot \mathbf{u}\right) 
 -\nabla \times \boldsymbol{\omega}\right] \nonumber \\
  & 
 + \dfrac{1}{\rho}  \nabla \times \left[ 2 \nabla \mu \cdot \nabla \mathbf{u} + \nabla \mu \times \boldsymbol{\omega} \right].  
\end{align}

Finally, using the scalings utilised in Section 2, the dimensionless form of the 2D vorticity equation is expressed by:
\begin{align}
\dfrac{D \boldsymbol{\omega}}{D t}  = & \ 
   \boldsymbol{\omega} \dfrac{Ma^2}{{\rho}} \dfrac{D {p}}{D {t}} 
   + \dfrac{1}{Re} \nabla^2 \boldsymbol{\omega}  
   + \dfrac{1}{Re} \dfrac{1}{\rho^2} \boldsymbol{\omega} \nabla \rho  \cdot \nabla \mu 
  \nonumber \\ & 
  + \dfrac{1}{\rho^2} \nabla \rho \times \left[\nabla p 
  + \dfrac{4 \, Ma^2}{3 \, Re} \mu \nabla \left(\dfrac{1}{\rho} \dfrac{D p}{D t}\right) 
  + \dfrac{1}{Re} \mu \nabla \times \boldsymbol{\omega}\right] 
  \nonumber \\ &
  -  \frac{1}{\rho^2}\nabla{\rho}\times\left[\dfrac{2}{Re} \nabla \mu \cdot \nabla \mathbf{u} 
  +  \dfrac{2 \, Ma^2}{3 \, \rho \, Re} \dfrac{D p}{D t} \nabla \mu \right] 
  \nonumber \\  & 
 - \dfrac{1}{Re} \dfrac{1}{\rho} \nabla \mu  \times \left[ \dfrac{2 \, Ma^2}{3} \nabla \left( \dfrac{1}{\rho}\dfrac{D p}{D t}\right) 
 +\nabla \times \boldsymbol{\omega}\right] \nonumber \\
  & 
 + \dfrac{1}{Re} \dfrac{1}{\rho}  \nabla \times \left[ 2 \nabla \mu \cdot \nabla \mathbf{u} + \nabla \mu \times \boldsymbol{\omega} \right].  
 \label{eq:app_2D_vorticity}
\end{align}
This equation provides the departure point for analysing the contributions associated with each term on the right-hand-side of
Eq. (\ref{eq:vorticity_balance}) 
to the production/damping of vorticity discussed in Section 3.2. 


%

\end{document}